\documentclass[fleqn,usenatbib]{mnras}

\usepackage{newtxtext,newtxmath}

\usepackage[T1]{fontenc}

\DeclareRobustCommand{\VAN}[3]{#2}
\let\VANthebibliography\thebibliography
\def\thebibliography{\DeclareRobustCommand{\VAN}[3]{##3}\VANthebibliography}

\usepackage{graphicx}	
\usepackage{amsmath}	
\usepackage{amssymb}	

\usepackage{wasysym}    
\usepackage[normalem]{ulem} 
\usepackage{mathtools}  
\usepackage{tikz}       
\usetikzlibrary{shapes.geometric, shapes.misc, arrows}

\tikzstyle{default} = [rectangle, rounded corners, minimum width=3cm, minimum height=1cm,text centered, draw=black,align = center, fill=gray!20]
\tikzstyle{datacyl} = [cylinder, text centered, shape aspect=0.2, shape border rotate=90, minimum width = 3cm, minimum height=1cm,text centered, draw=black, fill=gray!20,align = center]

\title[Impact of the Primordial Stellar IMF on the 21-cm Signal]{Impact of the Primordial Stellar Initial Mass Function  on the 21-cm Signal}

\author[T. Gessey-Jones et al.]{T. Gessey-Jones,$^{1}$\thanks{E-mail: tg400@cam.ac.uk}
N. S. Sartorio,$^{2}$
A. Fialkov,$^{2,3}$
G. M. Mirouh,$^{4,5}$
M. Magg,$^{6,7}$
R. G. Izzard,$^{4}$\newauthor
E. de Lera Acedo,$^{1,3}$
W. J. Handley,$^{1,3}$
and R. Barkana$^{8,9,10}$
\\
$^{1}$Astrophysics Group, Cavendish Laboratory, J. J. Thomson Avenue, Cambridge, CB3 0HE, UK\\
$^{2}$Institute of Astronomy, University of Cambridge, Madingley Road, Cambridge, CB3 0HA, UK\\
$^{3}$Kavli Institute for Cosmology, Madingley Road, Cambridge, CB3 0HA, UK\\
$^{4}$Astrophysics Research Group, Faculty of Engineering and Physical Sciences, University of Surrey, Guildford, GU2 7XH, UK\\
$^{5}$Departamento de Física Teórica y del Cosmos, Universidad de Granada, Campus de Fuentenueva s/n, 18071, Granada, Spain \\
$^{6}$Universit\"at Heidelberg, Zentrum f\"ur Astronomie, Institut f\"ur Theoretische Astrophysik, D-69120 Heidelberg, Germany\\
$^{7}$International Max Planck Research School for Astronomy and Cosmic Physics at the University of Heidelberg (IMPRS-HD)\\
$^{8}$School of Physics and Astronomy, Tel-Aviv University, Tel-Aviv 69978, Israel \\ 
$^{9}$Institute for Advanced Study, 1 Einstein Drive, Princeton, New Jersey 08540, USA \\
$^{10}$Department of Astronomy and Astrophysics, University of California, Santa Cruz, California 95064, USA
}

\date{Accepted XXX. Received YYY; in original form ZZZ}

\pubyear{2022}

\begin{document}
\label{firstpage}
\pagerange{\pageref{firstpage}--\pageref{lastpage}}
\maketitle

\begin{abstract}
Properties of the first generation of stars (Pop~III), such as their initial mass function (IMF), are poorly constrained by observations and have yet to converge between simulations. The cosmological 21-cm signal of neutral hydrogen is predicted to be sensitive to Lyman-band photons produced by these stars, thus providing a unique way to probe the first stellar population. 
In this paper, we investigate the impacts of the Pop~III IMF on the cosmic dawn 21-cm signal via the Wouthuysen-Field effect, Lyman-Werner feedback, Ly\,$\alpha$ heating, and CMB heating. We calculate the emission spectra of star-forming halos for different IMFs by integrating over individual metal-free stellar spectra, computed from a set of stellar evolution histories and stellar atmospheres, and taking into account variability of the spectra with stellar age. Through this study, we therefore relax two common assumptions: that the zero-age main sequence emission rate of a Pop~III star is representative of its lifetime mean emission rate, and that Pop~III emission can be treated as instantaneous.
Exploring a bottom-heavy, a top-heavy, and intermediate IMFs, we show that variations in the 21-cm signal are driven by stars lighter than $20$\,M$_{\astrosun}$. For the explored models we find maximum relative differences of $59$ per cent in the cosmic dawn global 21-cm signal, and $131$ per cent between power spectra.
Although this impact is modest,  precise modelling of the first stars and their evolution is necessary for accurate prediction and interpretation of the 21-cm signal.
\end{abstract}

\begin{keywords}
dark ages, reionization, first stars -- stars: Population III -- early Universe -- software: simulations
\end{keywords}



\section{Introduction}\label{sec:intro}

In 2018 the EDGES collaboration reported a detection of a sky-averaged (global) radio signal at 78 MHz which could be interpreted as the cosmological 21-cm signal of neutral hydrogen from $z\sim17$ \citep{EDGES}. If this signal truly is of cosmological origin, it provides evidence of star formation at cosmic dawn  \citep[e.g. ][]{Schauer_2019, Mirocha_2019, Fialkov_19, Reis_20, Mebane:2020}. Debate remains, however, over the true nature of the detected signal \citep{Hills_2018,Singh_2019,Bradley_2019, Sims_2020, SARAS3}. A multitude of low-frequency radio telescopes \citep{SARAS2, PRIZM, LOFAR, LEDA, HERA, NenuFar, Rahimi:2021, SARAS3} are attempting to detect or bound the sky-averaged 21-cm signal or the power spectrum of its variations in the hope of supporting or refuting the EDGES measurement. 

The numerous efforts to detect the 21-cm signal are reflective of the fact that the signal is expected to be a rich source of information about the Universe between recombination and reionization \citep[e.g.][]{Madau_1997, Furlanetto_2006,Pritchard_2012,Barkana_2016,Mesinger_2019}. The insights into the evolution and variation in the properties of the intergalactic medium (IGM) provided by  the 21-cm signal will, in turn, probe physical processes that impact the IGM during this period, such as the Wouthuysen-Field coupling \citep[WF,][]{Wouthuysen_1952,Field_1958}, heating by X-ray sources \citep[e.g.][]{Fialkov_2014, Pacucci:2014}, ionization \citep[e.g.][]{Park:2020}, and coupling between dark matter and gas \citep[e.g. ][]{Barkana:2018, Munoz_2018, Liu:2019}, as well as test the existence of any radio background in addition to the Cosmic Microwave Background (CMB) present at these high redshifts \citep[e.g. ][]{Feng_18, Ewall_18, Fialkov_19, Reis_20}.

The first generation of stars, commonly referred to as the population III (Pop~III) stars \citep{Bond_1981, Bromm_2004, Klessen:2019}, are expected to have a strong effect on the high-redshift IGM, and, thus, the 21-cm signal from cosmic dawn \citep{Cohen:2016, Mirocha_2018,Mebane_2018, Tanaka_2018,  Schauer_2019, Mebane:2020, Munoz_2021, Tanaka_2021, Hibbard:2022}. It has been shown that the efficiency and rate of star formation for the Pop~III stars strongly impact the depth of the 21-cm global signal absorption trough~\citep{Yajima_2015,Cohen_2018,Mebane_2018,Mirocha_2019,Schauer_2019,Chatterjee_2020,Munoz_2021,Reis_2021b}. Furthermore, recent studies of the transition between metal-free Pop~III star formation and the later generation of metal-poor Pop~II stars have predicted characteristic signatures of this transition in the  21-cm signal \citep{Mirocha_2018,Magg_2021}. The sensitivity of the 21-cm signal to the amount of, and the duration of, Pop~III star formation comes about due to the several mechanisms through which Pop~III stars affect the surrounding IGM. 

The Lyman-band emission, between Ly\,$\alpha$ and the Lyman limit, of these first stars gives rise to the WF effect:
Lyman line photons scattering off of atomic hydrogen couple the 21-cm spin temperature of hydrogen and the kinetic temperature of the IGM.
This effect is dominated by the Ly\,$\alpha$ line \citep[e.g. ][]{Pritchard_2006}. Such Ly\,$\alpha$ photons are both produced by redshifting of the continuum photons emitted between Ly\,$\alpha$ and Ly\,$\beta$, and as injected photons created in the decay of excited hydrogen atoms. Because Lyman-band photons emitted between Lyman lines contribute to the WF effect once they have redshifted into a Lyman line, the emission of the first stars enables the WF effect in a finite region around each star creating Ly\,$\alpha$  coupled bubbles \citep{Reis_2021}. Inside these bubbles the spin temperature tends to the IGM temperature rapidly, resulting in stronger absorption in the 21-cm signal within the bubble.
Consequently, a deeper global absorption signal and a corresponding peak in the power spectrum are expected to be seen around cosmic dawn as these first Ly\,$\alpha$ coupled bubbles form. 

In addition to the WF effect, scattering of the Lyman line photons transfers energy from these photons, and from the CMB, to the IGM, thus raising its kinetic temperature in processes called Ly\,$\alpha$ heating \citep{Madau_1997,Chuzhoy_2006, Reis_2021} and CMB heating \citep{Venumadhav_2018} respectively. Both of these significantly impact the IGM temperature and, thus, the 21-cm signal in scenarios where X-ray heating is weak \citep[e.g. ][]{Chuzhoy_2006, Ciardi:2010, Venumadhav_2018, Mittal_2021,  Reis_2021}.

Furthermore, higher energy Lyman-band emission of the Pop~III stars contributes to the Lyman-Werner (LW) feedback \citep{Haiman_2000}. LW photons have energies between $11.2$\,eV and $13.6$\,eV and can disassociate molecular hydrogen in the Solomon process \citep{Stecher_1967,Glover_2001}
\begin{equation}
\textrm{H}_2 + \gamma \rightarrow \textrm{H}_2^* \rightarrow 2\textrm{H}.   
\end{equation}
Since molecular hydrogen is the primary coolant of gas in the first star-forming halos  \citep{Haiman_1996}, a strong LW background has the potential to suppress this cooling, thus delaying further star formation and reducing the impact of the Pop~III stars on the 21-cm signal.  

Higher-energy UV photons emitted by the first stars will also ionize the hydrogen gas in a finite region around the stars, eliminating the 21-cm signal from this region \citep{Tanaka_2021}. However, at cosmic dawn, the impact of ionization on the large-scale 21-cm signal is expected to be subdominant. Furthermore, Pop~III stars keep affecting the IGM  even after the end of their lives via X-rays \citep{Ricotti_2004, Ricotti_2005, Yajima_2015} produced in supernovae explosions \citep{Jana_2019} and by Pop~III remnant X-ray binaries (XRBs). The X-ray luminosity of both Pop~III (Sartorio et al. in preparation) and Pop~II   \citep{Fragos:2013} XRBs is anticipated to be significant, potentially providing the dominant source of heating that drives the 21-cm signal into emission \citep{Fialkov_2014, Pacucci:2014}. 

Due to the multitude of processes through which the Pop~III stars affect the IGM and, thus, the 21-cm signal, modelling the high-redshift stellar populations is key to making realistic predictions for the 21-cm signal from cosmic dawn. However, much about these stars remains unknown with numerical simulations \citep{Hirano_2014,Greif_2015,Stacy_2016,Hirano_2017,Susa_2019,Sugimura_2020,Wollenberg_2020,Sharda_2020,Sharda_2021} and observational constraints \citep{de_Bennassuti_2014,Frebel_2015, Hartwig_2015,Fraser_2017,de_Bennassuti_2017,2018_Ishigaki, Magg_2019, Tarumi_2020} having yet to converge on the mass distribution of the first stars, i.e. their initial mass function (IMF). For now, the typical mass scale of the first generation of stars remains uncertain. 

In this paper, we aim to investigate how the uncertainties in the Pop~III stellar IMF propagate into the uncertainties in the predicted 21-cm signal, taking into account the dominant effects at cosmic dawn, namely the WF effect, Ly\,$\alpha$ heating, CMB heating, and Lyman-Werner feedback. As we are focusing on cosmic dawn, we ignore the impact of X-ray and ionizing photons, which are expected to have a delayed effect on the 21-cm signal \citep[e.g.][]{Cohen_2018}. In addition, we limit our investigation to $z \geq 12$, as at lower redshifts emission from Pop~II stars is likely to dominate the Lyman radiation fields and effects of X-ray and UV photons are expected to be significant. Throughout this work, we model the Pop~III to Pop~II transition in our simulations using the methodology of \citet{Magg_2021}. Importantly, through this investigation, we  relax two assumptions previously made in semi-numerical 21-cm simulations by our and other groups, namely:
that (i) the zero-age main sequence (ZAMS) emission rate of a Pop~III star is approximately equal to its lifetime mean emission rate, and that (ii) the lifetime of the Pop~III stars are short compared to cosmological time scales and so their emission can be modelled as instantaneous. Instead, we calculate the emission spectra of star-forming halos for different IMFs by integrating over individual metal-free stellar spectra, computed from a set of stellar evolution histories and
stellar atmospheres. We also implement finite stellar lifetimes in our modelling and take into account the variation of stellar spectra with age.

We begin in Sec.~\ref{sec:21cm} with a brief outline of both the theory of the 21-cm signal and our semi-numerical simulation code. Then in Sec.~\ref{sec:popIII_model} we discuss the assumptions previously made in 21-cm signal studies and detail our methodology for an improved model of Pop~III stellar emission spectra.
In Sec.~\ref{sec:radiation_fields}, we present our calculation of the Ly\,$\alpha$ and LW radiation fields, which are later used in our simulations to calculate the strength of the WF effect, LW feedback, Ly\,$\alpha$ heating, and CMB heating. In addition, we describe our new developments that extend these calculations to include the effect of arbitrary IMFs and the impact of finite stellar lifetimes. Using our extended radiation field calculation, in Sec.~\ref{sec:results} we investigate how important it is to properly model Pop~III star emission rates and lifetimes, before assessing the variation of the 21-cm signal with the Pop~III IMF.
We conclude in Sec.~\ref{sec:conc}.

\section{The 21-cm Signal}
\label{sec:21cm}

The 21-cm signal is produced by neutral hydrogen gas in the IGM.   At the hyperfine level, the $n = 1$ ground state of atomic hydrogen splits into a lower energy singlet state and a higher energy triplet of states. An atom undergoing the forbidden transition between these levels emits or absorbs a photon with a wavelength of 21\,cm. Whether a collection of hydrogen atoms will on net emit or absorb 21-cm photons is determined by the relative occupancy of the hyperfine levels typically quantified by the statistical spin temperature $T_{\textrm{S}}$ \citep{Scott_1990}:
If $T_{\textrm{S}}$ exceeds the background radiation temperature $T_\gamma$ (usually assumed to be the CMB temperature\footnote{However, models with excess radio background have been extensively discussed \citep[e.g. ][]{Feng_18, Ewall_18, Fialkov_19,  Reis_20}.}), then atomic hydrogen is a net emitter, otherwise hydrogen will on net absorb photons from the background radiation. 

Before reionization, the IGM is primarily composed of neutral atomic hydrogen and so 21-cm absorption/emission occurs throughout the Universe. Furthermore, due to the expansion of the Universe any photons emitted at the 21-cm line are over time redshifted out of the frequency range of the transition preventing re-absorption. Hence, the emission or absorption of 21-cm photons by the IGM at redshift $z$ can be seen today at a frequency
$  \nu_{\textrm{21,~obs}} = \nu_{\textrm{21}}/\left(1+z\right)$,
where $\nu_\textrm{21} = 1420$\,MHz is the intrinsic frequency of the 21-cm transition. 

Commonly, the excess or deficit of photons at frequency $\nu_{\textrm{21,~obs}}$ is quantified in terms of the differential brightness temperature, $T_\textrm{b}$, which measures  the difference between the observed radiation temperature at $\nu_{\textrm{21,~obs}}$ and the expected temperature of the background radiation.  By solving the equations of radiative transfer in an expanding universe, $T_\textrm{b}$ can be related to $T_{\textrm{S}}$ and $T_\gamma$ at the corresponding redshift 
\begin{equation}
T_{\textrm{b}}  = \left(1 - e^{-\tau_{\textrm{21}}} \right)\frac{T_{\textrm{S}} - T_\gamma}{1+ z}    ,
\label{eqn:Tb_equation}
\end{equation}
where $\tau_{\textrm{21}}$ is the 21-cm radiation optical depth at that redshift
\begin{equation}
\tau_{\textrm{21}} = \frac{3}{32 \pi} \frac{h c^3 A_\textrm{10}}{k_\mathrm{B} \nu_\textrm{21}^2} \left[\frac{x_{\textrm{HI}} n_{\textrm{H}}}{(1+z)^2 (dv_\parallel/dr_\parallel)} \right]\frac{1}{T_{\textrm{S}}},
\label{eqn:21cm_optical_depth}
\end{equation}
with $x_{\textrm{HI}}$ being the fraction of hydrogen in its atomic state, $n_{\textrm{H}}$ the number density of hydrogen, $A_\textrm{10} = 2.85 \times 10^{-15}$\,s$^{-1}$ the spontaneous emission rate of the 21-cm transition, and $dv_\parallel/dr_\parallel$ the proper velocity gradient along the line of sight including the Hubble flow.

The spin temperature at a given redshift is determined by the balance between the competing influences of the background radiation temperature and gas kinetic temperature, $T_\textrm{K}$. The former couples to the spin temperature via scattering of background photons, and the latter via collisional coupling and the WF effect. From the relative coupling strengths of these three processes ($x_{\textrm{CMB}}$, $x_\textrm{c}$, and $x_\alpha$) the spin temperature can be determined \citep{Furlanetto_2006,Venumadhav_2018}
\begin{equation}
    \frac{1}{T_{\textrm{S}}} = \frac{x_{\textrm{CMB}} T_\gamma^{-1} + x_\textrm{c} T_\textrm{K}^{-1} + x_\alpha T_\textrm{C}^{-1}}{x_{\textrm{CMB}} + x_\textrm{c} + x_\alpha},
\end{equation}
where  $T_\textrm{C}$ is the radiation temperature of the Ly\,$\alpha$ photons which mediate the WF effect.

Current generation experiments trying to detect the 21-cm signal typically fall into two categories, global signal experiments and power spectrum experiments. Global signal experiments attempt to detect the sky-averaged $\langle T_\textrm{b} \rangle$, whereas power spectrum experiments aim to detect the spatial variations of the 21-cm signal quantified in the 21-cm power spectrum $P_{\textrm{21}}(k,z)$ 
\begin{equation}
    \left\langle \tilde{T}_\textrm{b}\left(\mathbfit{k},z\right) \tilde{T}^*_\textrm{b}\left(\mathbfit{k}',z\right) \right\rangle = (2 \pi)^3 \delta^D\left(\mathbfit{k} - \mathbfit{k}'\right) P_{\textrm{21}}(k,z),
\end{equation}
with $\delta^D(\mathbfit{k} - \mathbfit{k}')$ being the 3D Dirac delta-function and $k$ the comoving wavevector. Throughout this paper we shall use the alternative form of the power spectrum
\begin{equation}
    \Delta ^2(k,z) =  \frac{k^3 P_{\textrm{21}}(k,z)}{2 \pi^2}.
\end{equation}

In this work we utilize our own semi-numerical code to compute the expected 21-cm signal \citep[e.g. ][ and other related works]{Visbal_2012, Fialkov_2014, Cohen:2016, Reis_2021}. Starting from initial conditions for density, velocity, and temperature fields set over large cosmic volumes of $384^3$ comoving Mpc$^3$, the simulation evolves radiative backgrounds and computes cubes of $T_{\textrm{b}}$ at specified redshifts. From these cubes, the global signal and power spectrum can then be evaluated. The simulation volume is split into $128^3$ cubic cells with side lengths of 3 comoving Mpc, containing on average $1.65 \times 10^{11}$\,M$_{\astrosun}$ of baryonic mass and $9.02 \times 10^{11}$\,M$_{\astrosun}$ of dark matter. Within each pixel halo formation is analytically modelled using the approach of \citet{Barkana_2004}, a hybrid between the \citet{Press_1974} and \citet{Sheth_1999} prescriptions. The halo mass function within each cell depends on the local  large-scale density and velocity fields \citep{Fialkov_2012, Visbal_2012}. Using a sub-grid prescription for star formation, the pixel-level halo distributions are then converted into star formation rates and ultimately radiation fields, which we describe in more detail in Sec.~\ref{ssec:sfr} and Sec.~\ref{sec:radiation_fields}.

As inputs the 21-cm simulation accepts values of several parameters which parameterize fundamental astrophysical processes driving the evolution of the 21-cm signal \citep[see][for recent summary]{Cohen_2020, Reis_2021}. Unless otherwise stated, in all of our simulation runs presented here we use a minimum virial circular velocity of halos for star formation of \mbox{$V_\mathrm{c} = 4.2$\,km\,s$^{-1}$} (corresponding to the molecular cooling threshold), ionizing efficiency of galactic UV emission of $\zeta = 15$, strong LW feedback \citep[see][for details]{Fialkov_2013}, and a mean free path of ionizing photons of $R_{\textrm{mfp}} = 50$\,cMpc (comoving Megaparsec). In addition, we assume no X-ray heating (by setting X-ray emission efficiency $f_X=0$); however, all the simulations include both Ly\,$\alpha$ heating and CMB heating  \citep[following the implementation of][respectively]{Reis_2021, Fialkov_19}. Furthermore, in our simulations, we model baryon dark matter relative velocities \citep{Tseliakhovich_2010, Fialkov_2012, Visbal_2012}, assume that the Pop~II star formation efficiency in molecular cooling halos (with $4.2<V_\mathrm{c}<16.5$\,km\,s$^{-1}$) is logarithmically suppressed  \citep{Fialkov_2013},  model Ly\,$\alpha$ multiple scattering and photoheating \citep{Reis_2021, Cohen:2016}, do not take Poisson fluctuations \citep[as introduced in][]{Reis_2021b} into account, and assume the CMB is the background radiation. Our assumptions for the stellar IMF are explained in the next section.

\section{Modelling Lyman Emission of  Pop~III Stars}\label{sec:popIII_model}

\begin{figure}
    \centering
    \begin{tikzpicture}[node distance=1.5cm]
    \node (mesa) [default] {Model evolution of 100 individual metal-free\\ stars using \textsc{mesa} (Sec.~\ref{ssec:popIII_evolution}).};
    \node (tlusty) [default, below of=mesa] {Compute stellar atmospheres and spectra for each \\ metal-free star using \textsc{tlusty} (Sec.~\ref{ssec:popIII_atmospheres}).};
    \node (synth) [default, below of=tlusty] {Combine evolution histories and spectra to find \\ lifetime photon-number emission spectra of each \\ metal-free star $\varepsilon(\nu;M)$ (Sec.~\ref{ssec:popIII_spectra}).};
    \node (data) [datacyl, yshift = -1cm, below of=synth] {Store lifetime spectra as lookup-table \\  available at  \href{https://zenodo.org/record/5553052}{10.5281/zenodo.5553052}.};
    \node (21cm) [default, yshift = -0.4cm, below of=data] {Integrate over $\varepsilon(\nu;M)$ weighted by the IMF \\ to compute the Pop~III photon emissivity per stellar baryon \\ $\varepsilon_\textrm{b}(\nu)$   (Sec. \ref{sec:radiation_fields}).};
    \draw [-stealth] (mesa) -- (tlusty);
    \draw [-stealth] (tlusty) -- (synth);
    \draw [-stealth] (synth) -- (data);
    \draw [-stealth] (data) -- (21cm);
    \end{tikzpicture}
    \caption{Schematic diagram outlining the key steps in our computation of the Pop~III Lyman-band photon emissivity $\varepsilon_\textrm{b}(\nu)$ for an arbitrary IMF. }
    \label{fig:flow}
\end{figure}
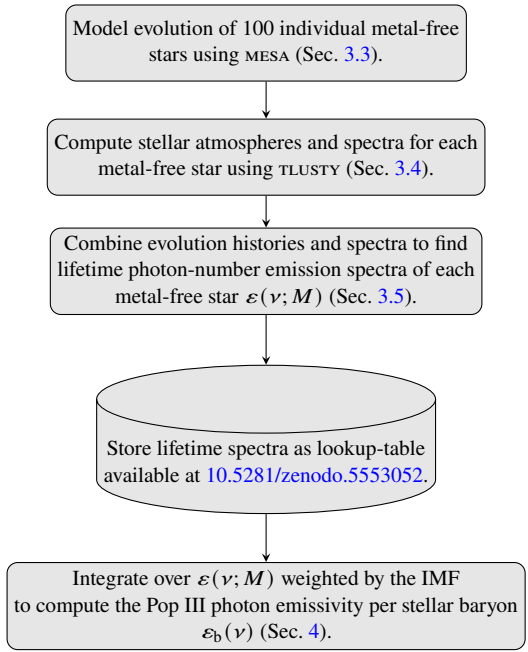

In this paper, we focus on the impact of the Pop~III star IMF on the 21-cm signal at the onset of star formation. As we are concentrating on cosmic dawn, we only consider effects introduced by Lyman-band photons and ignore the impact of ionizing and X-ray radiative backgrounds which we plan to model in future work. The processes triggered by the Lyman-band photons explored here (including WF effect, LW feedback, as well as Ly\,$\alpha$ and CMB heating) depend on the details of the Pop~III emission spectra in the Lyman band. Therefore, it is essential to model the impact of the IMF on the stellar spectra. Such a capability was lacking in our 21-cm code and we introduce it here.  The calculation is outlined in Fig.~\ref{fig:flow} and more details are provided in the rest of this section. In summary, we first simulated the stellar evolution history for an ensemble of individual metal-free stars of different masses using the \textsc{mesa} stellar evolution code version 12115 \citep[][and references therein]{mesa5} following the method outlined in \citet{mirouh_etal22}. We, then, computed the emission spectra of those individual stars throughout their lives using the stellar atmosphere code  \textsc{tlusty} version 205  \citep{TLUSTY0, TLUSTYI, TLUSTYII, TLUSTYIII, TLUSTYIV}. From these spectra and stellar evolution histories we synthesised a dataset of lifetime Lyman-band emission spectra for individual metal-free stars. We then integrate over these individual emission spectra weighted by the assumed IMF to compute the Lyman-band emission spectra per stellar baryon for the desired IMF. The resultant spectra are used in the 21-cm signal simulation to calculate the global signal and the large-scale power spectrum. 

\subsection{Initial mass function}\label{ssec:IMF}

The mass distribution of stars is commonly quantified in terms of the IMF, i.e. the number distribution of stellar masses ($dN/dM$) at ZAMS. Since we aim to investigate how 21-cm signal predictions vary with the IMF of the primordial stars, we upgraded our 21-cm code to admit any arbitrary IMF of Pop~III stars, and demonstrate the results for four example IMFs each of which is a truncated power-law
\begin{equation}
    \frac{dN}{dM} \propto M^{\alpha} \qquad M_{\textrm{min}} \leq M \leq M_{\textrm{max}},
\end{equation}
where $N$ is the number of stars with a given mass $M$, and $\alpha$ is the exponent of the power-law. As their names suggest, $M_{\textrm{min}}$ and $M_{\textrm{max}}$ bracket the range of possible stellar masses. The model parameters ($\alpha$, $M_{\textrm{min}}$, and $M_{\textrm{max}}$) for each one of the four example IMFs are given in Table~\ref{tab:fid_imfs}, along with the  characteristic  mass scale, $M_{50\%}$, defined so that 50 per cent of stellar mass is in stars of lower masses. Also shown for each IMF is its mean stellar mass $\bar{M}$, mass-weighted mean stellar lifetime $\bar{t}_{\textrm{life}}$, and the redshift $\bar{z}_{\textrm{death}}$ at which a star born at $z = 20$ and living for $\bar{t}_{\textrm{life}}$ would die.

\begin{table}
	\centering
	\begin{tabular}{cccccccc}
		\hline
		IMF &  $\alpha$ & $M_{\textrm{min}}$   & $M_{\textrm{max}}$    &  $M_{50\%}$   & $\bar{M}$ & $\bar{t}_{\textrm{life}}$ & $\bar{z}_{\textrm{death}}$\\
		\hline
		Salpeter & -2.35 & 0.8 & 250 & 4.1 & 2.7 & 1320 & 4.08\\
		Log-Flat & -1.0 & 0.8 & 150  & 75& 28.5 & 36 & 17.57\\
		Intermediate & -0.5 & 2.0 & 180 & 113& 67 & 4.3& 19.67\\
		Top-Heavy & 0.0 & 10 & 500 & 354& 255 & 2.2& 19.83\\
		\hline
	\end{tabular}
		\caption{Parameters  of the four example IMFs  used throughout the paper: slope $\alpha$, $M_{\textrm{min}}$ in units of M$_{\astrosun}$, and $M_{\textrm{max}}$ in M$_{\astrosun}$. For each IMF we also provide: the threshold mass scale, $M_{50\%}$, for which 50 per cent of stellar mass is in stars of lower masses, calculated in units of M$_{\astrosun}$; the mean stellar mass $\bar{M}$ in M$_{\astrosun}$; the mass-weighted mean stellar lifetime $\bar{t}_{\textrm{life}}$ in Myr; and the redshift $\bar{z}_{\textrm{death}}$ at which a star born at $z = 20$ and living for $\bar{t}_{\textrm{life}}$ would die. These values were calculated assuming the best-fit Planck~2018 $\Lambda$CDM cosmology \citep{Planck_VI}.}
			\label{tab:fid_imfs}
\end{table}

The four IMFs in Table \ref{tab:fid_imfs} have very different characteristics and are intended to span the large uncertainty in the properties of Pop~III stars and the resulting 21-cm signals. The Salpeter \citep{Salpeter_1955} IMF represents an extreme case in which the majority of stars are low-mass, with half of its stellar mass being in stars lighter than $4.1$\,M$_{\astrosun}$. Such stars are very long-lived even by cosmological standards, with the mass-weighted mean stellar lifetime of the Salpeter IMF being 1320 Myr, and the lowest-mass stars it contains having lifetimes of up to 14 billion years \citep{Marigo_2001}. In other words, if a population of such stars was formed at $z = 20$ the majority of the stellar mass would be in stars that survive until at least $z \approx 4.1$. The Salpeter IMF is similar to the one observed for the present-day Pop~I stars in the Milky Way. 

The Top-Heavy IMF represents the opposite extreme case with half of its stellar mass in stars heavier than 354 \,M$_{\astrosun}$. Such massive stars are short-lived  with a typical lifetime of just 2.2 Myr, corresponding to $\Delta z \approx 0.17$ at $z = 20$. We limit our investigation to  stars of at most $500$\,M$_{\astrosun}$, as present evolution models suggest that more massive  Pop~III  stars may experience an early onset of photo-disintegration in their core \citep{Heger_2003,Woosley_2007,Farmer_2019} and, thus, will not contribute significantly to the Lyman output.

The real Pop~III IMF is unknown and will likely lie between these two intentionally extreme cases. Therefore, we also explore the Log-Flat and the Intermediate IMF cases, which represent plausible Pop~III IMFs suggested by simulations \citep{Greif_2011,Dopcke_2013} and stellar archaeology studies of metal mixing \citep{Tarumi_2020} respectively. These two IMFs are composed of relatively massive stars  with   $M_{50\%} = 75$\,M$_{\astrosun}$ and $M_{50\%} = 113$\,M$_{\astrosun}$, respectively, with rather   short typical lifetimes of $\bar{t}_{\textrm{life}} =36$ Myr (corresponding to $\Delta z \approx 2.4$ at $z=20$) and $\bar{t}_{\textrm{life}}=4.3$ Myr  ($\Delta z \approx 0.33$ at $z=20$).

In this work, we model a gradual transition between Pop~III and Pop~II stars, see Sec.~\ref{ssec:sfr} for further details. When adding the contribution of Pop~II stars to the Lyman-band emission we follow our previous works and assume a Scalo IMF \citep{Scalo_1998} using the stellar spectra for stars of 1/20 solar metallicity computed in \citet{Leitherer_1999}.

\subsection{Assumptions in historic modelling of Pop~III stars}\label{ssec:popIII_historical}

Many previous 21-cm semi-numerical simulations \citep{Mesinger_2011,Visbal_2012,Fialkov_2014,Reis_20} have utilized the methods and results of \citet{Barkana_2005} for modelling the Lyman-band emission of the Pop~III stars, which are in turn based upon the stellar spectra calculated in \citet{Bromm_2001}. We refer to the combination of these methods and results as our {\it Legacy} model of Pop~III star emission. Built into this model are two assumptions that we need to reconsider when implementing a more general model for Pop~III IMFs: Firstly, it is usually assumed that the ZAMS emission rate of a Pop~III star is representative of its average lifetime emission rate (the {\it ZAMS assumption}). The second assumption is that the lives of the stars are short compared to the cosmological timescales of interest, and, thus, that the emission of the Pop~III stars can be taken as instantaneous (the {\it instantaneous emission} assumption). 

As we show below, to relax the first assumption we evolve stellar spectra throughout their lives. Our motivation is rooted in an early work by  \citet{Schaerer_2002} who showed that the ZAMS assumption could introduce errors into the computation of the ionizing fluxes of Pop~III stars. This claim suggests that similar errors may be introduced into the computation of Pop~III Lyman-band emission. If so, any under/overestimate would in turn propagate through our calculations of the Ly\,$\alpha$ and LW radiation fields into the magnitude of the various Lyman band mediated effects, and, thus, affect the 21-cm signal. 

We relax the instantaneous emission assumption in Sec.~\ref{ssec:non_instant_emission}.  In the derivation of our Legacy model, it was assumed that all Pop~III stars were $100$ solar masses. Such stars are expected to have lifetimes of $\approx 2.7$\,Myr \citep{Schaerer_2002} which indeed is a small fraction of the age of the Universe at $z = 20$, $180$\,Myr \citep{Wright_2006}. In this work, however, we are interested in a broad range of Pop~III stellar masses (between $0.8$\,M$_{\astrosun}$ and $500$\,M$_{\astrosun}$). As is shown in Table \ref{tab:fid_imfs}, the typical lifetime can be very long for the IMFs dominated by low-mass stars, and so the instantaneous emission assumption is expected to be only valid for IMFs dominated by very massive stars.

\subsection{Evolution of Pop~III stars}\label{ssec:popIII_evolution}

To calculate the Lyman-band emission from a population of primordial stars with a given IMF, we first need to calculate the Lyman-band emission spectra for individual metal-free stars with masses in the range between $M_{\textrm{min}}$ and $M_{\textrm{max}}$. Since stellar spectrum evolves over the lifetime of a star \citep{Schaerer_2002}, we need to trace the evolutionary histories of individual metal-free stars of different masses and then sum up the spectra weighted by IMF.

\begin{figure}
	\includegraphics[width=\columnwidth]{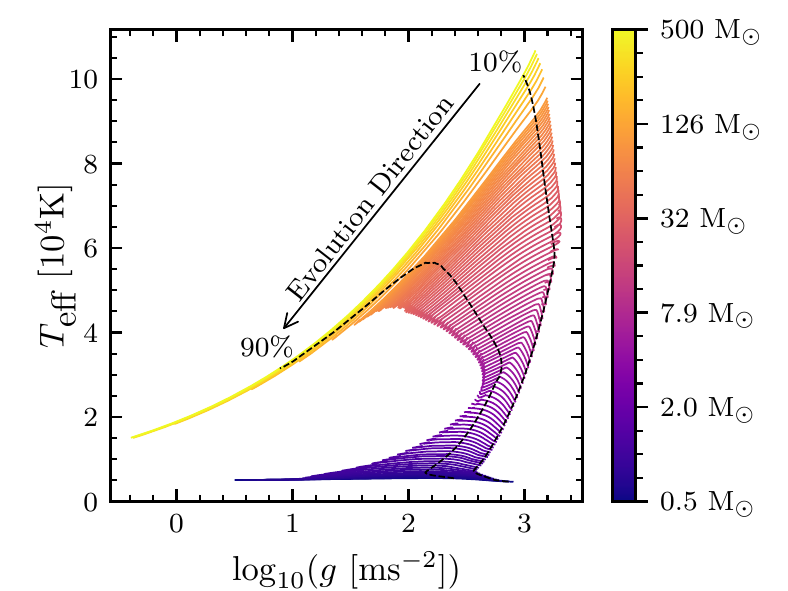}
    \caption{Evolutionary tracks of individual initially metal-free stars projected into effective temperature $T_{\textrm{eff}}$ vs logarithm of surface gravity $g$ space. Each line is colour-coded with the stellar mass (see colour-bar on the right). For stars of mass less than $310$\,M$_{\astrosun}$ these evolution tracks stretch from ZAMS to central hydrogen depletion, while for more massive stars the evolution tracks end when the star begins to photoevaporate. Also illustrated as black dashed lines are contours indicating 10 per cent and 90 per cent of the time along the evolutionary track. These serve to highlight the evolutionary direction of the stars and the portion of their evolutionary tracks that form the majority of their lives. We thus observe that the evolutionary tracks show that lighter stars initially increase in temperature before cooling toward the end of their lives, whereas for more massive stars temperature monotonically decreases. }
    \label{fig:evo_tracks}
\end{figure}

We use evolutionary histories for 100 metal-free stars with fixed, logarithmically-spaced masses $M$
ranging from $0.5$\,M$_{\astrosun}$ to $500$\,M$_{\astrosun}$ each simulated using the Modules for Experiments in Stellar Astrophysics
\textsc{mesa} stellar evolution code version 12115 \citep[][and references therein]{mesa5}. The evolution tracks of individual initially metal-free stars are illustrated in Fig.~\ref{fig:evo_tracks} where we show the effective temperature $T_{\textrm{eff}}$ versus the logarithm of surface gravity $g$ of the star. Technical details on the internal physics are provided in \citet{mirouh_etal22}. 

The stars were initialized with Big Bang nucleosynthesis (BBN) proportions of hydrogen and helium and then relaxed onto the ZAMS. The stars were then evolved assuming they were non-rotating, in isolation, and had no mass loss. The absence of metals results in these stars having peculiar nuclear reactions which make them
hotter and brighter \citep[for more details, we refer the reader to][]{Marigo_2001, lmurphy21}. Pop~III stars  thus exhibit shorter main-sequence lifetimes with respect to their more metallic counterparts.

From ZAMS stars with masses $M \leq 310$\,M$_{\astrosun}$ were simulated until hydrogen depletion in their core, after which the stars are expected to begin helium burning and expand, becoming red or blue giants. Truncating these stellar evolution tracks at hydrogen depletion is not anticipated to greatly impact the lifetime Lyman-band emission calculated for the stars, due to the giant branch representing a comparatively small portion of the life of a star immediately before its death. Our most massive stars ($M > 310$\,M$_{\astrosun}$) were instead evolved until the star began to photoevaporate. The assumption that the emission of massive stars while they are photoevaporating is negligible is not initially well motivated. Instead, this assumption is retroactively supported by our findings in Sec.~\ref{ssec:popIII_spectra} that for such stars the Lyman-band emission rapidly decreases before they begin photoevaporating. This is attributed to the effective temperature of these stars sharply dropping as their outer layers expand, for a more detailed discussion see appendix~\ref{app:app}.

\subsection{Stellar atmospheres and fluxes}\label{ssec:popIII_atmospheres}

\begin{figure}
	\includegraphics[width=\columnwidth]{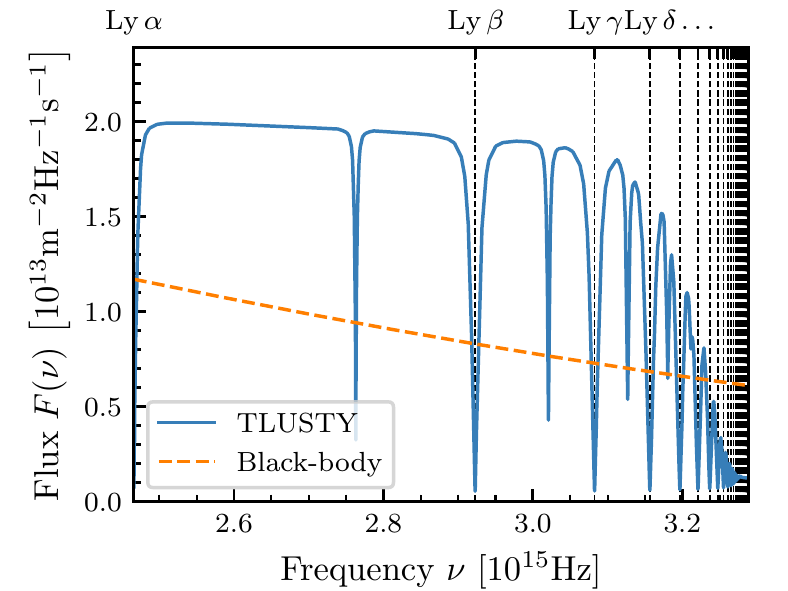}
    \caption{Comparison of stellar Lyman-band fluxes calculated using the black-body approximation (dashed orange) and the \textsc{tlusty} stellar atmosphere code (solid blue). Here we have taken $T_{\textrm{eff}} = 32,692$\,K and $\log(g) = 2.77$ (where $g$ is in units of  $\textrm{m\,s}^{-2}$), and dashed vertical lines indicate the Lyman lines. The black-body approximation fails to account for both  the shape and the magnitude of the stellar spectrum. Hence, in the rest of our work we use \textsc{tlusty} calculated stellar spectra. }
    \label{fig:BB_vs_NLTE}
\end{figure}

From the stellar evolutionary histories shown in Fig. \ref{fig:evo_tracks}, we need to calculate the spectral flux of the Pop~III stars at each point in their lives. Historically, two approaches have been used to address similar problems: either the spectra of the stars were assumed to have a black-body shape with effective temperature $T_\textrm{eff}$ \citep{Windhorst_2018}, or a stellar atmosphere code was used to calculate a detailed stellar spectrum \citep{Bromm_2001,Schaerer_2002}. A comparison of these two approaches is shown in Fig.~\ref{fig:BB_vs_NLTE}, from which we see that the black-body approximation does not match either the shape or the magnitude of the detailed spectra computed with a stellar atmosphere code. We, hence, adopted the latter approach since an accurate magnitude and shape of the Lyman-band emission is required for determining the strength and spatial distribution of the Ly\,$\alpha$ mediated effects of interest.

For our calculations, we employed the \textsc{tlusty} \citep{TLUSTY0,TLUSTYI,TLUSTYII,TLUSTYIII,TLUSTYIV} code version 205 to compute stellar atmospheres and the resulting spectra. This code was chosen due to it previously being successfully employed to model high-mass metal-free stars  \citep{Schaerer_2002,Lanz_2003,Lanz_2007}. To aid convergence of the stellar atmospheres we adopted the iterative approach of \citet{Auer_1969} calculating first an atmosphere with local thermodynamic equilibrium (LTE), then a non-local thermodynamic equilibrium (NLTE) atmosphere without lines, and finally an NLTE atmosphere with lines, with each step adopting as its initial atmosphere model the result of the previous step. 

Each \textsc{tlusty} atmosphere computation requires three physical parameters:  $T_{\textrm{eff}}$,  $\log(g)$, and chemical abundances. For the latter we adopt the BBN proportions of hydrogen and helium in all of our stellar atmospheres computations, because our \textsc{mesa} simulations predict negligible surface metallicity for all considered stars. This negligible metal mixing to the stellar surface is due to the structure of metal-free stars being primarily composed of radiative layers. All other \textsc{tlusty} settings, that determine the numeric routines used in the computations, were fixed to the fiducial values from \citet{Lanz_2003} for spectra at $T_{\textrm{eff}} > 35,000$\,K  and those of \citet{Lanz_2007} for spectra at $T_{\textrm{eff}} < 35,000$\,K.  

We computed the stellar flux for atmospheres on a 400 by 240 regular rectangular grid covering $T_{\textrm{eff}}$ from 4,000\,K to 110,000\,K and $\log(g)$ from -0.5 to 5.5. These values were chosen to fully cover the stellar evolutionary histories previously shown in Fig.~\ref{fig:evo_tracks}. Stellar fluxes at intermediate $T_{\textrm{eff}}$ and $\log(g)$ were then computed using linear  interpolation over the grid of fluxes (in log space). We verified by computing intermediate spectra using \textsc{tlusty} that this interpolation results in a root-mean-square error of $\leq 0.5$ per cent in the final stellar spectra. 

We note that not all the stellar atmospheres in the grid converged, with the failed atmospheres having been either super-Eddington or low-temperature ($T_{\textrm{eff}} < 15,000$\,K). However, out of the relevant parameter space in Fig. \ref{fig:evo_tracks}, only stars below $2.2$\,M$_{\astrosun}$ were affected. As we show below, such low-mass stars have negligible Lyman-band emission despite their long lifetime (shown in Table \ref{tab:fid_imfs} and Fig.~\ref{fig:ems_and_life}). Therefore, we neglect the Lyman-band emission from stars with masses less than $2.2$\,M$_{\astrosun}$ and the lack of convergence at low masses does not affect our results.

\subsection{Individual stellar spectra}\label{ssec:popIII_spectra}

Finally, by combining our stellar evolution tracks and grid of stellar fluxes we calculated the lifetime Lyman-band emission spectra of each individual star of mass $M$
\begin{equation}
    \varepsilon(\nu;M) = \int\limits_{\mathclap{\textrm{evolution track}}} dt  A(t)  F\left(\nu;T_{\textrm{eff}}(t),\log\left[g(t)\right]\right),  
\end{equation}
in units of photons\,Hz$^{-1}$ (see more details in Appendix \ref{app:app}). $A(t)$ is the area of the star (extracted from the evolutionary histories) and $F(\nu)$ the stellar flux which we calculate by interpolating our pre-computed grid for the star of mass $M$ to the required  values of $T_{\textrm{eff}}(t)$ and  $\log[g(t)]$. Similarly we can also find the emission rate (in units of photons\,Hz$^{-1}$\,s$^{-1}$) of each star at a time $t$ after it reaches ZAMS
\begin{equation} \label{eqn:rate_eq}
    \dot{\varepsilon}(\nu;M) = A(t) F\left(\nu;T_{\textrm{eff}}(t),\log[g(t)]\right).
\end{equation}
Hence the emission rate of a star across all frequencies scales linearly with its area. Similarly it was found increasing $T_{\rm eff}$ also increases the total emission rate, but this increase is chromatic with higher $T_{\rm eff}$ resulting in harder spectra. Conversely increasing $g$ produces softer Lyman-band spectra as well as  broadens spectral lines.

\begin{figure}
	\includegraphics[width=\columnwidth]{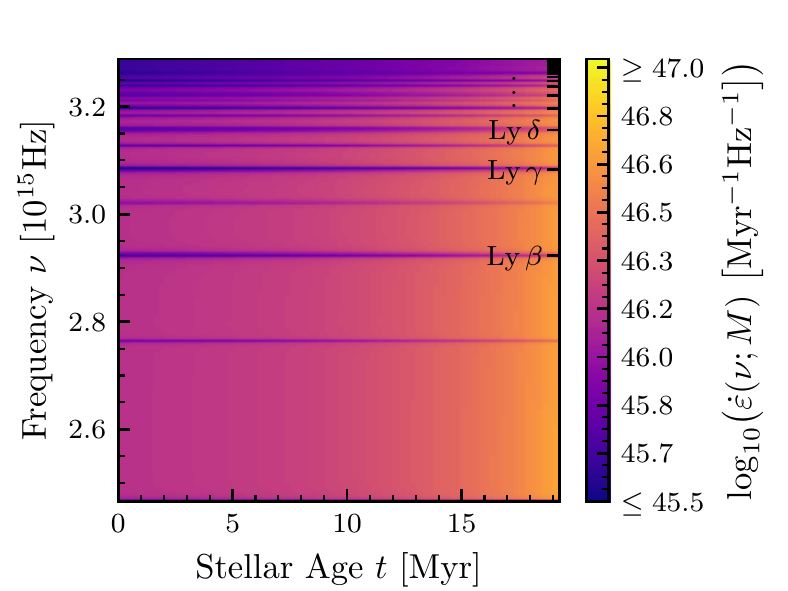}
    \caption{Spectral emission rate of a $10$\,M$_{\astrosun}$ Pop~III star over its 19.2\,Myr lifetime. The stellar emission rate increases substantially over its lifetime, being driven by the increasing area of the star. As an example, at a frequency of $3 \times 10^{15}$\,Hz the stellar emission rate increases by a factor of $3.28$ between ZAMS and hydrogen depletion. Also prominently visible are the H\,\textsc{i} Lyman and He\,\textsc{ii} Balmer lines, seen as thin lines of relatively low emission, the position of the Lyman lines are indicated by ticks on the right-hand side of the figure.}
    \label{fig:ems_in_time}
\end{figure}

\begin{figure}
	\includegraphics[width=\columnwidth]{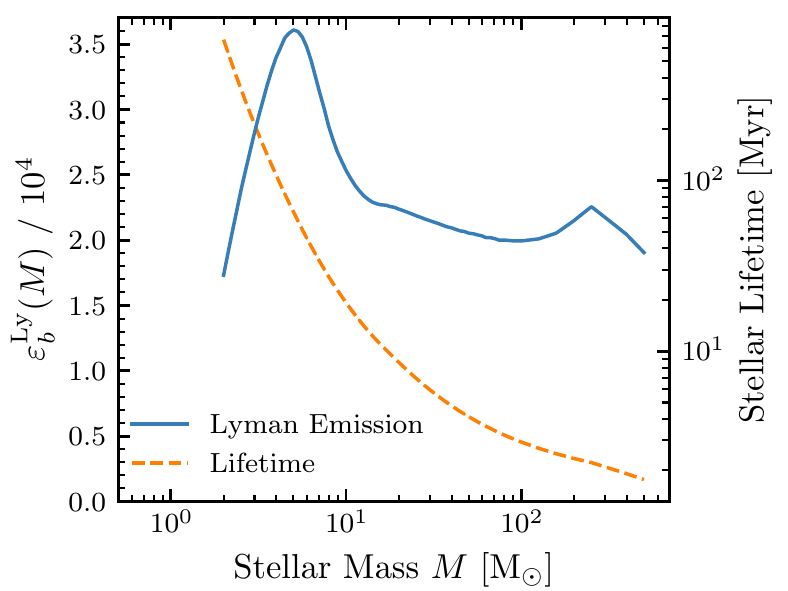}
    \caption{Total Lyman-band emission per stellar baryon $\varepsilon_\textrm{b}^{\textrm{Ly}}(M)$ of a Pop~III star of mass $M$ (solid blue) and its lifetime (dashed orange). Note the emission scale is shown on the left and is linear while the lifetime scale is shown on the right and is logarithmic. The lifetime emission of a star peaks at $5$\,M$_{\astrosun}$ with a sharp decay at lower masses and a slower decay for higher masses. Stellar lifetimes decrease by several orders of magnitude across the range of masses considered. The sharp decrease in emission and very large lifetimes of the low-mass stars leads us to assume that stars of mass less than $2.2$\,M$_{\astrosun}$ do not contribute to the cosmological Lyman-band radiation fields around cosmic dawn. }
    \label{fig:ems_and_life} 
\end{figure}

Two key results are apparent from our computed individual stellar spectra. Firstly, we find the Lyman-band emission rate of a Pop~III star increases over its life due to the area of the star increasing faster than its Lyman-band flux decreases. Thus taking the ZAMS emission as representative would result in an underestimation of the total stellar output. For example, for a  Pop III star of mass 10\,M$_{\astrosun}$, whose spectral emission rate is illustrated in Fig.~\ref{fig:ems_in_time}, we find that the emission rate increases by a factor of $3.28$ at a frequency of $3 \times 10^{15}$\,Hz over the stellar lifetime. Second, we find that the total Lyman-band emission per stellar baryon, $\varepsilon_{\textrm{b}}^{\textrm{Ly}}(M)$, of stars with masses below $\sim 5$\,M$_{\astrosun}$ decreases sharply with mass, while stellar lifetimes greatly increase, as is depicted in Fig.~\ref{fig:ems_and_life}.  Consequently, we confirm that low-mass stars ($\leq 3$\,$M_{\astrosun}$) contribute little to the cosmic dawn era Lyman radiation field, as their low emission, caused by their low effective temperatures, is spread out over a long time period ($\geq 215$\,Myr). This finding justifies our assumption made at the end of Sec. \ref{ssec:popIII_atmospheres} that low-mass stars ($M<2.2$\,M$_{\astrosun}$) do not significantly contribute to the Lyman-band emission.

It is worth noting here that we also find the peak in $\varepsilon_{\textrm{b}}^{\textrm{Ly}}(M)$ occurs for stars of mass $4$\,--\,$7$\,M$_{\astrosun}$.  At higher masses,   $\varepsilon_{\textrm{b}}^{\textrm{Ly}}$ decreases  as $M$ approaches $10$\,M$_{\astrosun}$, then remains roughly constant out to $500$\,M$_{\astrosun}$ with the only notable feature being a kink around $300$\,M$_{\astrosun}$, potentially due to the onset of photoevaporation (see discussion in Appendix \ref{app:app}). The plateau  in $\varepsilon_{\textrm{b}}^{\textrm{Ly}}(M)$  at 
 $10$\,--\,$500$\,M$_{\astrosun}$  suggests that the total Lyman-band emission from a population of such heavy stars will not strongly depend on the IMF. Based on this finding alone, we anticipate a small difference in the resulting 21-cm signal for the Log-Flat, Intermediate, and Top-Heavy IMFs when only the effects of Lyman-band photons are taken into account. However, the differences are expected to grow once the IMF-dependent impact of X-ray heating from Pop~III X-ray binaries is taken into account.

We discuss several additional trends in how metal-free stellar Lyman-band emission varies with stellar mass in appendix~\ref{app:app}.

\subsection{Star formation rate}\label{ssec:sfr}

Within our semi-numerical simulations we model the spatially varying Pop~II and Pop~III star formation rate densities (SFRD) via the methodology of \citet{Magg_2021}. In this star formation prescription each halo forms one generation of Pop~III stars when it first crosses the critical virial velocity for star formation, with efficiency $f_{*,{\rm III}}$. After this initial burst of star formation, it is assumed haloes take a time $t_{\rm recov}$ to recover from the ejection of material from Pop~III supernovae. We assume $t_{\rm recov} = 100$\,Myr which was shown by \citet{Magg_2021} to produce a Pop~III to Pop~II transition redshift qualitatively consistent with the FiBY simulations~\citep{Johnson_2013}. Once this time has passed, the halo  begins to form Pop~II stars with an efficiency $f_{*,{\rm II}}$. 

\begin{figure}
	\includegraphics[width=\columnwidth]{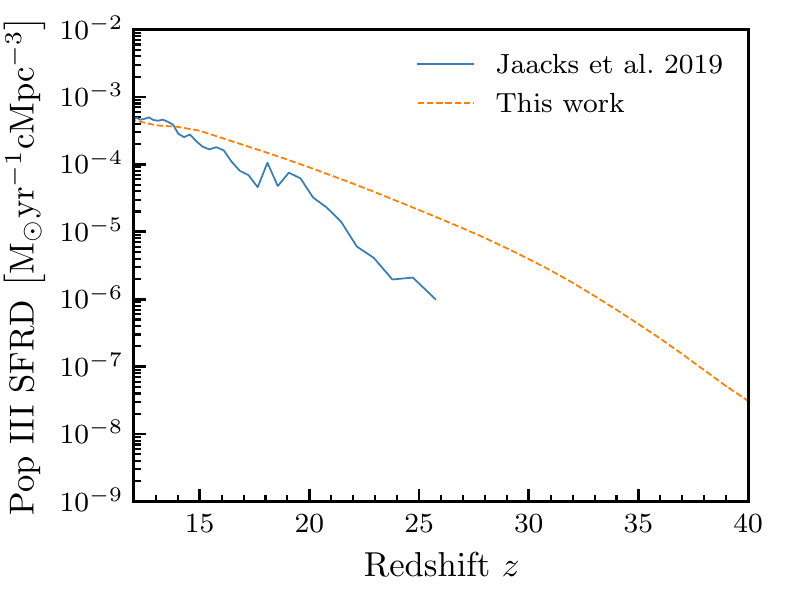}
    \caption{Comparison of our Pop III star formation rate density (dashed orange) to the numerical simulations of \citet{Jaacks_2019} (solid blue). We see that below $z = 20$ the two star formation rates are comparable, differing by at most a factor of 2.5. Due to our larger simulation boxes, which will include more rare overdensity peaks, our model predicts a greater rate of Pop III star formation at higher redshifts. The depicted SFRD is that of the Intermediate IMF model in Fig.~\ref{fig:IMF_comp}.}
    \label{fig:sfrd} 
\end{figure}

Unless otherwise stated throughout this paper we take $f_{*,{\rm II}} = 0.05$ and $f_{*,{\rm III}} = 0.002$ so that our Pop~III SFRDs (shown in Fig.~\ref{fig:sfrd}) are comparable within a factor of 2.5 to the simulations of \citet{Jaacks_2019} at $z < 20$. In this same redshift range the \citet{Jaacks_2019} predictions are in general agreement with other numerical studies of Pop III star formation \citep{Yoshida_2004, Greif_2006, Tornatore_2007, Wise_2012, Johnson_2013, Pallottini_2014, Xu_2016, Sarmento_2018}. At higher redshifts our model predicts more Pop III star formation than that of \citet{Jaacks_2019}. This is to be expected due to our larger simulation box sizes, side-length of $384$\,cMpc versus $5.9$\,cMpc in  \citet{Jaacks_2019}, meaning that our simulations contain more rare overdensity peaks that dominate Pop III star formation at the highest redshifts \citep{Bromm_2004, Bromm_2013}. 

An important caveat to our star-formation prescription is that it does not include external metal enrichment of halos. Detailed in an appendix of \citet{Magg_2021}, the authors found that introducing external enrichment had little impact on the Pop II and Pop III fractions found in their semi-numerical models. Similarly, \citet{Visbal_2018} found external metal enrichment to have a very small effect on the global star formation rates in their semi-analytic study of $z>20$ star formation. However, simulations that fully model the chemical and thermodynamic evolution of the star-forming gas find a significant degree of external metal enrichment \citep{Smith_2015, Chen_2017, Hicks_2021}. With externally enriched halos potentially being the site of formation for the first Pop II stars. It is thus probable that Pop III star formation in our simulations is overly clustered.

\section{Lyman-band Radiation Fields}\label{sec:radiation_fields}

Having calculated the Lyman-band emission of individual Pop~III stars we integrated these spectra into our semi-numerical simulations as a lookup table. Utilizing this lookup table we now extend the calculation of the spatially varying Ly\,$\alpha$ and LW radiation fields in our simulation to be IMF dependent.

\subsection{Ly\,$\alpha$ radiation field}\label{ssec:lyman_alpha}

The previous model employed for the Ly\,$\alpha$ radiation field in our simulation code was introduced in \citet{Barkana_2005}. The core idea of this model is that Lyman-band emitted photons can only contribute to the Ly\,$\alpha$ radiation field once redshifted into the Lyman line immediately below their emission frequency. Hence, emission at $z_{\textrm{em}}$ only affects the Ly\,$\alpha$ radiation field at $z_{\textrm{abs}}$ if it is emitted at a frequency
\begin{equation}
    \nu_n( z_{\textrm{em}},z_{\textrm{abs}}) = \left(\frac{1 +  z_{\textrm{em}}}{1 +  z_{\textrm{abs}}}\right) \nu^{\rm Ly}_n,
\end{equation}
for some $n$, where $\nu^{\rm Ly}_n$ is the frequency of the $n$\,th Lyman line. With the further condition that $\nu_n$ cannot exceed $\nu^{\rm Ly}_{n+1}$ as otherwise those photons would have already cascaded on redshifting into that higher line. As a result, Lyman-band photons emitted above the $n$\,th Lyman line have a maximum redshift from which they could have an impact on the Ly\,$\alpha$ radiation field at a given absorption redshift of
\begin{equation}
    1 + z_{\textrm{max},n}(z_{\textrm{abs}}) = \frac{\nu^{\rm Ly}_{n+1}}{\nu^{\rm Ly}_n}\left(1 + z_{\textrm{abs}}\right).
\end{equation}

Historically, it was assumed that between their emission and absorption any emitted photons travel in a straight line for a cosmology-dependent comoving distance $r(z_{\textrm{abs}},z_{\textrm{ems}})$. However,  this assumption was recently revised in our code by  \citet{Reis_2021} to incorporate multiple scattering of Ly\,$\alpha$ photons. When absorbed at a Lyman line, a photon cascades into a Ly\,$\alpha$ photon with a recycling efficiency $f_{\textrm{recycle}}(n)$. For Ly\,$\alpha$ continuum photons that redshift directly into the Ly\,$\alpha$ line, $f_{\textrm{recycle}}(1)$ is taken as 1. Under these assumptions, plus isotropic emission, it was shown by \citet{Barkana_2005} that the Lyman\,$\alpha$ radiation at the absorption redshift is
\begin{equation}
\begin{split}
    J_\alpha(z_{\textrm{abs}},\mathbfit{x}) = \sum_{n=1}^\infty f_{\textrm{recycle}}(n) \int_{z_{\textrm{abs}}}^{z_{\textrm{max},n}} \frac{c dz_{\textrm{em}}}{H(z_{\textrm{em}})} \frac{(1 + z_{\textrm{abs}})^2}{4 \pi} \\ 
    \times \left< \epsilon(\nu_n,z_{\textrm{em}})\right>_{r(z_{\textrm{abs}},z_{\textrm{ems}}),\mathbfit{x}}, 
\label{eqn:LyA_rad_field_calc}
\end{split}
\end{equation}
where $\left< \epsilon(\nu_n,z_{\textrm{em}})\right>_{r(z_{\textrm{abs}},z_{\textrm{ems}}),\mathbfit{x}}$  is the photon emissivity at the emission redshift averaged over a spherical shell of comoving distance $r$ about the point the field is being evaluated at $\mathbfit{x}$. In our 21-cm simulation code this spherical shell average was replaced by \citet{Reis_2021} with a broader window function to take into account multiple scattering of Ly\,$\alpha$ continuum photons.

Under the earlier discussed assumption that star formation is rapid and stellar lifetimes are short compared to the cosmological time-scales of interest it was also argued by \citet{Barkana_2005} that emissivity can be expressed as
\begin{equation}
    \epsilon(\nu,z_{\textrm{em}}) = \bar{n}^0_\textrm{b} f_* \frac{dF_{\rm coll}}{dt} \varepsilon_\textrm{b}(\nu),
    \label{eqn:epsilon_ingredients}
\end{equation}
where $\bar{n}^0_\textrm{b}$ is the mean baryon density, $f_*$ the star formation efficiency of the stellar population being modelled, $dF_{\rm coll}/dt$ the rate of change of gas collapse fraction into halos at that location, and $\varepsilon_\textrm{b}(\nu)$ the photon-number emissivity per stellar baryon.

For our purposes of generalising the calculation to an arbitrary IMF we need to revise the calculation of  $\varepsilon_\textrm{b}(\nu)$. In \citet{Barkana_2005} and subsequent studies,  $\varepsilon_\textrm{b}(\nu)$ is taken as a piece-wise power-law computed from the stellar spectra provided by \citet{Bromm_2001} under the assumption that all Pop~III stars were 100 solar masses (Legacy model).  In our extended simulations we instead calculate $\varepsilon_\textrm{b}(\nu)$ from an arbitrary IMF and our pre-calculated individual star lifetime photon-number emissivities $\varepsilon(\nu;M)$ as
\begin{equation}
    \varepsilon_\textrm{b}(\nu) = m_p \frac{\int  \varepsilon(\nu;M)\frac{dN}{dM} dM}{\int M \frac{dN}{dM} dM} ,
    \label{eqn:epsilon_calc}
\end{equation}
where $m_p$ is the mass of the proton. During implementation, it was found that sampling at 400 logarithmically spaced stellar masses was sufficient to ensure the integral in equation~\ref{eqn:epsilon_calc} converged to an accuracy greater than $1$ per cent, for all IMFs considered in this paper.

\begin{figure}
    \centering
    \includegraphics[scale = 1.08]{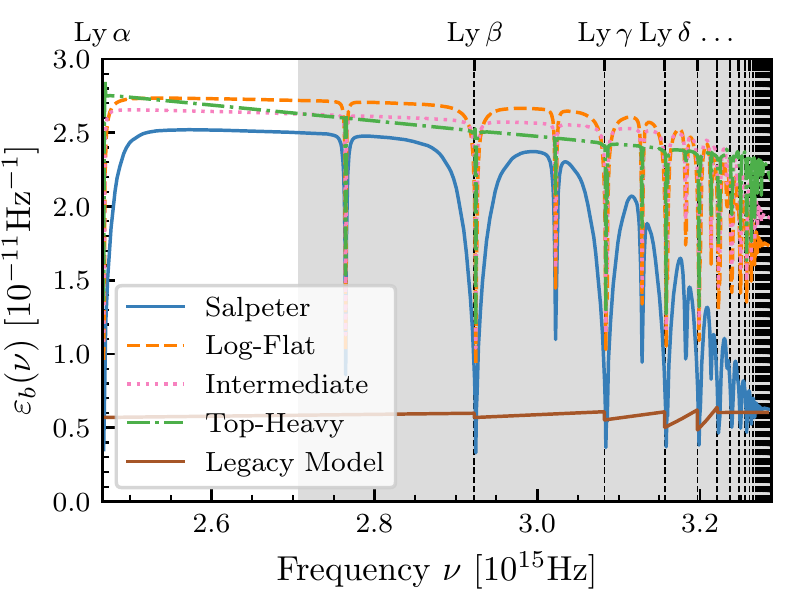}
    \caption{Photon-number emissivity per stellar baryon in the Lyman band for our example IMFs and our Legacy model from \citet{Barkana_2005}, where the model is indicated by line type and colour as detailed in the legend. The H\,\textsc{i} Lyman series is indicated by black dashed vertical lines, and the LW band via the grey region. The Log-Flat, Intermediate, and Top-Heavy IMF all produce very similar $\varepsilon_\textrm{b}(\nu)$ suggesting they may be indistinguishable via the cosmic dawn 21-cm signal. On the other hand, there are noticeable differences from the other IMFs in the overall magnitude and shape of the Salpeter photon emissivity. Lastly, our Legacy model leads to a comparatively low $\varepsilon_\textrm{b}(\nu)$, in part due to it only considering ZAMS emission.}
    \label{fig:epsilons}
\end{figure}

The resulting $\varepsilon_\textrm{b}(\nu)$ for our four example IMFs are shown in Fig.~\ref{fig:epsilons}. As expected from the flat shape of $\varepsilon_\textrm{b}^{\textrm{Ly}}(M)$ at high masses (Fig.~\ref{fig:ems_and_life}), we find the emissivity of the Log-Flat, Intermediate, and Top-Heavy IMFs to be very similar. Hence, we expect these three IMFs to produce similar 21-cm signals. Conversely, the Salpeter IMF $\varepsilon_\textrm{b}^{\textrm{Ly}}(M)$ is markedly different from the other three example IMFs, having lower emissivity at all frequencies, and a softer spectrum. Due to its lower $\varepsilon_\textrm{b}(\nu)$, we can anticipate that a Salpeter IMF would result in a delayed WF effect compared to the other three example IMFs, and hence a different 21-cm signal.

In addition to the overall intensity, the differences in the spectral shape of $\varepsilon_\textrm{b}(\nu)$ between IMFs can also affect the 21-cm signal. For example, the lower proportion of high-frequency emission seen in the Salpeter IMF leads to less injected Lyman cooling and thus stronger Lyman heating~\citep{Chuzhoy_2006, Reis_2021, Mittal_2021}, as well as reduced LW feedback. Furthermore, the spectral shape of $\varepsilon_\textrm{b}(\nu)$ impacts the spatial distribution of the \mbox{Ly-$\alpha$} radiation field and thus the WF coupling, potentially altering the 21-cm power spectra~\citep{Santos_2011}.  Hence, we anticipate that the 21-cm signal will vary with the Pop~III IMF, though it remains to be demonstrated whether these variations are measurable.

For comparison we also plot $\varepsilon_\textrm{b}(\nu)$ for our Legacy model in Fig.~\ref{fig:epsilons}. We find that the Lyman-band photon emissivity for our Legacy model is much lower than those for our example IMFs, with the mean $\varepsilon_\textrm{b}(\nu)$ being $0.26$ to $0.34$ times that of our example IMFs. By comparing our Legacy case to a model computed using our fully updated framework with the same delta-function Pop~III IMF composed purely of 100\,M$_{\astrosun}$ stars, we find that most of this difference is due to the relaxation of the ZAMS assumption and the changes in stellar modelling. This comparison shows the mean $\varepsilon_\textrm{b}(\nu)$ in our Legacy model is 72 per cent lower than the model using our methodology.  Recalculating $\varepsilon_\textrm{b}(\nu)$ using our stellar evolution histories, but now assuming that the ZAMS emission rate of a star is the same as its average lifetime emission, we find a 54 per cent drop in the emissivity. The remaining 18 per cent is thus due to differences in stellar modelling between our model and that of \citet{Bromm_2001}.  We, hence, see the ZAMS assumption leads to a significant underestimate of $\varepsilon_\textrm{b}(\nu)$, and, therefore, an underestimation of the magnitude of the Lyman mediated impacts on the IGM. In Sec.~\ref{ssec:zams} we go on to quantify the resulting impact on the 21-cm signal due to this underestimation and further discuss how it may have biased the 21-cm predictions of our Legacy model.

\subsection{Lyman-Werner feedback}\label{ssec:lyman_werner}

The LW radiation field, $J_{\textrm{LW}}$ \citep{Holzbauer_2012,Fialkov_2013}, is calculated in a similar manner to $J_{\alpha}$ (equation~\ref{eqn:LyA_rad_field_calc}), with a few key differences. These differences are driven by $J_{\textrm{LW}}$ being the LW band mean spectral intensity, rather than the photon radiance at the Ly\,$\alpha$ line, as is the case in the $J_{\alpha}$ calculation.

As the Ly\,$\alpha$ line is not in the LW band, cascading photons now leave the band rather than inject into the line of interest. Hence, the sum over cascading lines in equation~\ref{eqn:LyA_rad_field_calc} is replaced with a multiplicative modification factor $f_\textrm{mod}(z_{\textrm{abs}},z_{\textrm{em}})$ inside the integral. $f_\textrm{mod}(z_{\textrm{abs}},z_{\textrm{em}})$ accounts for both the loss of LW photons to cascading but also the relative effectiveness of the different LW lines at triggering H$_2$ disassociation. More formally, $f_\textrm{mod}(z_{\textrm{abs}},z_{\textrm{em}})$ is defined as the effectiveness of the surviving LW band emission at $z_{\textrm{abs}}$ to disassociate molecular hydrogen relative to the emission at the source. In our calculations we use $f_\textrm{mod}$ from \citet{Fialkov_2013}.

Because $J_{\textrm{LW}}$ is an averaged energetic radiometric quantity, the photon emissivity $\epsilon(\nu,z_{\textrm{em}})$ is replaced in the calculation of the radiative background by the mean LW emissivity, $\bar{\epsilon}(z_{\textrm{em}})$. Hence, $J_\textrm{LW}$ is given by
\begin{equation}
\begin{split}
    J_{\textrm{LW}}(z_{\textrm{abs}},\mathbfit{x}) =  \int_{z_{\textrm{abs}}}^{\infty} \frac{c dz_{\textrm{em}}}{H(z_{\textrm{em}})} \frac{(1 + z_{\textrm{abs}})^2}{4 \pi} f_\textrm{mod}(z_{\textrm{abs}},z_{\textrm{em}})\\ 
    \times \left< \bar{\epsilon}(z_{\textrm{em}})\right>_{r(z_{\textrm{abs}},z_{\textrm{ems}}),\mathbfit{x}}.
\label{eqn:LW_rad_field_calc}
\end{split}
\end{equation}
The quantity $\bar{\epsilon}(z_{\textrm{em}})$ can be computed in an analogous manner to equation  \ref{eqn:epsilon_ingredients} in which $\varepsilon_\textrm{b}(\nu)$ is replaced by the IMF-dependent  LW band averaged emissivity per stellar baryon, $\bar{\epsilon}_\textrm{b}$. 
In turn,  $\bar{\epsilon}_\textrm{b}$ is computed by taking a photon energy-weighted integral of $\varepsilon_\textrm{b}(\nu)$ given by equation~\ref{eqn:epsilon_calc} over the LW frequency  range. The $\bar{\epsilon}_\textrm{b}$ calculated for each of our four example IMFs and our Legacy model are given in Table~\ref{tab:bar_epsilons}. Since $\bar{\epsilon}_\textrm{b}$ is a weighted integral of a portion of $\varepsilon_\textrm{b}(\nu)$, the $\bar{\epsilon}_\textrm{b}$ values follow the same pattern as the magnitudes of $\varepsilon_\textrm{b}(\nu)$, with Legacy model yielding the lowest values followed by Salpeter, while the Log-Flat, Intermediate, and Top-Heavy IMFs result in the highest and approximately equal emissivites. 
\begin{table}
	\centering
	\begin{tabular}{cccc} 
		\hline
		IMF & $\bar{\epsilon}_\textrm{b}$  \\
		\hline
		Salpeter & $3.64 \times 10^{-22}$ \\
		Log-Flat & $4.89 \times 10^{-22}$ \\
		Intermediate & $4.86 \times 10^{-22}$ \\
		Top-Heavy & $4.85 \times 10^{-22}$  \\
		\hline
		Legacy & $1.12 \times 10^{-22}$ \\ 
		\hline
	\end{tabular}
	\caption{Mean LW band emissivities per stellar baryon  in units of erg\,Hz$^{-1}$ calculated for the four example IMFs alongside our Legacy value used in previous works.}	\label{tab:bar_epsilons}
\end{table}

Due to the large-scale overdensity and baryon-dark matter relative velocity differing between pixels in our simulations, the star formation rate and emissivity also spatially vary. Once convolved with the relevant window functions and integrated over time, as described by equations~\eqref{eqn:LyA_rad_field_calc} and \eqref{eqn:LW_rad_field_calc}, this in turn leads to spatial variations in the computed radiation fields. Inclusion of such variations is essential for accurate calculation of the 21-cm power spectrum with the resulting variations in the strength of the WF effect leading to the cosmic dawn power spectrum peak. Furthermore, the correlation between the strength of the LW radiation field and the matter overdensity in the early Universe is required to properly account for the global suppression of star formation via the LW feedback due to the clustering of the first radiative sources \citep{Ahn_2009}. Numerical simulations \citep[e.g.][]{Ahn_2009, Johnson_2013} show variations of the local LW radiation field on sub\,cMpc scales, an effect which we do not take into account in our work where the LW fluctuations are modelled on the scale of the simulation cell (3\,cMpc). These unresolved fluctuations in the LW background will contribute to the spatial variation of star formation rate, and, thus, could boost the 21-cm power spectrum on corresponding scales. However, \citet{Johnson_2013} found in their simulations that star formation rate density is mostly regulated by the background LW radiation field due to the large mean free path of LW photons. It is only in rare regions that the LW feedback is driven by the local small-scale star formation. Although the omission of these small-scale variation in the LW field and thus star formation rate is a limitation of our simulation, it is not expected to greatly affect our predictions for the large-scale observable signals targeted by the existing interferometers HERA~\citep{HERA}, MWA~\citep{MWA}, LOFAR~\citep{LOFAR} and the upcoming SKA~\citep{Koopmans_2015}.

\subsection{Non-instantaneous emission}\label{ssec:non_instant_emission}

We now relax the assumption of instantaneous stellar emission.
For non-instantaneous stellar emission the cosmology, radiative transfer, and atomic physics remain the same and so equations~\ref{eqn:LyA_rad_field_calc} and~\ref{eqn:LW_rad_field_calc} still hold. The differences in the non-instantaneous emission case instead enter into the calculation of the emissivity because the emission now depends not just on the stars forming at $z_{\textrm{em}}$ but those that formed prior and are still emitting at $z_{\textrm{em}}$. Consequently, while equation~\ref{eqn:epsilon_ingredients}  holds in the case of instantaneous emission it must be generalized to consider the star-formation history and the rate of stellar emission in the case of non-instantaneous emission.

\begin{figure*}
    \centering
    \includegraphics{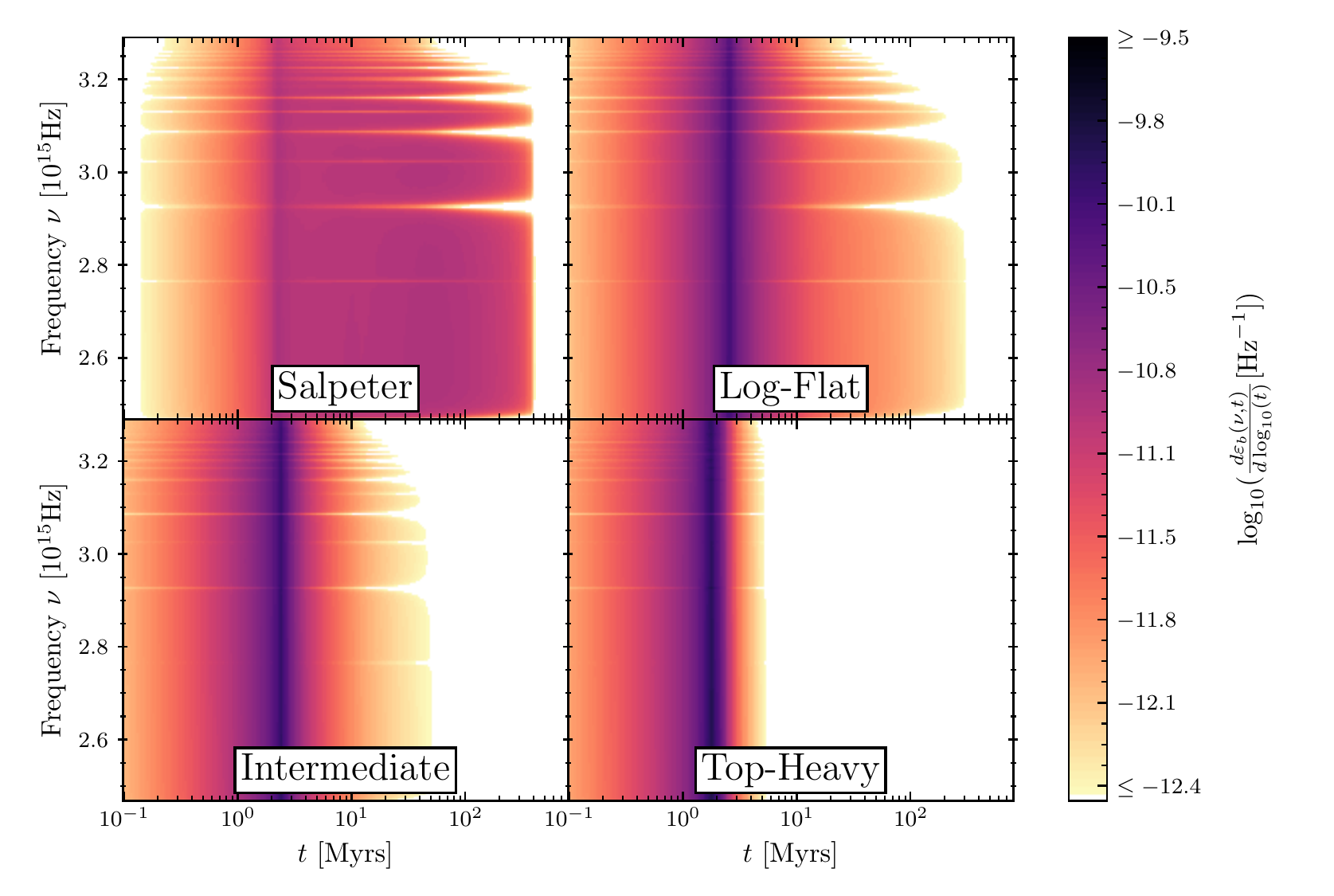}
    \caption{Photon emission rate per stellar baryon in the Lyman band for our example IMFs, as a function of frequency and time since star formation. The time axis is logarithmic and the emission rate is per unit log time. The emission of the Salpeter IMF spreads over hundreds of millions of years, whereas with the other IMFs it is concentrated in the first few million years after star formation. As the age of the Universe at $z = 20$ is only 180\,Myr the assumption of instantaneous stellar emission for the Salpeter IMF immediately seems likely to be a poor approximation.}
    \label{fig:epsilon_dots}
\end{figure*}

Using our results for the emissivity rate of individual stars (equation~\ref{eqn:rate_eq}) we derive a population photon emissivity rate per stellar baryon
\begin{equation}
    \dot{\varepsilon}_\textrm{b}(\nu,t) = m_p \frac{\int   \dot{\varepsilon}(\nu,t;M)\frac{dN}{dM} dM}{\int M \frac{dN}{dM} dM},
    \label{eqn:epsilon_dot_calc}
\end{equation}
where $\dot{\varepsilon}(\nu,t;M)$ is 0 for $t$ larger than the lifetime of a Pop~III star of mass $M$. The results are shown in  Fig.~\ref{fig:epsilon_dots}.  

Considering first the Salpeter IMF we find an initial increase in the emission rate before a plateau stretching from a few\,Myr to over 300\,Myr, and finally a sharp decrease in emission. This plateau shows that a significant proportion of the emission from the low-mass dominated Salpeter IMF is spread out over hundreds of millions of years. Because, in the Planck best-fit cosmological concordance model \citep{Planck_VI}, redshift $z = 20$ corresponds to 180\,Myr after the Big Bang, the emission from stars in the Salpeter case is certainly non-instantaneous. This extended emission is expected to slow down the build-up of the cosmological Ly\,$\alpha$ and LW radiation fields. As a result, the associated impacts on the IGM of these stars are anticipated to be delayed, leading to the characteristic feature of cosmic dawn in the 21-cm signal appearing at later times. 

For our other example IMFs emission is concentrated during the first 2 to 3\,Myr after star formation, with little emission beyond 10\,Myr. The characteristic time-scales of emission for the Log-Flat, Intermediate, and Top-Heavy IMFs are, hence, much shorter than that for the Salpeter IMF. As a result, we expect the instantaneous emission assumption to be a better approximation for these IMFs.

The photon emissivity at a location $\vec{x}$ in the simulation box can hence be calculated from $\dot{\varepsilon}_\textrm{b}(\nu,t)$ by integrating over the previous rate of star formation at that position
\begin{equation}
    \epsilon(\nu,z_{\textrm{em}})_{\vec{x}} = \bar{n}^0_\textrm{b} f_* \int_0^{t_{\textrm{em}}} dt' \left(\frac{dF_{\rm coll}}{dt}\left[t_{\textrm{em}} - t'\right]\right) \dot{\varepsilon}_\textrm{b}\left(\nu,t'\right)_{\vec{x}}.
\end{equation}
Here $t'$ is the time between the onset of star formation and the moment at which the emission rate is sampled, and $t_{\textrm{em}}$ is the time corresponding to the emission redshift $z_{\textrm{em}}$. In the limiting case of $\dot{\varepsilon}_\textrm{b}(\nu,t')$ tending to a delta-function, the instantaneous emission limit is recovered (equation~\ref{eqn:epsilon_ingredients}).  The non-instantaneous stellar emission version of $\bar{\epsilon}(z_{\textrm{em}})_{\vec{x}}$ used for the LW radiation field calculation can be computed in an analogous manner to $\epsilon(\nu,z_{\textrm{em}})_{\vec{x}}$. Alternatively $\bar{\epsilon}(z_{\textrm{em}})_{\vec{x}}$ can be found by averaging $h \nu \epsilon(\nu,z_{\textrm{em}})_{\vec{x}}$ over the LW band. Ultimately we are interested in the 21-cm signal so we quantify the significance of non-instantaneous emission in more detail in the next section using the results of our extended semi-numerical simulations.

\section{Predicted 21-cm Signal}\label{sec:results}

In this section, we present our results for our full calculation of the 21-cm signal.
We start by assessing the impacts of the highlighted assumptions on the 21-cm signal (Sec.~\ref{ssec:assumptions}). We then explore the sensitivity of the 21-cm signal to the IMF of Pop~III stars. Firstly, we show the 21-cm signal for a range of models in which all stars are of equal masses (Sec.~\ref{ssec:mono}). This allows us to determine which stellar masses will create distinct signatures in the observable signal. Second, we compare the 21-cm signal for our four example IMFs (Sec.~\ref{ssec:Fid_21cm}) under a range of star formation models. Finally, we briefly address potential parameter degeneracy  in Sec.~\ref{ssec:ftar_degneracy}.

\subsection{Testing assumptions}\label{ssec:assumptions}

\subsubsection{Impacts of the ZAMS  assumption}\label{ssec:zams}

\begin{figure*}
    \centering
    \includegraphics{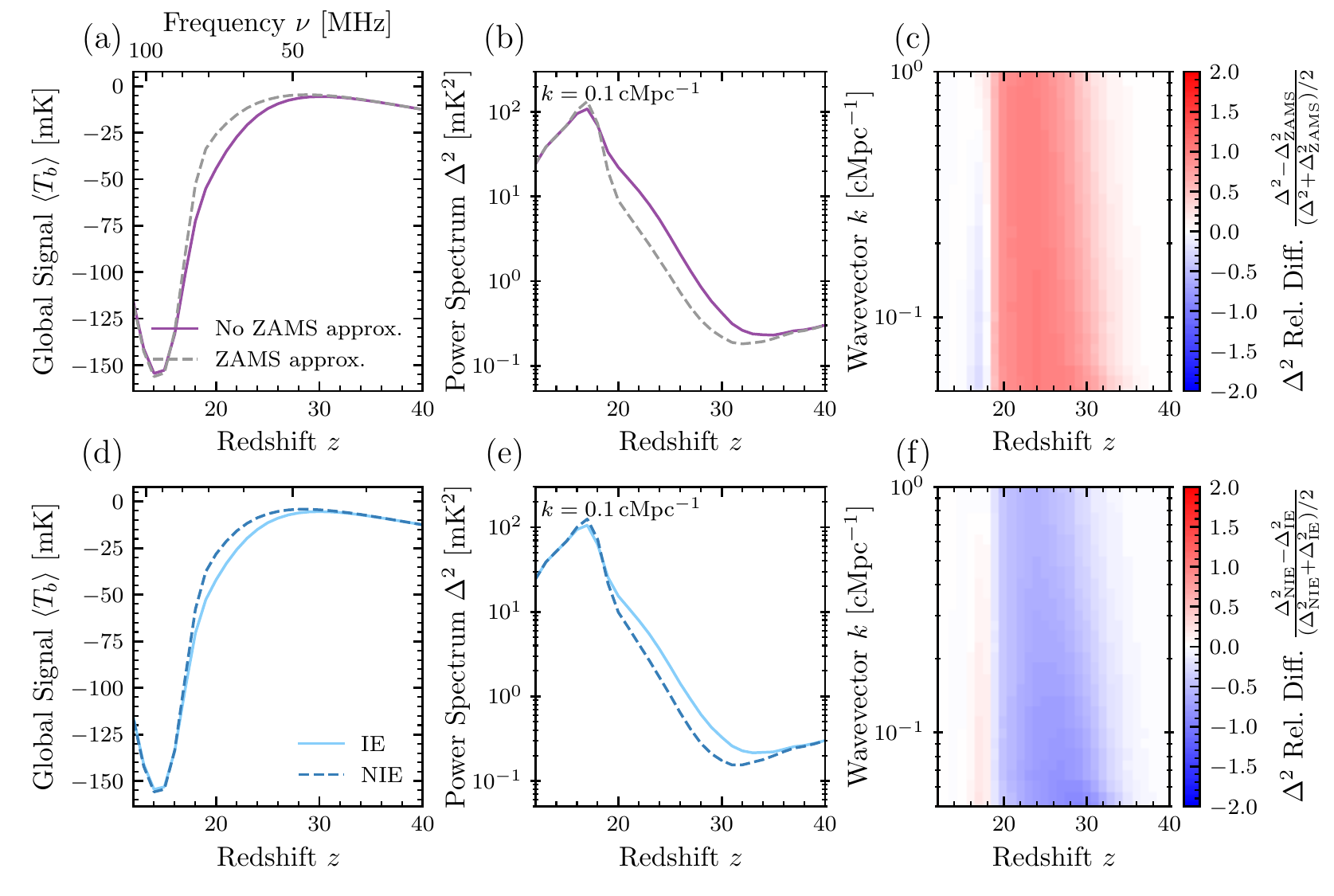}
    \caption{Testing the ZAMS assumption (top) and instantaneous emission assumption (bottom). We show the global 21-cm signals (left, panels  a and d),  power spectra at $k = 0.1$\,cMpc$^{-1}$ (middle, panels b and e), and the relative differences in the power spectra (calculated as the difference divided by the mean of the two power spectra, right, panels c and f). {\bf Top:} The 21-cm signals  predicted for an IMF of purely 100\,$M_{\astrosun}$ stars, with (dashed) and without (solid) the ZAMS approximation. With the ZAMS assumption, the 21-cm signals are shifted  by $\Delta z \approx 1$ to lower redshifts, compared to the model without. These shifts produce relative differences of up $56$ per cent (at $z = 23$) in the  global 21-cm signals, and relative differences between the power spectra reaching $103$ per cent at $z=28$ for $k=0.022$\,cMpc$^{-1}$. {\bf Bottom:} The 21-cm signals predicted for the Salpeter IMF under the instantaneous (solid) and non-instantaneous stellar emission (dashed) models.  Relaxing the instantaneous emission assumption is found to shift the signals to lower redshifts by $\Delta z \approx 1$, resulting in the maximal change of  up to 53 per cent at $z=24$ in the global signal, and a maximal relative difference of 119 per cent  between the power spectra (at $z=31$). A similar comparison for our other IMFs is shown in Fig.~\ref{fig:IE_NIE_comp_full}.}
    \label{fig:assumptions}
\end{figure*}

As discussed in Sec.~\ref{ssec:lyman_alpha}, the assumption that stellar emission at the ZAMS well represents stellar lifetime emission leads to an erroneously low photon-number emissivity with a 54 per cent reduction in mean $\varepsilon_\textrm{b}(\nu)$ for a Pop~III IMF composed of only 100\,$M_{\astrosun}$ stars. The ZAMS assumption thus reduces the WF effect, Ly\,$\alpha$ heating, CMB heating, and LW feedback. To quantify the implications of this assumption on the 21-cm signal we run our semi-numerical 21-cm simulation adopting an IMF composed of only 100\,$M_{\astrosun}$ Pop~III stars with the ZAMS assumption on and off.  The results are shown in the top row of Fig.~\ref{fig:assumptions}.

The global 21-cm signal predicted for both cases is shown in the top left panel (a) of Fig.~\ref{fig:assumptions}. As expected the suppressed Lyman-band emission under the ZAMS assumption shifts the onset of the global signal absorption trough to higher frequencies (lower redshift), by $\Delta z \approx 2$. With a resulting maximum relative difference (calculated as the difference divided by the mean of the two quantities) between the fixed redshift global 21-cm signals of $56$ per cent, at $z = 23$. The dominant mechanism causing this shift is the decreased WF effect, which is partially compensated for by the weakened LW feedback.

Furthermore, in Fig.~\ref{fig:assumptions} we depict the 21-cm power spectra (panel b) and the resulting differences between the power spectra for the two cases (panel c). Similarly to the global signal, the ZAMS assumption leads to a shift of the power spectrum cosmic dawn peak to lower redshifts by $\Delta z \approx 1$. This shift leads to large relative differences between the power spectra computed with and without the ZAMS assumption, the maximal value of which is $103$ per cent at $z=28$ for $k=0.022$\,cMpc$^{-1}$. Owing to this impact, we relax the ZAMS assumption in all our subsequent calculations.

As our Legacy model employs the ZAMS assumption, we anticipate that previous 21-cm signal predictions utilizing this model have thus been biased.  While the exact magnitude of the effect depends on the specific astrophysical parameters,  we generically expect a deficit in Lyman radiation which, in the cases explored here, leads to a shift of the high-redshift part of the 21-cm signal to lower redshifts by $\Delta z$ of a few.

\subsubsection{Instantaneous versus non-instantaneous emission}\label{ssec:IE_vs_NIE}

We now explore the implications of the instantaneous stellar emission assumption for the cosmic dawn 21-cm signal showing the results in the bottom row of Fig.~\ref{fig:assumptions}. We compare the signal calculated with this assumption against the case in which this assumption is relaxed for each one of our four example IMFs. As in the case of non-instantaneous stellar emission a star's radiation output is spread over its lifetime, and stellar Lyman-band emission increases over said lifetime, \textit{ab initio} we would expect the resulting Lyman-band radiation fields to take longer to reach the same magnitude when compared to instantaneous stellar emission. Consequently, we anticipate a delay in all Lyman-band mediated effects, and so a later global signal absorption trough and cosmic dawn power spectrum peak in the non-instantaneous emission case. Indeed, a mild shift of the signals to lower redshifts is observed in our results  (Fig.~\ref{fig:assumptions} panels d-f and Fig.~\ref{fig:IE_NIE_comp_full}).

We find that the signal shift is modest for all our explored IMFs, with the largest impact being on the Salpeter IMF (Fig.~\ref{fig:assumptions} panels d-f) where the proportion of low-mass long-lived stars is greatest. In this case, both the global signal absorption trough and the cosmic dawn peak in the power spectrum shift to lower redshift by $\Delta z \approx 1$ when non-instantaneous stellar emission is modelled.
This shift results in a difference between the fixed redshift global 21-cm signal of at most 53 per cent (at $z=24$), and maximum relative difference between the power spectrum of 119 per cent measured at $z=31$.

With our other example IMFs, dominated by more massive and thus short-lived stars, the effect is much weaker (Fig.~\ref{fig:IE_NIE_comp_full}) with the maximal relative differences in the global 21-cm signals reaching 11 per cent at $z=27$, 6.6 per cent at $z=28$, and 4.1 per cent at $z=29$ for the Log-Flat, Intermediate, and Top-Heavy IMFs respectively. The corresponding relative differences in the respective power spectra have maximum values of 41 per cent at $z=33$, 31 per cent at $z=33$, and 22 per cent at $z=33$.

Our results suggest that for the global signal instantaneous emission is a reasonable approximation for IMFs dominated by high-mass stars ($\gtrsim 70$\,M$_{\astrosun}$, Table~\ref{tab:fid_imfs}), while the IMFs dominated by lower-mass stars  ($< 5$\,M$_{\astrosun}$) require non-instantaneous stellar emission to be modelled.  However,  the effect on the power spectra is much stronger ($\geq22$ per cent) for all our considered IMFs. Hence, including non-instantaneous stellar emission in our semi-numerical simulations is necessary if we hope to accurately predict the 21-cm power spectrum during the early phases of cosmic dawn. These results, therefore, motivate us to utilize the fully-modelled non-instantaneous stellar emission in all of our subsequent simulations. 

In all our previous papers we used our Legacy model, which is based upon an IMF composed of 100\,$M_{\astrosun}$ stars and thus is an example of a high-mass dominated IMF. Therefore we do not anticipate the instantaneous stellar emission assumption to significantly alter our previous conclusions\footnote{At high redshifts, an order 10 per cent effect in the power spectra is expected, while the impact on the global signal is smaller.}. Hence, while the effects of the ZAMS and instantaneous emission assumptions on the  21-cm signal are in opposite directions, they do not cancel out for our Legacy IMF as the ZAMS assumption impact is much greater. Therefore, previous predictions using our Legacy model were biased, under-predicting the Lyman radiation fields in the early Universe and producing lower-redshift  cosmic dawn 21-cm signal features due to the ZAMS assumption.

\subsection{Single-mass Pop~III IMFs}\label{ssec:mono}

We now consider the case of delta-function IMFs, each composed of stars of a single mass $M$. Such simplified IMFs allow us to explore the implications of different stellar masses on the 21-cm signal and so aid in the interpretation of the effects of our more complicated four example IMFs which we show in the next subsection. 

\begin{figure*}
    \centering
    \includegraphics{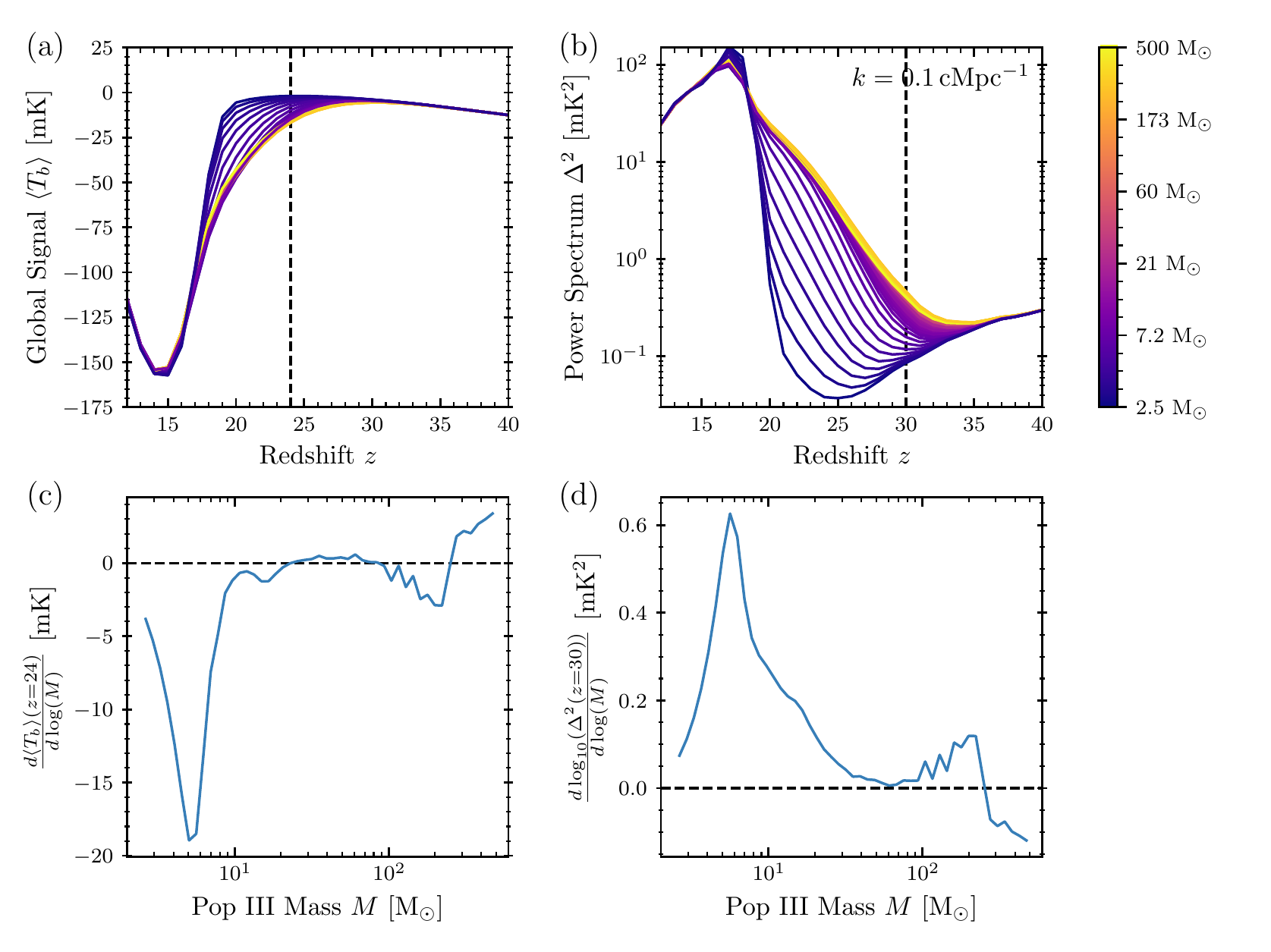}
    \caption{21-cm global signal (top left, a) and power spectrum at $k = 0.1$\,cMpc$^{-1}$ (top right, b) for single-mass Pop~III IMFs composed of stars with mass $M$, indicated by the colour of the line (see colour-bar on the right). The derivative of the global signal and log power spectrum with respect to the logarithm of $M$ are shown on the bottom plots (panels c and d) at fixed redshift 24 and 30 respectively (indicated by black dashed lines in panels (a) and (b)). A generic high-mass IMF behaviour is apparent in panels (a) and (b) where IMFs composed of stars above $\approx 20$\,M$_{\astrosun}$ all produce very similar 21-cm signals. For lower $M$ the power spectra and global signals show greater variation with $M$. This increased sensitivity of the 21-cm signal to $M$ occurs for $M < 10$\,M$_{\astrosun}$ in the global signal and for $M < 20$\,M$_{\astrosun}$ in the power spectrum, as can be seen from the bottom panels. }
    \label{fig:mono}
\end{figure*}

The 21-cm global signal and power spectrum at fixed \mbox{$k=0.1$ cMpc$^{-1}$} are shown in Fig.~\ref{fig:mono} (panels a and b respectively) for logarithmically spaced values of $M$ with masses in the range from $2.5$ to $500$\,M$_{\astrosun}$. A visual inspection of these results (as well as power spectra at a wide range of scales $0.05$\,--\,$1.0$\,cMpc$^{-1}$, not shown) suggests that high-mass dominated IMFs produce very similar 21-cm signals with little difference between the models with $M> 20$\,M$_{\astrosun}$. As $M$ decreases, divergence from the generic high-mass behaviour is first seen in the power spectrum, and later in the global signals as well.  In both the power spectrum and global signals these deviations manifest themselves primarily at high redshifts $z > 20$. However, the lower $M$, the stronger the impact on the late-time signals.

For IMFs with lower $M$ the global signal absorption trough shifts to lower redshifts and the onset of the cosmic dawn power spectrum peak is seen at later times compared to the generic high-mass 21-cm signal. We also observe a slight increase in the magnitude of the power spectrum peak. The delay in the signals as we move to lower $M$ is attributed to the weaker WF effect, which stems from lower values of the Lyman-band radiation fields. As we have seen above, the dominant mechanism for the latter is the sharp increase in the lifetime of the Pop~III stars at lower $M$ which leads to the emission being spread over a longer time period. Another mechanism that leads to the reduction in the Lyman-band fields is the inefficiency (over the lifetime) of stars lower in mass than $4$\,M$_{\astrosun}$ at production of Lyman-band photons. However, this is a secondary effect, as even at the masses which yield the highest total number of photons per stellar baryon ($4$\,--\,$7$\,M$_{\astrosun}$, Fig.~\ref{fig:ems_and_life}) the 21-cm signals are still delayed compared to the cases with less efficient higher mass stars.

By taking the derivative (at a fixed redshift) of both the 21-cm global signal and log power spectrum with respect to $\log(M)$, we can identify the threshold mass at which the signals start to deviate from the generic high-mass behaviour. Such derivatives evaluated for the global signal at $z = 24$ and power spectrum at $z = 30$ are shown in panels (c) and (d), respectively, of Fig.~\ref{fig:mono}. Both derivatives fluctuate around zero at the high $M$ end, confirming that the signal does not vary with $M$ for the high-mass dominated cases. At lower masses, however, large peaks in the magnitudes of both derivatives can be seen, indicating the mass range at which the 21-cm signal becomes sensitive to the mass scale $M$ of the Pop~III IMF. For the global signal at $z = 24$  we find the derivative to be large (in absolute value) at $M\lesssim  10$\,M$_{\astrosun}$, and for the  power spectrum (at $z = 30$) the derivative grows at $M\lesssim 20$\,M$_{\astrosun}$. 

In summary, through the suite of the single-mass test cases, we showed that the 21-cm signal is not sensitive to the shape of the Pop~III IMF in the high-mass regime,  $M\gtrsim  20$\,M$_{\astrosun}$. Therefore, if the majority of the Pop~III stellar mass was in stars more massive than $20$\,M$_{\astrosun}$ as some simulations suggest  \citep{Hirano_2014,Hirano_2017}, the 21-cm signal would not be able to probe the shape of the Pop~III IMF in detail. However, we should still be able to test whether or not the first stellar population was composed of predominantly massive stars, by checking if the signal traces the generic high-mass solution. Conversely, if the majority of the Pop~III stellar mass is in stars below the $20$\,M$_{\astrosun}$ threshold, the 21-cm signal from cosmic dawn can be used to probe the Pop~III IMF in more detail. However, degeneracies with other parameters, such as the Pop III star formation efficiency $f_{*,{\rm III}}$ which we consider in Sec.~\ref{ssec:ftar_degneracy}, as well as the instrumental noise and systematics, may weaken the constraints or measurements in practice.

\subsection{21-cm signal with realistic Pop~III IMFs}\label{ssec:Fid_21cm}

\begin{figure*}
    \centering
    \includegraphics{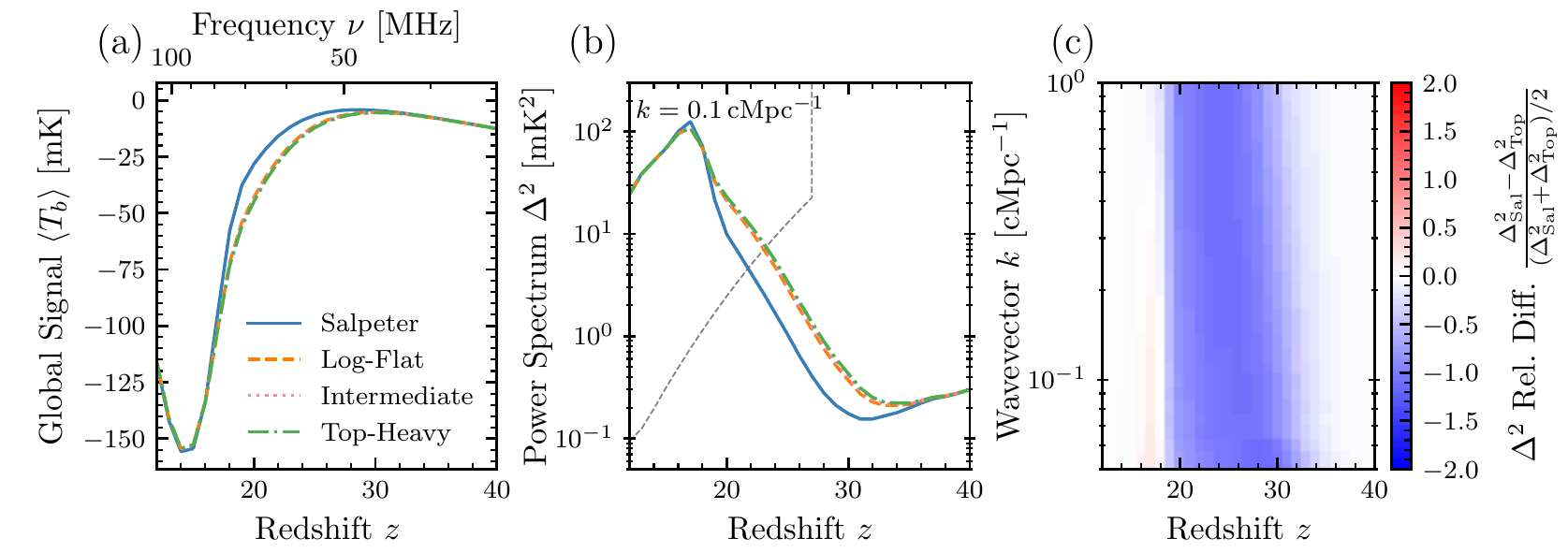}
    \caption{21-cm signal with our four example IMFs.
    We show a detailed comparison of the predicted global 21-cm signals (left, a), and $k = 0.1$\,cMpc$^{-1}$ power spectrum (middle, b) between our four example IMFs. Also shown is the relative difference between the two most disparate power spectra (right, c), those of the Salpeter and Top-Heavy IMFs. In panel (b) the projected sensitivity of 1,000 hours of SKA1-Low single beam observations with $10$\,MHz bandwidth is shown as a grey dashed line. The global signal absorption trough and power spectrum cosmic dawn peak are found to occur later by $\Delta z \approx 1$ for the Salpeter IMF due to its lower Lyman-band emissivities and being dominated by long-lived stars. This shift results in fixed redshift global signal relative differences of up to $59$ per cent at $z = 24$ and large power spectrum relative differences reaching a peak of 131 per cent at $z = 30$ between the Salpeter and Top-Heavy IMF.}
    \label{fig:IMF_comp}
\end{figure*}

We now consider the variations in the 21-cm signal between our example Pop~III IMFs (Table \ref{tab:fid_imfs}), showing the results in Fig.~\ref{fig:IMF_comp}. As opposed to our test cases in the previous subsection, each example IMF represents a distribution of masses and, thus, the effect on the 21-cm signal is more complicated. However, owing to the typical masses of stars in the case of the Intermediate, Log-Flat, and Top-Heavy IMFs being very high (with $M_{50\%} =$ 75, 113, and 354 \,M$_{\astrosun}$, respectively), we expect the 21-cm signals of these three IMFs to closely follow the generic high-mass behaviour. This is indeed what we find from our simulations (Fig.~\ref{fig:IMF_comp}). On the other hand, the Salpeter IMF is dominated by low-mass stars (with  $M_{50\%}=4.1$\,M$_{\astrosun}$). Therefore, we expect it to have a delayed 21-cm signal compared to the other cases.  This is exactly what we see in Fig.~\ref{fig:IMF_comp}, where both the power spectrum and the global signal corresponding to the Salpeter IMF are shifted to lower redshifts by  ${\Delta z \approx 1}$ compared to the more massive-star dominated cases.

Comparing the two IMFs with the greatest difference in cosmic dawn 21-cm signals, namely Salpeter and Top-Heavy, we find power spectrum relative differences of up to 131 per cent at ${z = 30}$, shown in the panel (c) of Fig.~\ref{fig:IMF_comp}. With relative difference $> 80\%$ down to ${z = 21}$. These large differences in the power spectrum predicted between the two IMFs suggest that the 21-cm power spectrum from $z = 20$\,--\,$35$ is potentially a useful observable for constraining the Pop~III IMF if observations are sufficiently precise. Putting this into an experimental context, the redshift $z = 21$ power spectrum differences of 10.2\,mK$^2$ seen between these two IMFs exceeds the thermal-noise sensitivity predicted for 1000 hrs of SKA1-Low observation~\citep{Koopmans_2015} by a factor of 2.9. 

Furthermore, the shift in redshift of $\Delta z \approx 1$ seen in the 21-cm global signal absorption trough between the two most extreme IMFs leads to a difference of up to $18.1$\,mK in the global signal at {$z = 19$}. The EDGES collaboration reported root-mean-square residuals of $25$\,mK \citep{EDGES} for their existing experiment and the upcoming global signal experiment REACH is forecast to be able to reach a sensitivity of $5$\,mK using $\sim 2500$ hrs of observations~\citep{Anstey_2021, REACH}. Hence, the differences seen between the predicted global signals are comparable to the sensitivity of current global signal experiments and above the sensitivity forecast for near-future experiments.

The above suggests that at least in some scenarios the differences between 21-cm signals caused by variations in the Pop~III IMF are measurable. However, so far all of the above has assumed one set of astrophysical simulation parameters that lead to the Pop~III star formation rate density (SFRD) depicted in Fig.~\ref{fig:sfrd}. While this SFRD is consistent with some of the existing numerical simulations, we are ignorant of the real star formation history at cosmic dawn due to the lack of  observations. Therefore, it is reasonable to consider a series of alternative star formation models. 

\begin{figure*}
    \centering
    \includegraphics{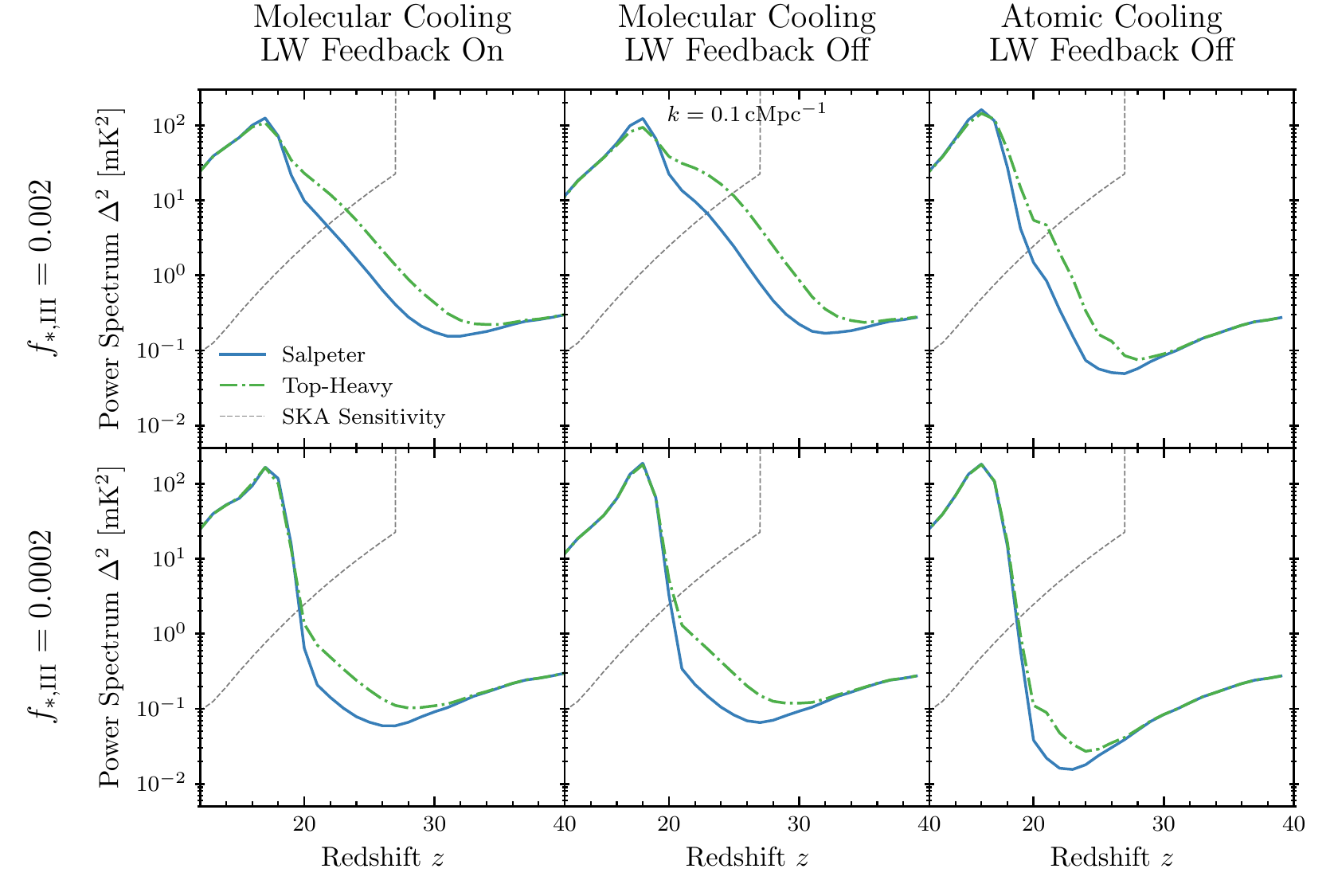}
    \caption{21-cm power spectra for our different star formation models.
    Each panel shows a comparison of the $k = 0.1$\,cMpc$^{-1}$ 21-cm power spectrum predicted for the Salpeter and Top-Heavy Pop III IMF, with each different panel representing a different star formation model. Also shown as a grey dashed line is the anticipated SKA1-Low sensitivity for 1000 hours of observation. Panels in the top row have our standard $f_{*,{\rm III}} = 0.002$ star formation efficiency, and those in the bottom row a reduced value of $f_{*,{\rm III}} = 0.0002$. Models in the first column have Pop~III stars forming in molecular cooling halos ($V_\mathrm{c} = 4.2$\,km\,s$^{-1}$) subject to strong LW feedback, whereas the second column assumes negligible LW feedback (i.e. disabled). The final column considers the case of Pop~III stars forming in atomic cooling halos ($V_\mathrm{c} = 16.5$\,km\,s$^{-1}$). We see in all cases the power spectra differ by around a factor of 2 between $z = 20$ and $z = 25$, with differences seen beyond $z = 30$ in some cases. A confident 21-cm power spectrum signal detection in this redshift range would, hence, be able to distinguish between the two IMF models.}
    \label{fig:alt_sf_modes}
\end{figure*}

The star formation model considered so far assumed Pop III stars form in molecular cooling halos (critical virial velocity for star formation $V_\mathrm{c} = 4.2$\,km\,s$^{-1}$), subject to a strong LW feedback \citep{Fialkov_2013}, with a star formation efficiency of $f_{*,{\rm III}} = 0.002$. Let us now consider alternative star formation histories with a lower star formation efficiency $f_{*,{\rm III}} = 0.0002$. We also consider Pop III star formation in atomic cooling halos ($V_\mathrm{c} = 16.5$\,km\,s$^{-1}$) as well as the case with negligible LW feedback in molecular cooling halos\footnote{Recent work by \citet{Latif_2019}  suggested that the neglection of H$_2$ self-shielding in \citet{Machacek_2001} has led to an overestimate of the strength of the LW feedback  \citep[including in our own simulations][]{Fialkov_2013}.}  (potentially due to halo self-shielding). The resulting  21-cm power spectra for the Salpeter and Top-Heavy IMFs at a comoving wavevector of \mbox{$k = 0.1$\,cMpc$^{-1}$} are shown in Fig.~\ref{fig:alt_sf_modes}.  In all of the considered cases we find differences in the power spectrum greater than a factor of 2 above $z \approx 20$, and so a confident detection of the 21-cm power spectrum in the range $z = 20$ to $z = 25$ would allow the two IMFs to be distinguished. Comparing to the expected sensitivity for 1000\,hrs of SKA1-LOW observation \citep{Koopmans_2015, NenuFar}, we see that the differences are detectable in all the considered cases with  $f_{*,{\rm III}} = 0.002$. In contrast, for $f_{*,{\rm III}} = 0.0002$ the 21-cm signals are considerably weaker and the IMF signature is below the sensitivity of the  SKA1-LOW. Additionally, we find that self-shielding of molecular hydrogen in minihalos could result in enhanced differences between the 21-cm power spectrum of the two IMFs. This is attributable to the enhanced Pop~III star formation rate owing to the negligible LW feedback.

\subsection{$f_{*,{\rm III}}$ degeneracy}\label{ssec:ftar_degneracy}

Our simulations  predict potentially observable differences between the 21-cm global signals and power spectra of high and low-mass dominated Pop~III IMFs. If the 21-cm signal alone is to constrain the Pop~III IMF, an important question that needs to be addressed is whether the  signature in the 21-cm signal of the IMF is degenerate with other astrophysical processes.

We found earlier that the differences in signal between IMFs arise because of the lower intensity of the cosmic-dawn Lyman-band radiation field, with the Salpeter IMF resulting in a lower intensity compared to the IMFs dominated by more massive stars. Although details of the IMF signature in the 21-cm signal could be unique, a general suppression in   Lyman-band radiation could also be  caused by a lower Pop III star formation efficiency $f_{*,{\rm III}}$ or a larger $V_\mathrm{c}$\footnote{ Note that while some correlations between $f_{*,{\rm III}}$, $V_\mathrm{c}$, and IMF are expected due to radiative feedback cutting off star formation, it is impossible to model all these effects from first principles. Therefore, here we adopt an observation-driven approach in which $f_{*,{\rm III}}$, $V_\mathrm{c}$, and IMF are modelled separately. Such an approach allows for data to distinguish between processes and point out degeneracies.}.  

\begin{figure*}
    \centering
    \includegraphics{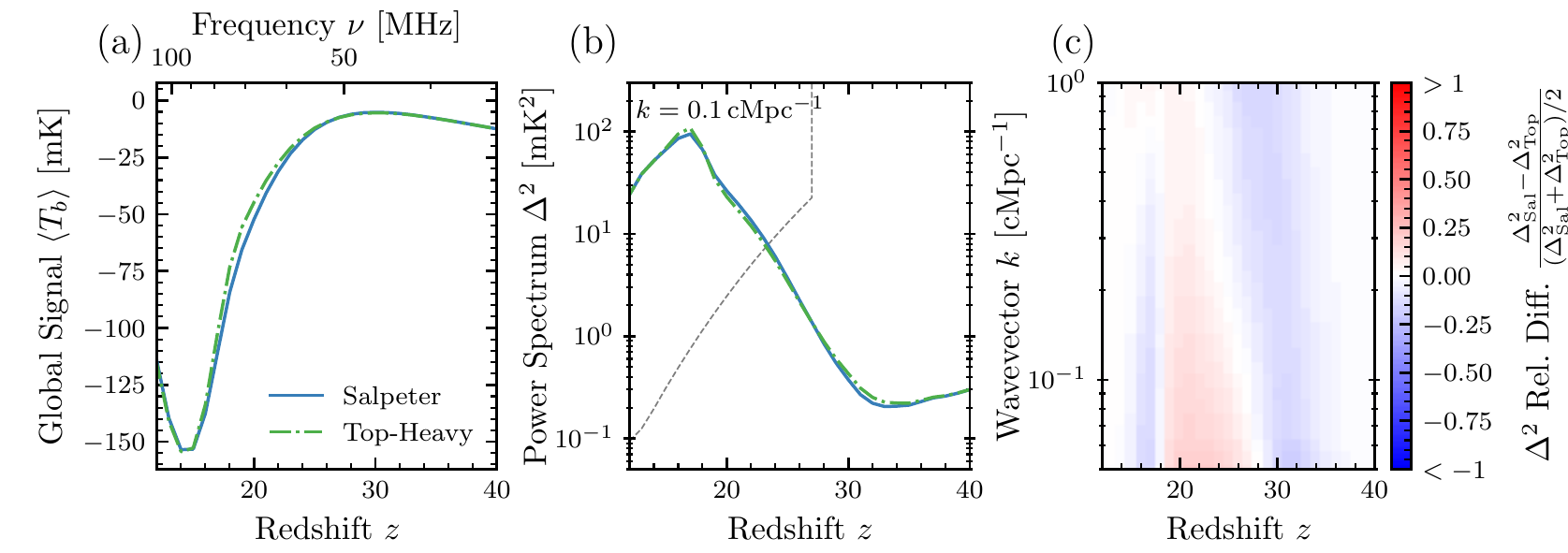}
    \caption{21-cm signal star formation parameters degeneracy. We show the 21-cm global signals (left, a),  power spectrum at $k = 0.1$\,cMpc$^{-1}$ (middle, b), and power spectrum relative difference (right, c) predicted for the Salpeter IMF with $f_{*,{\rm III}} = 0.00528$ and Top-Heavy IMF with $f_{*,{\rm III}} = 0.002$. The projected 1000 hour SKA1-Low sensitivity is shown in panel (b) as a grey gashed line. While the positions of the global signal minima coincide, the shapes of the global signals are slightly different, with a broader absorption trough for the Salpeter IMF. These small variations in shape leave residual differences of $7.5$\,mK between the $z = 20$ global signals. Also the shapes of the power spectra differ with a potentially measurable $13.3$\,mK$^2$ difference seen between the two IMFs at $z = 17$ $k = 0.1$\,cMpc$^{-1}$. Together these differences indicate that the effects of  $f_{*,{\rm III}}$ and the IMF shape are not fully degenerate.}
    \label{fig:fstar}
\end{figure*}

Here, to demonstrate the degree of degeneracy, we concentrate on an illustrative example leaving  a full analysis of the parameter space of Pop~III star formation models to future work. We take the case of the Top-Heavy IMF with \mbox{$f_{*,{\rm III}} = 0.002$} as a reference. Next, we calculate a series of 21-cm signals for the Salpeter IMF varying $f_{*,{\rm III}}$ values and keeping all the other model parameters fixed to their reference case values. We then identify the value of $f_{*,{\rm III}}$ which minimizes the maximum relative difference in the power spectra between the Salpeter case and the reference case. Through this process we find that the minimal maximum relative difference is obtained for $f_{*,{\rm III}} = 0.00528$ and is 28 per cent,  down from 131 per cent when the models with the different IMFs have the same $f_{*,{\rm III}}$. A visual comparison of the 21-cm signal for the Salpeter IMF with ${f_{*,{\rm III}} = 0.00528}$ and the Top-Heavy IMF with ${f_{*,{\rm III}} = 0.002}$ is shown in Fig.~\ref{fig:fstar}. 

We find a small but non-negligible difference between the global signals and the power spectra suggesting that the effects of $f_{*,{\rm III}}$ and the shape of IMF are not fully degenerate. The larger $f_{*,{\rm III}}$ in the Salpeter case required to match the 21-cm signal for the Top-Heavy IMF is expected because in the case of the Salpeter IMF stellar emission is spread over a longer time period and, thus, is effectively less efficient. With this boosted $f_{*,{\rm III}}$, the Salpeter case develops a strong global 21-cm signal faster, being 7.5\,mK deeper than the Top-Heavy IMF case at $z = 20$. Variations are also seen between the two power spectra, the maximum relative difference of 28 per cent is found at $z = 23$, with differences of a similar magnitude (17 per cent) at the power spectrum peak at $z = 17$. This $13.3$\,mK$^2$ difference seen between the power spectra of the two IMFs at $z = 17$ \mbox{$k = 0.1$\,cMpc$^{-1}$} is $17.6$ times the projected SKA1-Low thermal-noise sensitivity, demonstrating that, in this case, the partial $f_*$ degeneracy does not eliminate the possibility of constraining the Pop~III IMF using the SKA.

\section{Conclusions}\label{sec:conc}

In this work we explore the impact of the IMF of Pop~III stars on the cosmic dawn 21-cm signal focusing on the effects mediated by photons in the Lyman band. To bracket the uncertainty in Pop~III IMF we consider four different scenarios which include extreme cases as well as IMFs motivated by simulations of star formation. As we focus on the early stages of cosmic dawn, our simulations of the 21-cm signal include  WF coupling and LW  feedback, as well as several subtle effects such as Ly\,$\alpha$ and CMB heating, multiple scattering of Ly\,$\alpha$  photons, and a gradual transition from Pop~III to Pop~II star formation. Throughout this work we ignore X-ray heating and ionization which are expected to drive the 21-cm signal at later times. 

To rigorously model stellar emissivity in the Lyman band, we simulate stellar evolution histories for a range of initially metal-free stars using the stellar evolution code \textsc{mesa}. We compute the emission spectra of these stars throughout their lives using a stellar atmosphere code  \textsc{tlusty}. Finally, to compute the Lyman band emission spectra for a population of stars with each considered IMF, we integrate over the individual stellar-emission spectra weighted by the stellar-mass distribution. The resultant spectra are used in our 21-cm signal simulations to calculate the global signals and the large-scale power spectra. 

Through this work, we also relax two assumptions usually made in the literature, namely that (1) the mean emission rate of a Pop~III star is represented by its ZAMS emission rate and (2) Pop~III stellar emission can be treated as instantaneous. These assumptions were found to introduce errors of up to 56 per cent and 53 per cent in the global signals and 103 per cent and 119 per cent in the power spectra, respectively. Because such errors are compatible with the expected precision of the upcoming experiments, our results show that accurately modelling the evolution of Pop~III stars and their lifetimes is required for precision analysis and interpretation of the cosmic-dawn 21-cm signal. Our models rely on the following assumptions: all stars are assumed to form in isolation, be non-rotating, and have no mass loss on the main sequence. These three assumptions are not fully supported by simulations \citep{Marigo_2003,Stacy_2010,Stacy_2013,Sugimura_2020,Murphy_2021} and, thus, could be further relaxed in future work. 

Using the methods developed in this work, we find that, in the absence of X-ray heating and ionization, the 21-cm signal is sensitive to the Pop~III IMF and can probe the typical mass of stars if the IMF is dominated by stars lighter than $20$\,M$_{\astrosun}$. In contrast,  the 21-cm signal is not sensitive to the details of the IMF if it is dominated by heavier stars. By considering two extreme examples of a Salpeter IMF dominated by low-mass stars (50 per cent of mass is in stars lighter than $4.1$\,M$_{\astrosun}$)  and a Top-Heavy IMF dominated by stars with $M > 354$\,M$_{\astrosun}$, we find relative differences of up to 131 per cent in the power spectrum and $59$ per cent in the global signal. These factor of 2 differences in the power spectra were shown to be robust against changes in the star formation model parameters used in our simulations. By considering the degeneracy between the effects of the IMF and those of other parameters on the 21-cm signal such as Pop III star formation efficiency, we find that only a partial degeneracy is present in the signal. However, a more quantitative exploration of degeneracy is required and is left for future work. Adding the dependence of X-ray heating and ionization on stellar IMF will increase the discrepancies between different scenarios and help remove the degeneracy between the effects of the IMF and other model parameters.

The signatures introduced by the IMF in the 21-cm signal are found to be at the noise level of existing 21-cm global instruments. Future generations of both interferometers and radiometers are expected to have lower noise, thus allowing us to probe properties of the first stellar population via the 21-cm signal.

\section*{Acknowledgements}
We would like to express  gratitude to Jason Jaacks for providing us with the Pop~III star formation rate densities computed in \citet{Jaacks_2019}. We also thank the anonymous referee for their insightful comments that helped improving our paper. TGJ thanks the Science and Technology Facilities Council (STFC) for their support through grant number ST/V506606/1. NSS and AF are supported by AF's Royal Society University Research Fellowship (\#181073 and \#180523 respectively).
GMM acknowledges funding by the STFC consolidated grant ST/R000603/1. 
MM was supported by the Max-Planck-Gesellschaft via the fellowship of the International Max Planck Research School for Astronomy and Cosmic Physics at the University of Heidelberg (IMPRS-HD).
RGI thanks the STFC for funding his Rutherford fellowship under grant ST/L003910/1 and helping him through the difficult Covid-19 period with grant ST/R000603/1.
EdLA acknowledges the support of the Science and Technology Facilities Council (UK) to this work. 
WJH was supported by a Royal Society University Research Fellowship. 
RB acknowledges the support of the Israel Science Foundation (grant No.\ 2359/20), the Ambrose Monell Foundation, the Institute for Advanced Study, the Vera Rubin Presidential Chair in Astronomy, and the Packard Foundation.

\section*{Data Availability}

The Lyman-band lifetime emission spectra and emission rates of individual Pop~III stars computed in this work are made available at \href{https://zenodo.org/record/5553052}{10.5281/zenodo.5553052}, alongside pre-computed photon emissivities for each of our example IMFs. \textsc{mesa} inlists are available upon request.



\bibliographystyle{mnras}
\bibliography{IMF_Impacts_Via_Ly} 

\begin{thebibliography}{}
\makeatletter
\relax
\def\mn@urlcharsother{\let\do\@makeother \do\$\do\&\do\#\do\^\do\_\do\%\do\~}
\def\mn@doi{\begingroup\mn@urlcharsother \@ifnextchar [ {\mn@doi@}
  {\mn@doi@[]}}
\def\mn@doi@[#1]#2{\def\@tempa{#1}\ifx\@tempa\@empty \href
  {http://dx.doi.org/#2} {doi:#2}\else \href {http://dx.doi.org/#2} {#1}\fi
  \endgroup}
\def\mn@eprint#1#2{\mn@eprint@#1:#2::\@nil}
\def\mn@eprint@arXiv#1{\href {http://arxiv.org/abs/#1} {{\tt arXiv:#1}}}
\def\mn@eprint@dblp#1{\href {http://dblp.uni-trier.de/rec/bibtex/#1.xml}
  {dblp:#1}}
\def\mn@eprint@#1:#2:#3:#4\@nil{\def\@tempa {#1}\def\@tempb {#2}\def\@tempc
  {#3}\ifx \@tempc \@empty \let \@tempc \@tempb \let \@tempb \@tempa \fi \ifx
  \@tempb \@empty \def\@tempb {arXiv}\fi \@ifundefined
  {mn@eprint@\@tempb}{\@tempb:\@tempc}{\expandafter \expandafter \csname
  mn@eprint@\@tempb\endcsname \expandafter{\@tempc}}}

\bibitem[\protect\citeauthoryear{{Abdurashidova} et~al.,}{{Abdurashidova}
  et~al.}{2022}]{HERA}
{Abdurashidova} Z.,  et~al., 2022, \mn@doi [\apj] {10.3847/1538-4357/ac1c78},
  \href {https://ui.adsabs.harvard.edu/abs/2022ApJ...925..221A} {925, 221}

\bibitem[\protect\citeauthoryear{{Ahn}, {Shapiro}, {Iliev}, {Mellema}  \&
  {Pen}}{{Ahn} et~al.}{2009}]{Ahn_2009}
{Ahn} K.,  {Shapiro} P.~R.,  {Iliev} I.~T.,  {Mellema} G.,   {Pen} U.-L.,
  2009, \mn@doi [\apj] {10.1088/0004-637X/695/2/1430}, \href
  {https://ui.adsabs.harvard.edu/abs/2009ApJ...695.1430A} {695, 1430}

\bibitem[\protect\citeauthoryear{{Anstey}, {Cumner}, {de Lera Acedo}  \&
  {Handley}}{{Anstey} et~al.}{2022}]{Anstey_2021}
{Anstey} D.,  {Cumner} J.,  {de Lera Acedo} E.,   {Handley} W.,  2022, \mn@doi
  [\mnras] {10.1093/mnras/stab3211}, \href
  {https://ui.adsabs.harvard.edu/abs/2022MNRAS.509.4679A} {509, 4679}

\bibitem[\protect\citeauthoryear{{Auer} \& {Mihalas}}{{Auer} \&
  {Mihalas}}{1969}]{Auer_1969}
{Auer} L.~H.,  {Mihalas} D.,  1969, \mn@doi [\apj] {10.1086/150226}, \href
  {https://ui.adsabs.harvard.edu/abs/1969ApJ...158..641A} {158, 641}

\bibitem[\protect\citeauthoryear{{Barkana}}{{Barkana}}{2016}]{Barkana_2016}
{Barkana} R.,  2016, \mn@doi [\physrep] {10.1016/j.physrep.2016.06.006}, \href
  {https://ui.adsabs.harvard.edu/abs/2016PhR...645....1B} {645, 1}

\bibitem[\protect\citeauthoryear{{Barkana}}{{Barkana}}{2018}]{Barkana:2018}
{Barkana} R.,  2018, \mn@doi [\nat] {10.1038/nature25791}, \href
  {https://ui.adsabs.harvard.edu/abs/2018Natur.555...71B} {555, 71}

\bibitem[\protect\citeauthoryear{{Barkana} \& {Loeb}}{{Barkana} \&
  {Loeb}}{2004}]{Barkana_2004}
{Barkana} R.,  {Loeb} A.,  2004, \mn@doi [\apj] {10.1086/421079}, \href
  {https://ui.adsabs.harvard.edu/abs/2004ApJ...609..474B} {609, 474}

\bibitem[\protect\citeauthoryear{{Barkana} \& {Loeb}}{{Barkana} \&
  {Loeb}}{2005}]{Barkana_2005}
{Barkana} R.,  {Loeb} A.,  2005, \mn@doi [\apj] {10.1086/429954}, \href
  {https://ui.adsabs.harvard.edu/abs/2005ApJ...626....1B} {626, 1}

\bibitem[\protect\citeauthoryear{{Bond}}{{Bond}}{1981}]{Bond_1981}
{Bond} H.~E.,  1981, \mn@doi [\apj] {10.1086/159186}, \href
  {https://ui.adsabs.harvard.edu/abs/1981ApJ...248..606B} {248, 606}

\bibitem[\protect\citeauthoryear{{Bowman}, {Rogers}, {Monsalve}, {Mozdzen}  \&
  {Mahesh}}{{Bowman} et~al.}{2018}]{EDGES}
{Bowman} J.~D.,  {Rogers} A. E.~E.,  {Monsalve} R.~A.,  {Mozdzen} T.~J.,
  {Mahesh} N.,  2018, \mn@doi [\nat] {10.1038/nature25792}, \href
  {https://ui.adsabs.harvard.edu/abs/2018Natur.555...67B} {555, 67}

\bibitem[\protect\citeauthoryear{{Bradley}, {Tauscher}, {Rapetti}  \&
  {Burns}}{{Bradley} et~al.}{2019}]{Bradley_2019}
{Bradley} R.~F.,  {Tauscher} K.,  {Rapetti} D.,   {Burns} J.~O.,  2019, \mn@doi
  [\apj] {10.3847/1538-4357/ab0d8b}, \href
  {https://ui.adsabs.harvard.edu/abs/2019ApJ...874..153B} {874, 153}

\bibitem[\protect\citeauthoryear{{Bromm}}{{Bromm}}{2013}]{Bromm_2013}
{Bromm} V.,  2013, \mn@doi [Reports on Progress in Physics]
  {10.1088/0034-4885/76/11/112901}, \href
  {https://ui.adsabs.harvard.edu/abs/2013RPPh...76k2901B} {76, 112901}

\bibitem[\protect\citeauthoryear{{Bromm} \& {Larson}}{{Bromm} \&
  {Larson}}{2004}]{Bromm_2004}
{Bromm} V.,  {Larson} R.~B.,  2004, \mn@doi [\araa]
  {10.1146/annurev.astro.42.053102.134034}, \href
  {https://ui.adsabs.harvard.edu/abs/2004ARA&A..42...79B} {42, 79}

\bibitem[\protect\citeauthoryear{{Bromm}, {Kudritzki}  \& {Loeb}}{{Bromm}
  et~al.}{2001}]{Bromm_2001}
{Bromm} V.,  {Kudritzki} R.~P.,   {Loeb} A.,  2001, \mn@doi [\apj]
  {10.1086/320549}, \href
  {https://ui.adsabs.harvard.edu/abs/2001ApJ...552..464B} {552, 464}

\bibitem[\protect\citeauthoryear{{Chatterjee}, {Dayal}, {Choudhury}  \&
  {Schneider}}{{Chatterjee} et~al.}{2020}]{Chatterjee_2020}
{Chatterjee} A.,  {Dayal} P.,  {Choudhury} T.~R.,   {Schneider} R.,  2020,
  \mn@doi [\mnras] {10.1093/mnras/staa1609}, \href
  {https://ui.adsabs.harvard.edu/abs/2020MNRAS.496.1445C} {496, 1445}

\bibitem[\protect\citeauthoryear{{Chen}, {Whalen}, {Wollenberg}, {Glover}  \&
  {Klessen}}{{Chen} et~al.}{2017}]{Chen_2017}
{Chen} K.-J.,  {Whalen} D.~J.,  {Wollenberg} K. M.~J.,  {Glover} S. C.~O.,
  {Klessen} R.~S.,  2017, \mn@doi [\apj] {10.3847/1538-4357/aa7b34}, \href
  {https://ui.adsabs.harvard.edu/abs/2017ApJ...844..111C} {844, 111}

\bibitem[\protect\citeauthoryear{{Chuzhoy} \& {Shapiro}}{{Chuzhoy} \&
  {Shapiro}}{2006}]{Chuzhoy_2006}
{Chuzhoy} L.,  {Shapiro} P.~R.,  2006, \mn@doi [\apj] {10.1086/507670}, \href
  {https://ui.adsabs.harvard.edu/abs/2006ApJ...651....1C} {651, 1}

\bibitem[\protect\citeauthoryear{{Ciardi}, {Salvaterra}  \& {Di
  Matteo}}{{Ciardi} et~al.}{2010}]{Ciardi:2010}
{Ciardi} B.,  {Salvaterra} R.,   {Di Matteo} T.,  2010, \mn@doi [\mnras]
  {10.1111/j.1365-2966.2009.15843.x}, \href
  {https://ui.adsabs.harvard.edu/abs/2010MNRAS.401.2635C} {401, 2635}

\bibitem[\protect\citeauthoryear{{Cohen}, {Fialkov}  \& {Barkana}}{{Cohen}
  et~al.}{2016}]{Cohen:2016}
{Cohen} A.,  {Fialkov} A.,   {Barkana} R.,  2016, \mn@doi [\mnras]
  {10.1093/mnrasl/slw047}, \href
  {https://ui.adsabs.harvard.edu/abs/2016MNRAS.459L..90C} {459, L90}

\bibitem[\protect\citeauthoryear{{Cohen}, {Fialkov}  \& {Barkana}}{{Cohen}
  et~al.}{2018}]{Cohen_2018}
{Cohen} A.,  {Fialkov} A.,   {Barkana} R.,  2018, \mn@doi [\mnras]
  {10.1093/mnras/sty1094}, \href
  {https://ui.adsabs.harvard.edu/abs/2018MNRAS.478.2193C} {478, 2193}

\bibitem[\protect\citeauthoryear{{Cohen}, {Fialkov}, {Barkana}  \&
  {Monsalve}}{{Cohen} et~al.}{2020}]{Cohen_2020}
{Cohen} A.,  {Fialkov} A.,  {Barkana} R.,   {Monsalve} R.~A.,  2020, \mn@doi
  [\mnras] {10.1093/mnras/staa1530}, \href
  {https://ui.adsabs.harvard.edu/abs/2020MNRAS.495.4845C} {495, 4845}

\bibitem[\protect\citeauthoryear{{Dopcke}, {Glover}, {Clark}  \&
  {Klessen}}{{Dopcke} et~al.}{2013}]{Dopcke_2013}
{Dopcke} G.,  {Glover} S. C.~O.,  {Clark} P.~C.,   {Klessen} R.~S.,  2013,
  \mn@doi [\apj] {10.1088/0004-637X/766/2/103}, \href
  {https://ui.adsabs.harvard.edu/abs/2013ApJ...766..103D} {766, 103}

\bibitem[\protect\citeauthoryear{{Ewall-Wice} et~al.,}{{Ewall-Wice}
  et~al.}{2016}]{MWA}
{Ewall-Wice} A.,  et~al., 2016, \mn@doi [\mnras] {10.1093/mnras/stw1022}, \href
  {https://ui.adsabs.harvard.edu/abs/2016MNRAS.460.4320E} {460, 4320}

\bibitem[\protect\citeauthoryear{{Ewall-Wice}, {Chang}, {Lazio}, {Dor{\'e}},
  {Seiffert}  \& {Monsalve}}{{Ewall-Wice} et~al.}{2018}]{Ewall_18}
{Ewall-Wice} A.,  {Chang} T.~C.,  {Lazio} J.,  {Dor{\'e}} O.,  {Seiffert} M.,
  {Monsalve} R.~A.,  2018, \mn@doi [\apj] {10.3847/1538-4357/aae51d}, \href
  {https://ui.adsabs.harvard.edu/abs/2018ApJ...868...63E} {868, 63}

\bibitem[\protect\citeauthoryear{{Farmer}, {Renzo}, {de Mink}, {Marchant}  \&
  {Justham}}{{Farmer} et~al.}{2019}]{Farmer_2019}
{Farmer} R.,  {Renzo} M.,  {de Mink} S.~E.,  {Marchant} P.,   {Justham} S.,
  2019, \mn@doi [\apj] {10.3847/1538-4357/ab518b}, \href
  {https://ui.adsabs.harvard.edu/abs/2019ApJ...887...53F} {887, 53}

\bibitem[\protect\citeauthoryear{{Feng} \& {Holder}}{{Feng} \&
  {Holder}}{2018}]{Feng_18}
{Feng} C.,  {Holder} G.,  2018, \mn@doi [\apjl] {10.3847/2041-8213/aac0fe},
  \href {https://ui.adsabs.harvard.edu/abs/2018ApJ...858L..17F} {858, L17}

\bibitem[\protect\citeauthoryear{{Fialkov} \& {Barkana}}{{Fialkov} \&
  {Barkana}}{2019}]{Fialkov_19}
{Fialkov} A.,  {Barkana} R.,  2019, \mn@doi [\mnras] {10.1093/mnras/stz873},
  \href {https://ui.adsabs.harvard.edu/abs/2019MNRAS.486.1763F} {486, 1763}

\bibitem[\protect\citeauthoryear{{Fialkov}, {Barkana}, {Tseliakhovich}  \&
  {Hirata}}{{Fialkov} et~al.}{2012}]{Fialkov_2012}
{Fialkov} A.,  {Barkana} R.,  {Tseliakhovich} D.,   {Hirata} C.~M.,  2012,
  \mn@doi [\mnras] {10.1111/j.1365-2966.2012.21318.x}, \href
  {https://ui.adsabs.harvard.edu/abs/2012MNRAS.424.1335F} {424, 1335}

\bibitem[\protect\citeauthoryear{{Fialkov}, {Barkana}, {Visbal},
  {Tseliakhovich}  \& {Hirata}}{{Fialkov} et~al.}{2013}]{Fialkov_2013}
{Fialkov} A.,  {Barkana} R.,  {Visbal} E.,  {Tseliakhovich} D.,   {Hirata}
  C.~M.,  2013, \mn@doi [\mnras] {10.1093/mnras/stt650}, \href
  {https://ui.adsabs.harvard.edu/abs/2013MNRAS.432.2909F} {432, 2909}

\bibitem[\protect\citeauthoryear{{Fialkov}, {Barkana}  \& {Visbal}}{{Fialkov}
  et~al.}{2014}]{Fialkov_2014}
{Fialkov} A.,  {Barkana} R.,   {Visbal} E.,  2014, \mn@doi [\nat]
  {10.1038/nature12999}, \href
  {https://ui.adsabs.harvard.edu/abs/2014Natur.506..197F} {506, 197}

\bibitem[\protect\citeauthoryear{{Field}}{{Field}}{1958}]{Field_1958}
{Field} G.~B.,  1958, \mn@doi [Proceedings of the IRE]
  {10.1109/JRPROC.1958.286741}, \href
  {https://ui.adsabs.harvard.edu/abs/1958PIRE...46..240F} {46, 240}

\bibitem[\protect\citeauthoryear{{Fragos}, {Lehmer}, {Naoz}, {Zezas}  \&
  {Basu-Zych}}{{Fragos} et~al.}{2013}]{Fragos:2013}
{Fragos} T.,  {Lehmer} B.~D.,  {Naoz} S.,  {Zezas} A.,   {Basu-Zych} A.,  2013,
  \mn@doi [\apjl] {10.1088/2041-8205/776/2/L31}, \href
  {https://ui.adsabs.harvard.edu/abs/2013ApJ...776L..31F} {776, L31}

\bibitem[\protect\citeauthoryear{{Fraser}, {Casey}, {Gilmore}, {Heger}  \&
  {Chan}}{{Fraser} et~al.}{2017}]{Fraser_2017}
{Fraser} M.,  {Casey} A.~R.,  {Gilmore} G.,  {Heger} A.,   {Chan} C.,  2017,
  \mn@doi [\mnras] {10.1093/mnras/stx480}, \href
  {https://ui.adsabs.harvard.edu/abs/2017MNRAS.468..418F} {468, 418}

\bibitem[\protect\citeauthoryear{{Frebel} \& {Norris}}{{Frebel} \&
  {Norris}}{2015}]{Frebel_2015}
{Frebel} A.,  {Norris} J.~E.,  2015, \mn@doi [\araa]
  {10.1146/annurev-astro-082214-122423}, \href
  {https://ui.adsabs.harvard.edu/abs/2015ARA&A..53..631F} {53, 631}

\bibitem[\protect\citeauthoryear{{Furlanetto}, {Oh}  \& {Briggs}}{{Furlanetto}
  et~al.}{2006}]{Furlanetto_2006}
{Furlanetto} S.~R.,  {Oh} S.~P.,   {Briggs} F.~H.,  2006, \mn@doi [\physrep]
  {10.1016/j.physrep.2006.08.002}, \href
  {https://ui.adsabs.harvard.edu/abs/2006PhR...433..181F} {433, 181}

\bibitem[\protect\citeauthoryear{{Garsden} et~al.,}{{Garsden}
  et~al.}{2021}]{LEDA}
{Garsden} H.,  et~al., 2021, \mn@doi [\mnras] {10.1093/mnras/stab1671}, \href
  {https://ui.adsabs.harvard.edu/abs/2021MNRAS.506.5802G} {506, 5802}

\bibitem[\protect\citeauthoryear{{Glover} \& {Brand}}{{Glover} \&
  {Brand}}{2001}]{Glover_2001}
{Glover} S.~C.~O.,  {Brand} P.~W.~J.~L.,  2001, \mn@doi [\mnras]
  {10.1046/j.1365-8711.2001.03993.x}, \href
  {https://ui.adsabs.harvard.edu/abs/2001MNRAS.321..385G} {321, 385}

\bibitem[\protect\citeauthoryear{{Greif}}{{Greif}}{2015}]{Greif_2015}
{Greif} T.~H.,  2015, \mn@doi [Computational Astrophysics and Cosmology]
  {10.1186/s40668-014-0006-2}, \href
  {https://ui.adsabs.harvard.edu/abs/2015ComAC...2....3G} {2, 3}

\bibitem[\protect\citeauthoryear{{Greif} \& {Bromm}}{{Greif} \&
  {Bromm}}{2006}]{Greif_2006}
{Greif} T.~H.,  {Bromm} V.,  2006, \mn@doi [\mnras]
  {10.1111/j.1365-2966.2006.11017.x}, \href
  {https://ui.adsabs.harvard.edu/abs/2006MNRAS.373..128G} {373, 128}

\bibitem[\protect\citeauthoryear{{Greif}, {Springel}, {White}, {Glover},
  {Clark}, {Smith}, {Klessen}  \& {Bromm}}{{Greif} et~al.}{2011}]{Greif_2011}
{Greif} T.~H.,  {Springel} V.,  {White} S. D.~M.,  {Glover} S. C.~O.,  {Clark}
  P.~C.,  {Smith} R.~J.,  {Klessen} R.~S.,   {Bromm} V.,  2011, \mn@doi [\apj]
  {10.1088/0004-637X/737/2/75}, \href
  {https://ui.adsabs.harvard.edu/abs/2011ApJ...737...75G} {737, 75}

\bibitem[\protect\citeauthoryear{{Haiman}, {Thoul}  \& {Loeb}}{{Haiman}
  et~al.}{1996}]{Haiman_1996}
{Haiman} Z.,  {Thoul} A.~A.,   {Loeb} A.,  1996, \mn@doi [\apj]
  {10.1086/177343}, \href
  {https://ui.adsabs.harvard.edu/abs/1996ApJ...464..523H} {464, 523}

\bibitem[\protect\citeauthoryear{{Haiman}, {Abel}  \& {Rees}}{{Haiman}
  et~al.}{2000}]{Haiman_2000}
{Haiman} Z.,  {Abel} T.,   {Rees} M.~J.,  2000, \mn@doi [\apj]
  {10.1086/308723}, \href
  {https://ui.adsabs.harvard.edu/abs/2000ApJ...534...11H} {534, 11}

\bibitem[\protect\citeauthoryear{{Hartwig}, {Bromm}, {Klessen}  \&
  {Glover}}{{Hartwig} et~al.}{2015}]{Hartwig_2015}
{Hartwig} T.,  {Bromm} V.,  {Klessen} R.~S.,   {Glover} S. C.~O.,  2015,
  \mn@doi [\mnras] {10.1093/mnras/stu2740}, \href
  {https://ui.adsabs.harvard.edu/abs/2015MNRAS.447.3892H} {447, 3892}

\bibitem[\protect\citeauthoryear{{Heger}, {Fryer}, {Woosley}, {Langer}  \&
  {Hartmann}}{{Heger} et~al.}{2003}]{Heger_2003}
{Heger} A.,  {Fryer} C.~L.,  {Woosley} S.~E.,  {Langer} N.,   {Hartmann} D.~H.,
   2003, \mn@doi [\apj] {10.1086/375341}, \href
  {https://ui.adsabs.harvard.edu/abs/2003ApJ...591..288H} {591, 288}

\bibitem[\protect\citeauthoryear{{Hibbard}, {Mirocha}, {Rapetti}, {Bassett},
  {Burns}  \& {Tauscher}}{{Hibbard} et~al.}{2022}]{Hibbard:2022}
{Hibbard} J.~J.,  {Mirocha} J.,  {Rapetti} D.,  {Bassett} N.,  {Burns} J.~O.,
  {Tauscher} K.,  2022, \mn@doi [\apj] {10.3847/1538-4357/ac5ea3}, \href
  {https://ui.adsabs.harvard.edu/abs/2022ApJ...929..151H} {929, 151}

\bibitem[\protect\citeauthoryear{{Hicks}, {Wells}, {Norman}, {Wise}, {Smith}
  \& {O'Shea}}{{Hicks} et~al.}{2021}]{Hicks_2021}
{Hicks} W.~M.,  {Wells} A.,  {Norman} M.~L.,  {Wise} J.~H.,  {Smith} B.~D.,
  {O'Shea} B.~W.,  2021, \mn@doi [\apj] {10.3847/1538-4357/abda3a}, \href
  {https://ui.adsabs.harvard.edu/abs/2021ApJ...909...70H} {909, 70}

\bibitem[\protect\citeauthoryear{{Hills}, {Kulkarni}, {Meerburg}  \&
  {Puchwein}}{{Hills} et~al.}{2018}]{Hills_2018}
{Hills} R.,  {Kulkarni} G.,  {Meerburg} P.~D.,   {Puchwein} E.,  2018, \mn@doi
  [\nat] {10.1038/s41586-018-0796-5}, \href
  {https://ui.adsabs.harvard.edu/abs/2018Natur.564E..32H} {564, E32}

\bibitem[\protect\citeauthoryear{{Hirano} \& {Bromm}}{{Hirano} \&
  {Bromm}}{2017}]{Hirano_2017}
{Hirano} S.,  {Bromm} V.,  2017, \mn@doi [\mnras] {10.1093/mnras/stx1220},
  \href {https://ui.adsabs.harvard.edu/abs/2017MNRAS.470..898H} {470, 898}

\bibitem[\protect\citeauthoryear{{Hirano}, {Hosokawa}, {Yoshida}, {Umeda},
  {Omukai}, {Chiaki}  \& {Yorke}}{{Hirano} et~al.}{2014}]{Hirano_2014}
{Hirano} S.,  {Hosokawa} T.,  {Yoshida} N.,  {Umeda} H.,  {Omukai} K.,
  {Chiaki} G.,   {Yorke} H.~W.,  2014, \mn@doi [\apj]
  {10.1088/0004-637X/781/2/60}, \href
  {https://ui.adsabs.harvard.edu/abs/2014ApJ...781...60H} {781, 60}

\bibitem[\protect\citeauthoryear{{Holzbauer} \& {Furlanetto}}{{Holzbauer} \&
  {Furlanetto}}{2012}]{Holzbauer_2012}
{Holzbauer} L.~N.,  {Furlanetto} S.~R.,  2012, \mn@doi [\mnras]
  {10.1111/j.1365-2966.2011.19752.x}, \href
  {https://ui.adsabs.harvard.edu/abs/2012MNRAS.419..718H} {419, 718}

\bibitem[\protect\citeauthoryear{{Hubeny}}{{Hubeny}}{1988}]{TLUSTY0}
{Hubeny} I.,  1988, \mn@doi [Computer Physics Communications]
  {10.1016/0010-4655(88)90177-4}, \href
  {https://ui.adsabs.harvard.edu/abs/1988CoPhC..52..103H} {52, 103}

\bibitem[\protect\citeauthoryear{{Hubeny} \& {Lanz}}{{Hubeny} \&
  {Lanz}}{2017a}]{TLUSTYI}
{Hubeny} I.,  {Lanz} T.,  2017a, arXiv e-prints, \href
  {https://ui.adsabs.harvard.edu/abs/2017arXiv170601859H} {p. arXiv:1706.01859}

\bibitem[\protect\citeauthoryear{{Hubeny} \& {Lanz}}{{Hubeny} \&
  {Lanz}}{2017b}]{TLUSTYII}
{Hubeny} I.,  {Lanz} T.,  2017b, arXiv e-prints, \href
  {https://ui.adsabs.harvard.edu/abs/2017arXiv170601935H} {p. arXiv:1706.01935}

\bibitem[\protect\citeauthoryear{{Hubeny} \& {Lanz}}{{Hubeny} \&
  {Lanz}}{2017c}]{TLUSTYIII}
{Hubeny} I.,  {Lanz} T.,  2017c, arXiv e-prints, \href
  {https://ui.adsabs.harvard.edu/abs/2017arXiv170601937H} {p. arXiv:1706.01937}

\bibitem[\protect\citeauthoryear{{Hubeny}, {Allende Prieto}, {Osorio}  \&
  {Lanz}}{{Hubeny} et~al.}{2021}]{TLUSTYIV}
{Hubeny} I.,  {Allende Prieto} C.,  {Osorio} Y.,   {Lanz} T.,  2021, arXiv
  e-prints, \href {https://ui.adsabs.harvard.edu/abs/2021arXiv210402829H} {p.
  arXiv:2104.02829}

\bibitem[\protect\citeauthoryear{{Ishigaki}, {Tominaga}, {Kobayashi}  \&
  {Nomoto}}{{Ishigaki} et~al.}{2018}]{2018_Ishigaki}
{Ishigaki} M.~N.,  {Tominaga} N.,  {Kobayashi} C.,   {Nomoto} K.,  2018,
  \mn@doi [\apj] {10.3847/1538-4357/aab3de}, \href
  {https://ui.adsabs.harvard.edu/abs/2018ApJ...857...46I} {857, 46}

\bibitem[\protect\citeauthoryear{{Jaacks}, {Finkelstein}  \& {Bromm}}{{Jaacks}
  et~al.}{2019}]{Jaacks_2019}
{Jaacks} J.,  {Finkelstein} S.~L.,   {Bromm} V.,  2019, \mn@doi [\mnras]
  {10.1093/mnras/stz1529}, \href
  {https://ui.adsabs.harvard.edu/abs/2019MNRAS.488.2202J} {488, 2202}

\bibitem[\protect\citeauthoryear{{Jana}, {Nath}  \& {Biermann}}{{Jana}
  et~al.}{2019}]{Jana_2019}
{Jana} R.,  {Nath} B.~B.,   {Biermann} P.~L.,  2019, \mn@doi [\mnras]
  {10.1093/mnras/sty3426}, \href
  {https://ui.adsabs.harvard.edu/abs/2019MNRAS.483.5329J} {483, 5329}

\bibitem[\protect\citeauthoryear{{Johnson}, {Dalla Vecchia}  \&
  {Khochfar}}{{Johnson} et~al.}{2013}]{Johnson_2013}
{Johnson} J.~L.,  {Dalla Vecchia} C.,   {Khochfar} S.,  2013, \mn@doi [\mnras]
  {10.1093/mnras/sts011}, \href
  {https://ui.adsabs.harvard.edu/abs/2013MNRAS.428.1857J} {428, 1857}

\bibitem[\protect\citeauthoryear{{Klessen}}{{Klessen}}{2019}]{Klessen:2019}
{Klessen} R.,  2019, {Formation of the first stars}.
pp 67--97, \mn@doi{10.1142/9789813227958\_0004}

\bibitem[\protect\citeauthoryear{{Koopmans} et~al.,}{{Koopmans}
  et~al.}{2015}]{Koopmans_2015}
{Koopmans} L.,  et~al., 2015, in Advancing Astrophysics with the Square
  Kilometre Array (AASKA14). p.~1 (\mn@eprint {arXiv} {1505.07568}),
  \mn@doi{10.22323/1.215.0001}

\bibitem[\protect\citeauthoryear{{Lanz} \& {Hubeny}}{{Lanz} \&
  {Hubeny}}{2003}]{Lanz_2003}
{Lanz} T.,  {Hubeny} I.,  2003, \mn@doi [\apjs] {10.1086/374373}, \href
  {https://ui.adsabs.harvard.edu/abs/2003ApJS..146..417L} {146, 417}

\bibitem[\protect\citeauthoryear{{Lanz} \& {Hubeny}}{{Lanz} \&
  {Hubeny}}{2007}]{Lanz_2007}
{Lanz} T.,  {Hubeny} I.,  2007, \mn@doi [\apjs] {10.1086/511270}, \href
  {https://ui.adsabs.harvard.edu/abs/2007ApJS..169...83L} {169, 83}

\bibitem[\protect\citeauthoryear{{Latif} \& {Khochfar}}{{Latif} \&
  {Khochfar}}{2019}]{Latif_2019}
{Latif} M.~A.,  {Khochfar} S.,  2019, \mn@doi [\mnras] {10.1093/mnras/stz2812},
  \href {https://ui.adsabs.harvard.edu/abs/2019MNRAS.490.2706L} {490, 2706}

\bibitem[\protect\citeauthoryear{{Leitherer} et~al.,}{{Leitherer}
  et~al.}{1999}]{Leitherer_1999}
{Leitherer} C.,  et~al., 1999, \mn@doi [\apjs] {10.1086/313233}, \href
  {https://ui.adsabs.harvard.edu/abs/1999ApJS..123....3L} {123, 3}

\bibitem[\protect\citeauthoryear{{Liu}, {Outmezguine}, {Redigolo}  \&
  {Volansky}}{{Liu} et~al.}{2019}]{Liu:2019}
{Liu} H.,  {Outmezguine} N.~J.,  {Redigolo} D.,   {Volansky} T.,  2019, \mn@doi
  [\prd] {10.1103/PhysRevD.100.123011}, \href
  {https://ui.adsabs.harvard.edu/abs/2019PhRvD.100l3011L} {100, 123011}

\bibitem[\protect\citeauthoryear{{Machacek}, {Bryan}  \& {Abel}}{{Machacek}
  et~al.}{2001}]{Machacek_2001}
{Machacek} M.~E.,  {Bryan} G.~L.,   {Abel} T.,  2001, \mn@doi [\apj]
  {10.1086/319014}, \href
  {https://ui.adsabs.harvard.edu/abs/2001ApJ...548..509M} {548, 509}

\bibitem[\protect\citeauthoryear{{Madau}, {Meiksin}  \& {Rees}}{{Madau}
  et~al.}{1997}]{Madau_1997}
{Madau} P.,  {Meiksin} A.,   {Rees} M.~J.,  1997, \mn@doi [\apj]
  {10.1086/303549}, \href
  {https://ui.adsabs.harvard.edu/abs/1997ApJ...475..429M} {475, 429}

\bibitem[\protect\citeauthoryear{{Magg}, {Klessen}, {Glover}  \& {Li}}{{Magg}
  et~al.}{2019}]{Magg_2019}
{Magg} M.,  {Klessen} R.~S.,  {Glover} S. C.~O.,   {Li} H.,  2019, \mn@doi
  [\mnras] {10.1093/mnras/stz1210}, \href
  {https://ui.adsabs.harvard.edu/abs/2019MNRAS.487..486M} {487, 486}

\bibitem[\protect\citeauthoryear{{Magg} et~al.,}{{Magg}
  et~al.}{2022}]{Magg_2021}
{Magg} M.,  et~al., 2022, \mn@doi [\mnras] {10.1093/mnras/stac1664}, \href
  {https://ui.adsabs.harvard.edu/abs/2022MNRAS.514.4433M} {514, 4433}

\bibitem[\protect\citeauthoryear{{Marigo}, {Girardi}, {Chiosi}  \&
  {Wood}}{{Marigo} et~al.}{2001}]{Marigo_2001}
{Marigo} P.,  {Girardi} L.,  {Chiosi} C.,   {Wood} P.~R.,  2001, \mn@doi [\aap]
  {10.1051/0004-6361:20010309}, \href
  {https://ui.adsabs.harvard.edu/abs/2001A&A...371..152M} {371, 152}

\bibitem[\protect\citeauthoryear{{Marigo}, {Chiosi}  \& {Kudritzki}}{{Marigo}
  et~al.}{2003}]{Marigo_2003}
{Marigo} P.,  {Chiosi} C.,   {Kudritzki} R.~P.,  2003, \mn@doi [\aap]
  {10.1051/0004-6361:20021756}, \href
  {https://ui.adsabs.harvard.edu/abs/2003A&A...399..617M} {399, 617}

\bibitem[\protect\citeauthoryear{{Mebane}, {Mirocha}  \& {Furlanetto}}{{Mebane}
  et~al.}{2018}]{Mebane_2018}
{Mebane} R.~H.,  {Mirocha} J.,   {Furlanetto} S.~R.,  2018, \mn@doi [\mnras]
  {10.1093/mnras/sty1833}, \href
  {https://ui.adsabs.harvard.edu/abs/2018MNRAS.479.4544M} {479, 4544}

\bibitem[\protect\citeauthoryear{{Mebane}, {Mirocha}  \& {Furlanetto}}{{Mebane}
  et~al.}{2020}]{Mebane:2020}
{Mebane} R.~H.,  {Mirocha} J.,   {Furlanetto} S.~R.,  2020, \mn@doi [\mnras]
  {10.1093/mnras/staa280}, \href
  {https://ui.adsabs.harvard.edu/abs/2020MNRAS.493.1217M} {493, 1217}

\bibitem[\protect\citeauthoryear{{Mertens} et~al.,}{{Mertens}
  et~al.}{2020}]{LOFAR}
{Mertens} F.~G.,  et~al., 2020, \mn@doi [\mnras] {10.1093/mnras/staa327}, \href
  {https://ui.adsabs.harvard.edu/abs/2020MNRAS.493.1662M} {493, 1662}

\bibitem[\protect\citeauthoryear{{Mertens}, {Semelin}  \& {Koopmans}}{{Mertens}
  et~al.}{2021}]{NenuFar}
{Mertens} F.~G.,  {Semelin} B.,   {Koopmans} L.~V.~E.,  2021, in {Siebert} A.,
  et~al., eds, SF2A-2021: Proceedings of the Annual meeting of the French
  Society of Astronomy and Astrophysics. pp 211--214 (\mn@eprint {arXiv}
  {2109.10055})

\bibitem[\protect\citeauthoryear{{Mesinger}}{{Mesinger}}{2019}]{Mesinger_2019}
{Mesinger} A.,  2019, {The Cosmic 21-cm Revolution; Charting the first billion
  years of our universe}, \mn@doi{10.1088/2514-3433/ab4a73.
}

\bibitem[\protect\citeauthoryear{{Mesinger}, {Furlanetto}  \& {Cen}}{{Mesinger}
  et~al.}{2011}]{Mesinger_2011}
{Mesinger} A.,  {Furlanetto} S.,   {Cen} R.,  2011, \mn@doi [\mnras]
  {10.1111/j.1365-2966.2010.17731.x}, \href
  {https://ui.adsabs.harvard.edu/abs/2011MNRAS.411..955M} {411, 955}

\bibitem[\protect\citeauthoryear{{Mirocha} \& {Furlanetto}}{{Mirocha} \&
  {Furlanetto}}{2019}]{Mirocha_2019}
{Mirocha} J.,  {Furlanetto} S.~R.,  2019, \mn@doi [\mnras]
  {10.1093/mnras/sty3260}, \href
  {https://ui.adsabs.harvard.edu/abs/2019MNRAS.483.1980M} {483, 1980}

\bibitem[\protect\citeauthoryear{{Mirocha}, {Mebane}, {Furlanetto}, {Singal}
  \& {Trinh}}{{Mirocha} et~al.}{2018}]{Mirocha_2018}
{Mirocha} J.,  {Mebane} R.~H.,  {Furlanetto} S.~R.,  {Singal} K.,   {Trinh} D.,
   2018, \mn@doi [\mnras] {10.1093/mnras/sty1388}, \href
  {https://ui.adsabs.harvard.edu/abs/2018MNRAS.478.5591M} {478, 5591}

\bibitem[\protect\citeauthoryear{{Mirouh}, {Hendriks}, {Dykes}, {Moe}  \&
  {Izzard}}{{Mirouh} et~al.}{submitted}]{mirouh_etal22}
{Mirouh} G.~M.,  {Hendriks} D.~D.,  {Dykes} S.,  {Moe} M.,   {Izzard} R.~G.,
  submitted, \mnras

\bibitem[\protect\citeauthoryear{{Mittal} \& {Kulkarni}}{{Mittal} \&
  {Kulkarni}}{2021}]{Mittal_2021}
{Mittal} S.,  {Kulkarni} G.,  2021, \mn@doi [\mnras] {10.1093/mnras/staa3811},
  \href {https://ui.adsabs.harvard.edu/abs/2021MNRAS.503.4264M} {503, 4264}

\bibitem[\protect\citeauthoryear{{Mu{\~n}oz}, {Dvorkin}  \& {Loeb}}{{Mu{\~n}oz}
  et~al.}{2018}]{Munoz_2018}
{Mu{\~n}oz} J.~B.,  {Dvorkin} C.,   {Loeb} A.,  2018, \mn@doi [\prl]
  {10.1103/PhysRevLett.121.121301}, \href
  {https://ui.adsabs.harvard.edu/abs/2018PhRvL.121l1301M} {121, 121301}

\bibitem[\protect\citeauthoryear{{Mu{\~n}oz}, {Qin}, {Mesinger}, {Murray},
  {Greig}  \& {Mason}}{{Mu{\~n}oz} et~al.}{2022}]{Munoz_2021}
{Mu{\~n}oz} J.~B.,  {Qin} Y.,  {Mesinger} A.,  {Murray} S.~G.,  {Greig} B.,
  {Mason} C.,  2022, \mn@doi [\mnras] {10.1093/mnras/stac185}, \href
  {https://ui.adsabs.harvard.edu/abs/2022MNRAS.511.3657M} {511, 3657}

\bibitem[\protect\citeauthoryear{{Murphy} et~al.,}{{Murphy}
  et~al.}{2021a}]{lmurphy21}
{Murphy} L.~J.,  et~al., 2021a, \mn@doi [\mnras] {10.1093/mnras/staa3803},
  \href {https://ui.adsabs.harvard.edu/abs/2021MNRAS.501.2745M} {501, 2745}

\bibitem[\protect\citeauthoryear{{Murphy}, {Groh}, {Farrell}, {Meynet},
  {Ekstr{\"o}m}, {Tsiatsiou}, {Hackett}  \& {Martinet}}{{Murphy}
  et~al.}{2021b}]{Murphy_2021}
{Murphy} L.~J.,  {Groh} J.~H.,  {Farrell} E.,  {Meynet} G.,  {Ekstr{\"o}m} S.,
  {Tsiatsiou} S.,  {Hackett} A.,   {Martinet} S.,  2021b, \mn@doi [\mnras]
  {10.1093/mnras/stab2073}, \href
  {https://ui.adsabs.harvard.edu/abs/2021MNRAS.506.5731M} {506, 5731}

\bibitem[\protect\citeauthoryear{{Pacucci}, {Mesinger}, {Mineo}  \&
  {Ferrara}}{{Pacucci} et~al.}{2014}]{Pacucci:2014}
{Pacucci} F.,  {Mesinger} A.,  {Mineo} S.,   {Ferrara} A.,  2014, \mn@doi
  [\mnras] {10.1093/mnras/stu1240}, \href
  {https://ui.adsabs.harvard.edu/abs/2014MNRAS.443..678P} {443, 678}

\bibitem[\protect\citeauthoryear{{Pallottini}, {Ferrara}, {Gallerani},
  {Salvadori}  \& {D'Odorico}}{{Pallottini} et~al.}{2014}]{Pallottini_2014}
{Pallottini} A.,  {Ferrara} A.,  {Gallerani} S.,  {Salvadori} S.,   {D'Odorico}
  V.,  2014, \mn@doi [\mnras] {10.1093/mnras/stu451}, \href
  {https://ui.adsabs.harvard.edu/abs/2014MNRAS.440.2498P} {440, 2498}

\bibitem[\protect\citeauthoryear{{Park}, {Gillet}, {Mesinger}  \&
  {Greig}}{{Park} et~al.}{2020}]{Park:2020}
{Park} J.,  {Gillet} N.,  {Mesinger} A.,   {Greig} B.,  2020, \mn@doi [\mnras]
  {10.1093/mnras/stz3278}, \href
  {https://ui.adsabs.harvard.edu/abs/2020MNRAS.491.3891P} {491, 3891}

\bibitem[\protect\citeauthoryear{{Paxton} et~al.,}{{Paxton}
  et~al.}{2019}]{mesa5}
{Paxton} B.,  et~al., 2019, \mn@doi [\apjs] {10.3847/1538-4365/ab2241}, 243, 10

\bibitem[\protect\citeauthoryear{{Philip} et~al.,}{{Philip}
  et~al.}{2019}]{PRIZM}
{Philip} L.,  et~al., 2019, \mn@doi [Journal of Astronomical Instrumentation]
  {10.1142/S2251171719500041}, \href
  {https://ui.adsabs.harvard.edu/abs/2019JAI.....850004P} {8, 1950004}

\bibitem[\protect\citeauthoryear{{Planck Collaboration} et~al.,}{{Planck
  Collaboration} et~al.}{2020}]{Planck_VI}
{Planck Collaboration} et~al., 2020, \mn@doi [\aap]
  {10.1051/0004-6361/201833910}, \href
  {https://ui.adsabs.harvard.edu/abs/2020A&A...641A...6P} {641, A6}

\bibitem[\protect\citeauthoryear{{Press} \& {Schechter}}{{Press} \&
  {Schechter}}{1974}]{Press_1974}
{Press} W.~H.,  {Schechter} P.,  1974, \mn@doi [\apj] {10.1086/152650}, \href
  {https://ui.adsabs.harvard.edu/abs/1974ApJ...187..425P} {187, 425}

\bibitem[\protect\citeauthoryear{{Pritchard} \& {Furlanetto}}{{Pritchard} \&
  {Furlanetto}}{2006}]{Pritchard_2006}
{Pritchard} J.~R.,  {Furlanetto} S.~R.,  2006, \mn@doi [\mnras]
  {10.1111/j.1365-2966.2006.10028.x}, \href
  {https://ui.adsabs.harvard.edu/abs/2006MNRAS.367.1057P} {367, 1057}

\bibitem[\protect\citeauthoryear{{Pritchard} \& {Loeb}}{{Pritchard} \&
  {Loeb}}{2012}]{Pritchard_2012}
{Pritchard} J.~R.,  {Loeb} A.,  2012, \mn@doi [Reports on Progress in Physics]
  {10.1088/0034-4885/75/8/086901}, \href
  {https://ui.adsabs.harvard.edu/abs/2012RPPh...75h6901P} {75, 086901}

\bibitem[\protect\citeauthoryear{{Rahimi} et~al.,}{{Rahimi}
  et~al.}{2021}]{Rahimi:2021}
{Rahimi} M.,  et~al., 2021, \mn@doi [\mnras] {10.1093/mnras/stab2918}, \href
  {https://ui.adsabs.harvard.edu/abs/2021MNRAS.508.5954R} {508, 5954}

\bibitem[\protect\citeauthoryear{{Reis}, {Fialkov}  \& {Barkana}}{{Reis}
  et~al.}{2020}]{Reis_20}
{Reis} I.,  {Fialkov} A.,   {Barkana} R.,  2020, \mn@doi [\mnras]
  {10.1093/mnras/staa3091}, \href
  {https://ui.adsabs.harvard.edu/abs/2020MNRAS.499.5993R} {499, 5993}

\bibitem[\protect\citeauthoryear{{Reis}, {Fialkov}  \& {Barkana}}{{Reis}
  et~al.}{2021}]{Reis_2021}
{Reis} I.,  {Fialkov} A.,   {Barkana} R.,  2021, \mn@doi [\mnras]
  {10.1093/mnras/stab2089}, \href
  {https://ui.adsabs.harvard.edu/abs/2021MNRAS.506.5479R} {506, 5479}

\bibitem[\protect\citeauthoryear{{Reis}, {Barkana}  \& {Fialkov}}{{Reis}
  et~al.}{2022}]{Reis_2021b}
{Reis} I.,  {Barkana} R.,   {Fialkov} A.,  2022, \mn@doi [\mnras]
  {10.1093/mnras/stac411}, \href
  {https://ui.adsabs.harvard.edu/abs/2022MNRAS.511.5265R} {511, 5265}

\bibitem[\protect\citeauthoryear{{Ricotti} \& {Ostriker}}{{Ricotti} \&
  {Ostriker}}{2004}]{Ricotti_2004}
{Ricotti} M.,  {Ostriker} J.~P.,  2004, \mn@doi [\mnras]
  {10.1111/j.1365-2966.2004.07942.x}, \href
  {https://ui.adsabs.harvard.edu/abs/2004MNRAS.352..547R} {352, 547}

\bibitem[\protect\citeauthoryear{{Ricotti}, {Ostriker}  \& {Gnedin}}{{Ricotti}
  et~al.}{2005}]{Ricotti_2005}
{Ricotti} M.,  {Ostriker} J.~P.,   {Gnedin} N.~Y.,  2005, \mn@doi [\mnras]
  {10.1111/j.1365-2966.2004.08623.x}, \href
  {https://ui.adsabs.harvard.edu/abs/2005MNRAS.357..207R} {357, 207}

\bibitem[\protect\citeauthoryear{{Salpeter}}{{Salpeter}}{1955}]{Salpeter_1955}
{Salpeter} E.~E.,  1955, \mn@doi [\apj] {10.1086/145971}, \href
  {https://ui.adsabs.harvard.edu/abs/1955ApJ...121..161S} {121, 161}

\bibitem[\protect\citeauthoryear{{Santos}, {Silva}, {Pritchard}, {Cen}  \&
  {Cooray}}{{Santos} et~al.}{2011}]{Santos_2011}
{Santos} M.~G.,  {Silva} M.~B.,  {Pritchard} J.~R.,  {Cen} R.,   {Cooray} A.,
  2011, \mn@doi [\aap] {10.1051/0004-6361/201015695}, \href
  {https://ui.adsabs.harvard.edu/abs/2011A&A...527A..93S} {527, A93}

\bibitem[\protect\citeauthoryear{{Sarmento}, {Scannapieco}  \&
  {Cohen}}{{Sarmento} et~al.}{2018}]{Sarmento_2018}
{Sarmento} R.,  {Scannapieco} E.,   {Cohen} S.,  2018, \mn@doi [\apj]
  {10.3847/1538-4357/aa989a}, \href
  {https://ui.adsabs.harvard.edu/abs/2018ApJ...854...75S} {854, 75}

\bibitem[\protect\citeauthoryear{{Scalo}}{{Scalo}}{1998}]{Scalo_1998}
{Scalo} J.,  1998, in {Gilmore} G.,  {Howell} D.,  eds,  Astronomical Society
  of the Pacific Conference Series Vol. 142, The Stellar Initial Mass Function
  (38th Herstmonceux Conference). p.~201 (\mn@eprint {arXiv}
  {astro-ph/9712317})

\bibitem[\protect\citeauthoryear{{Schaerer}}{{Schaerer}}{2002}]{Schaerer_2002}
{Schaerer} D.,  2002, \mn@doi [\aap] {10.1051/0004-6361:20011619}, \href
  {https://ui.adsabs.harvard.edu/abs/2002A&A...382...28S} {382, 28}

\bibitem[\protect\citeauthoryear{{Schauer}, {Liu}  \& {Bromm}}{{Schauer}
  et~al.}{2019}]{Schauer_2019}
{Schauer} A. T.~P.,  {Liu} B.,   {Bromm} V.,  2019, \mn@doi [\apjl]
  {10.3847/2041-8213/ab1e51}, \href
  {https://ui.adsabs.harvard.edu/abs/2019ApJ...877L...5S} {877, L5}

\bibitem[\protect\citeauthoryear{{Scott} \& {Rees}}{{Scott} \&
  {Rees}}{1990}]{Scott_1990}
{Scott} D.,  {Rees} M.~J.,  1990, \mnras, \href
  {https://ui.adsabs.harvard.edu/abs/1990MNRAS.247..510S} {247, 510}

\bibitem[\protect\citeauthoryear{{Sharda}, {Federrath}  \& {Krumholz}}{{Sharda}
  et~al.}{2020}]{Sharda_2020}
{Sharda} P.,  {Federrath} C.,   {Krumholz} M.~R.,  2020, \mn@doi [\mnras]
  {10.1093/mnras/staa1926}, \href
  {https://ui.adsabs.harvard.edu/abs/2020MNRAS.497..336S} {497, 336}

\bibitem[\protect\citeauthoryear{{Sharda}, {Federrath}, {Krumholz}  \&
  {Schleicher}}{{Sharda} et~al.}{2021}]{Sharda_2021}
{Sharda} P.,  {Federrath} C.,  {Krumholz} M.~R.,   {Schleicher} D. R.~G.,
  2021, \mn@doi [\mnras] {10.1093/mnras/stab531}, \href
  {https://ui.adsabs.harvard.edu/abs/2021MNRAS.503.2014S} {503, 2014}

\bibitem[\protect\citeauthoryear{{Sheth} \& {Tormen}}{{Sheth} \&
  {Tormen}}{1999}]{Sheth_1999}
{Sheth} R.~K.,  {Tormen} G.,  1999, \mn@doi [\mnras]
  {10.1046/j.1365-8711.1999.02692.x}, \href
  {https://ui.adsabs.harvard.edu/abs/1999MNRAS.308..119S} {308, 119}

\bibitem[\protect\citeauthoryear{{Sims} \& {Pober}}{{Sims} \&
  {Pober}}{2020}]{Sims_2020}
{Sims} P.~H.,  {Pober} J.~C.,  2020, \mn@doi [\mnras] {10.1093/mnras/stz3388},
  \href {https://ui.adsabs.harvard.edu/abs/2020MNRAS.492...22S} {492, 22}

\bibitem[\protect\citeauthoryear{{Singh} \& {Subrahmanyan}}{{Singh} \&
  {Subrahmanyan}}{2019}]{Singh_2019}
{Singh} S.,  {Subrahmanyan} R.,  2019, \mn@doi [\apj]
  {10.3847/1538-4357/ab2879}, \href
  {https://ui.adsabs.harvard.edu/abs/2019ApJ...880...26S} {880, 26}

\bibitem[\protect\citeauthoryear{{Singh} et~al.,}{{Singh}
  et~al.}{2018}]{SARAS2}
{Singh} S.,  et~al., 2018, \mn@doi [\apj] {10.3847/1538-4357/aabae1}, \href
  {https://ui.adsabs.harvard.edu/abs/2018ApJ...858...54S} {858, 54}

\bibitem[\protect\citeauthoryear{{Singh} et~al.,}{{Singh}
  et~al.}{2022}]{SARAS3}
{Singh} S.,  et~al., 2022, \mn@doi [Nature Astronomy]
  {10.1038/s41550-022-01610-5}, \href
  {https://ui.adsabs.harvard.edu/abs/2022NatAs...6..607S} {6, 607}

\bibitem[\protect\citeauthoryear{{Smith}, {Wise}, {O'Shea}, {Norman}  \&
  {Khochfar}}{{Smith} et~al.}{2015}]{Smith_2015}
{Smith} B.~D.,  {Wise} J.~H.,  {O'Shea} B.~W.,  {Norman} M.~L.,   {Khochfar}
  S.,  2015, \mn@doi [\mnras] {10.1093/mnras/stv1509}, \href
  {https://ui.adsabs.harvard.edu/abs/2015MNRAS.452.2822S} {452, 2822}

\bibitem[\protect\citeauthoryear{{Stacy}}{{Stacy}}{2013}]{Stacy_2013}
{Stacy} A.,  2013, in American Astronomical Society Meeting Abstracts \#221. p.
  442.02

\bibitem[\protect\citeauthoryear{{Stacy}, {Greif}  \& {Bromm}}{{Stacy}
  et~al.}{2010}]{Stacy_2010}
{Stacy} A.,  {Greif} T.~H.,   {Bromm} V.,  2010, \mn@doi [\mnras]
  {10.1111/j.1365-2966.2009.16113.x}, \href
  {https://ui.adsabs.harvard.edu/abs/2010MNRAS.403...45S} {403, 45}

\bibitem[\protect\citeauthoryear{{Stacy}, {Bromm}  \& {Lee}}{{Stacy}
  et~al.}{2016}]{Stacy_2016}
{Stacy} A.,  {Bromm} V.,   {Lee} A.~T.,  2016, \mn@doi [\mnras]
  {10.1093/mnras/stw1728}, \href
  {https://ui.adsabs.harvard.edu/abs/2016MNRAS.462.1307S} {462, 1307}

\bibitem[\protect\citeauthoryear{{Stecher} \& {Williams}}{{Stecher} \&
  {Williams}}{1967}]{Stecher_1967}
{Stecher} T.~P.,  {Williams} D.~A.,  1967, \mn@doi [\apjl] {10.1086/180047},
  \href {https://ui.adsabs.harvard.edu/abs/1967ApJ...149L..29S} {149, L29}

\bibitem[\protect\citeauthoryear{{Sugimura}, {Matsumoto}, {Hosokawa}, {Hirano}
  \& {Omukai}}{{Sugimura} et~al.}{2020}]{Sugimura_2020}
{Sugimura} K.,  {Matsumoto} T.,  {Hosokawa} T.,  {Hirano} S.,   {Omukai} K.,
  2020, \mn@doi [\apjl] {10.3847/2041-8213/ab7d37}, \href
  {https://ui.adsabs.harvard.edu/abs/2020ApJ...892L..14S} {892, L14}

\bibitem[\protect\citeauthoryear{{Susa}}{{Susa}}{2019}]{Susa_2019}
{Susa} H.,  2019, \mn@doi [\apj] {10.3847/1538-4357/ab1b6f}, \href
  {https://ui.adsabs.harvard.edu/abs/2019ApJ...877...99S} {877, 99}

\bibitem[\protect\citeauthoryear{{Tanaka} \& {Hasegawa}}{{Tanaka} \&
  {Hasegawa}}{2021}]{Tanaka_2021}
{Tanaka} T.,  {Hasegawa} K.,  2021, \mn@doi [\mnras] {10.1093/mnras/stab072},
  \href {https://ui.adsabs.harvard.edu/abs/2021MNRAS.502..463T} {502, 463}

\bibitem[\protect\citeauthoryear{{Tanaka}, {Hasegawa}, {Yajima}, {Kobayashi}
  \& {Sugiyama}}{{Tanaka} et~al.}{2018}]{Tanaka_2018}
{Tanaka} T.,  {Hasegawa} K.,  {Yajima} H.,  {Kobayashi} M. I.~N.,   {Sugiyama}
  N.,  2018, \mn@doi [\mnras] {10.1093/mnras/sty1967}, \href
  {https://ui.adsabs.harvard.edu/abs/2018MNRAS.480.1925T} {480, 1925}

\bibitem[\protect\citeauthoryear{{Tarumi}, {Hartwig}  \& {Magg}}{{Tarumi}
  et~al.}{2020}]{Tarumi_2020}
{Tarumi} Y.,  {Hartwig} T.,   {Magg} M.,  2020, \mn@doi [\apj]
  {10.3847/1538-4357/ab960d}, \href
  {https://ui.adsabs.harvard.edu/abs/2020ApJ...897...58T} {897, 58}

\bibitem[\protect\citeauthoryear{{Tornatore}, {Ferrara}  \&
  {Schneider}}{{Tornatore} et~al.}{2007}]{Tornatore_2007}
{Tornatore} L.,  {Ferrara} A.,   {Schneider} R.,  2007, \mn@doi [\mnras]
  {10.1111/j.1365-2966.2007.12215.x}, \href
  {https://ui.adsabs.harvard.edu/abs/2007MNRAS.382..945T} {382, 945}

\bibitem[\protect\citeauthoryear{{Tseliakhovich} \& {Hirata}}{{Tseliakhovich}
  \& {Hirata}}{2010}]{Tseliakhovich_2010}
{Tseliakhovich} D.,  {Hirata} C.,  2010, \mn@doi [\prd]
  {10.1103/PhysRevD.82.083520}, \href
  {https://ui.adsabs.harvard.edu/abs/2010PhRvD..82h3520T} {82, 083520}

\bibitem[\protect\citeauthoryear{{Venumadhav}, {Dai}, {Kaurov}  \&
  {Zaldarriaga}}{{Venumadhav} et~al.}{2018}]{Venumadhav_2018}
{Venumadhav} T.,  {Dai} L.,  {Kaurov} A.,   {Zaldarriaga} M.,  2018, \mn@doi
  [\prd] {10.1103/PhysRevD.98.103513}, \href
  {https://ui.adsabs.harvard.edu/abs/2018PhRvD..98j3513V} {98, 103513}

\bibitem[\protect\citeauthoryear{{Visbal}, {Barkana}, {Fialkov},
  {Tseliakhovich}  \& {Hirata}}{{Visbal} et~al.}{2012}]{Visbal_2012}
{Visbal} E.,  {Barkana} R.,  {Fialkov} A.,  {Tseliakhovich} D.,   {Hirata}
  C.~M.,  2012, \mn@doi [\nat] {10.1038/nature11177}, \href
  {https://ui.adsabs.harvard.edu/abs/2012Natur.487...70V} {487, 70}

\bibitem[\protect\citeauthoryear{{Visbal}, {Haiman}  \& {Bryan}}{{Visbal}
  et~al.}{2018}]{Visbal_2018}
{Visbal} E.,  {Haiman} Z.,   {Bryan} G.~L.,  2018, \mn@doi [\mnras]
  {10.1093/mnras/sty142}, \href
  {https://ui.adsabs.harvard.edu/abs/2018MNRAS.475.5246V} {475, 5246}

\bibitem[\protect\citeauthoryear{{Windhorst} et~al.,}{{Windhorst}
  et~al.}{2018}]{Windhorst_2018}
{Windhorst} R.~A.,  et~al., 2018, \mn@doi [\apjs] {10.3847/1538-4365/aaa760},
  \href {https://ui.adsabs.harvard.edu/abs/2018ApJS..234...41W} {234, 41}

\bibitem[\protect\citeauthoryear{{Wise}, {Turk}, {Norman}  \& {Abel}}{{Wise}
  et~al.}{2012}]{Wise_2012}
{Wise} J.~H.,  {Turk} M.~J.,  {Norman} M.~L.,   {Abel} T.,  2012, \mn@doi
  [\apj] {10.1088/0004-637X/745/1/50}, \href
  {https://ui.adsabs.harvard.edu/abs/2012ApJ...745...50W} {745, 50}

\bibitem[\protect\citeauthoryear{{Wollenberg}, {Glover}, {Clark}  \&
  {Klessen}}{{Wollenberg} et~al.}{2020}]{Wollenberg_2020}
{Wollenberg} K. M.~J.,  {Glover} S. C.~O.,  {Clark} P.~C.,   {Klessen} R.~S.,
  2020, \mn@doi [\mnras] {10.1093/mnras/staa289}, \href
  {https://ui.adsabs.harvard.edu/abs/2020MNRAS.494.1871W} {494, 1871}

\bibitem[\protect\citeauthoryear{{Woosley} \& {Heger}}{{Woosley} \&
  {Heger}}{2007}]{Woosley_2007}
{Woosley} S.~E.,  {Heger} A.,  2007, \mn@doi [\physrep]
  {10.1016/j.physrep.2007.02.009}, \href
  {https://ui.adsabs.harvard.edu/abs/2007PhR...442..269W} {442, 269}

\bibitem[\protect\citeauthoryear{{Wouthuysen}}{{Wouthuysen}}{1952}]{Wouthuysen_1952}
{Wouthuysen} S.~A.,  1952, \mn@doi [\aj] {10.1086/106661}, \href
  {https://ui.adsabs.harvard.edu/abs/1952AJ.....57R..31W} {57, 31}

\bibitem[\protect\citeauthoryear{{Wright}}{{Wright}}{2006}]{Wright_2006}
{Wright} E.~L.,  2006, \mn@doi [\pasp] {10.1086/510102}, \href
  {https://ui.adsabs.harvard.edu/abs/2006PASP..118.1711W} {118, 1711}

\bibitem[\protect\citeauthoryear{{Xu}, {Ahn}, {Norman}, {Wise}  \&
  {O'Shea}}{{Xu} et~al.}{2016}]{Xu_2016}
{Xu} H.,  {Ahn} K.,  {Norman} M.~L.,  {Wise} J.~H.,   {O'Shea} B.~W.,  2016,
  \mn@doi [\apjl] {10.3847/2041-8205/832/1/L5}, \href
  {https://ui.adsabs.harvard.edu/abs/2016ApJ...832L...5X} {832, L5}

\bibitem[\protect\citeauthoryear{{Yajima} \& {Khochfar}}{{Yajima} \&
  {Khochfar}}{2015}]{Yajima_2015}
{Yajima} H.,  {Khochfar} S.,  2015, \mn@doi [\mnras] {10.1093/mnras/stu2687},
  \href {https://ui.adsabs.harvard.edu/abs/2015MNRAS.448..654Y} {448, 654}

\bibitem[\protect\citeauthoryear{{Yoshida}, {Bromm}  \& {Hernquist}}{{Yoshida}
  et~al.}{2004}]{Yoshida_2004}
{Yoshida} N.,  {Bromm} V.,   {Hernquist} L.,  2004, \mn@doi [\apj]
  {10.1086/382499}, \href
  {https://ui.adsabs.harvard.edu/abs/2004ApJ...605..579Y} {605, 579}

\bibitem[\protect\citeauthoryear{{de Bennassuti}, {Schneider}, {Valiante}  \&
  {Salvadori}}{{de Bennassuti} et~al.}{2014}]{de_Bennassuti_2014}
{de Bennassuti} M.,  {Schneider} R.,  {Valiante} R.,   {Salvadori} S.,  2014,
  \mn@doi [\mnras] {10.1093/mnras/stu1962}, \href
  {https://ui.adsabs.harvard.edu/abs/2014MNRAS.445.3039D} {445, 3039}

\bibitem[\protect\citeauthoryear{{de Bennassuti}, {Salvadori}, {Schneider},
  {Valiante}  \& {Omukai}}{{de Bennassuti} et~al.}{2017}]{de_Bennassuti_2017}
{de Bennassuti} M.,  {Salvadori} S.,  {Schneider} R.,  {Valiante} R.,
  {Omukai} K.,  2017, \mn@doi [\mnras] {10.1093/mnras/stw2687}, \href
  {https://ui.adsabs.harvard.edu/abs/2017MNRAS.465..926D} {465, 926}

\bibitem[\protect\citeauthoryear{{de Lera Acedo} et~al.,}{{de Lera Acedo}
  et~al.}{2022}]{REACH}
{de Lera Acedo} E.,  et~al., 2022, \mn@doi [Nature Astronomy]
  {10.1038/s41550-022-01709-9}

\makeatother
\end{thebibliography}



%
%

\appendix

\section{Variation of Lyman-Band Emission with Stellar Mass}\label{app:app}

In the main body of this paper, we present our key findings concerning the Lyman-band emission of individual Pop~III stars. Here we continue this discussion in more detail, and show how the emission rate of these stars changes with time and exploring the variation in Lyman-band spectra between stars of different masses. 

\begin{figure}
    \centering
    \includegraphics{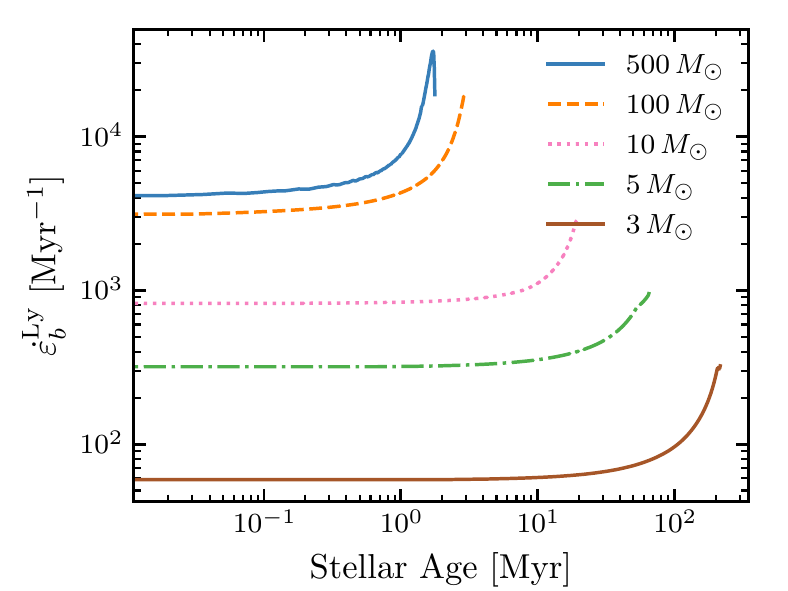}
    \caption{Lyman-band photon emission rate per stellar baryon for five Pop~III stars of different masses (as indicated in the legend). As the area of the star increases with stellar age, so does the emission rate, as can be seen in the figure. The increase in emission between zero-age main sequence and hydrogen depletion of the star is larger than a factor of 5 in some cases. The $500$\,M$_{\astrosun}$ star shows a peak and a sharp drop in the emission rate in the final period of its life. This sharp drop is believed to be due to the rapid decrease in the effective temperatures of such massive stars as they approach photoevaporation. The low temperature in the stellar atmosphere exponentially suppresses the Lyman flux of the star, resulting in the observed drop in its overall Lyman emission rate.}
    \label{fig:emission_over_lives}
\end{figure}

In Fig.~\ref{fig:emission_over_lives} we show the rate of emission of Lyman-band photons for five Pop~III stars of different masses. We see that the emission rate sharply rises with stellar age. In the explored scenarios, the increase in the emission rate is driven by the increasing area of the stars as they expand when approaching hydrogen depletion or photoevaporation. For some stellar masses the increase is quite large, with the maximum emission rate exceeding 5 times the emission rate at ZAMS (i.e., the emission rate at the beginning of stellar life). The simultaneous decrease in effective temperature (and, hence, reduced Lyman-band flux) seen in these stars as they expand is subdominant to the impacts of the increasing stellar area. Thus, the overall trend of increasing Lyman-band emission is observed. 

The one exception to the overall observed increase in the emission rate with stellar age is the sharp drop seen for the most massive stars ($\geq 310$\,M$_{\astrosun}$) at the end of their lives. For the $500$\,M$_{\astrosun}$ star, illustrated in the figure, the Lyman-band emission rate decreases from its peak to 52 per cent of that value in the final 53,000 years before the star begins to photoevaporate. As the surface area of this massive star is still increasing during this emission drop, the observed drop must be caused by a strong decrease in the Lyman flux of the star. This occurs due to the effective temperature of the star becoming low enough that all layers of the stellar atmosphere that are contributing to its emergent spectra are significantly cooler than the characteristic temperature of Ly\,$\alpha$ photons. For any such layer of the stellar atmosphere the thermal Lyman-band flux is suppressed exponentially with temperature; whereas if the effective temperature is above this threshold, the flux varies linearly with $T_{\rm eff}$. Therefore, as the effective temperature of such a massive star rapidly decreases at the end of its life (as illustrated in Fig.~\ref{fig:evo_tracks}) the system enters the exponential regime, and, therefore, the stellar Lyman flux is greatly suppressed. As this suppression is exponential, it dominates the increasing stellar surface area leading to the overall sharp decline in stellar Lyman-band emission. Owing to this rapid decrease in the Lyman-band emissivity of massive stars with age, we can somewhat justify our earlier assumption that truncating stellar evolution tracks at photoevaporation will not greatly impact stellar lifetime Lyman emissivity. An improved model would incorporate this evaporation by considering stars of non-fixed mass and would thus verify whether this truncation is indeed a good approximation. 

\begin{figure}
    \centering
    \includegraphics{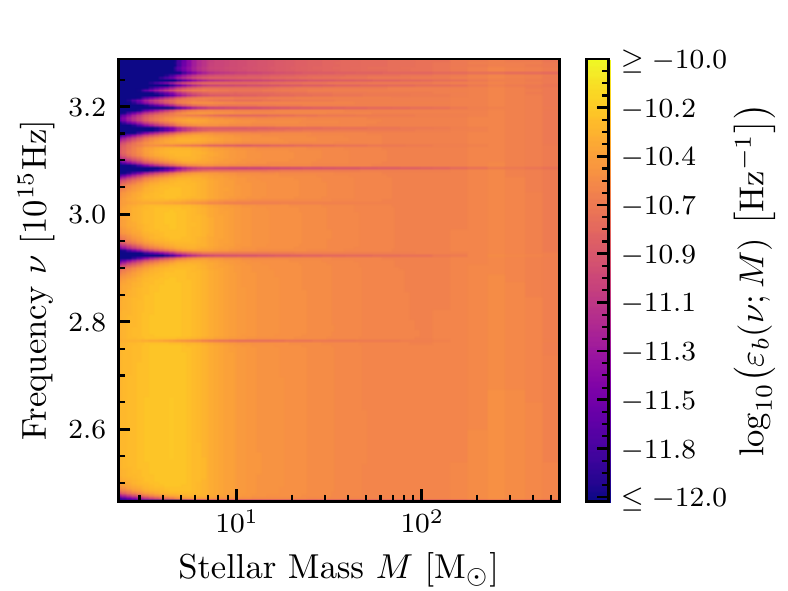}
    \caption{Lifetime photon emissivity per baryon for a range of individual Pop~III stellar masses. Higher mass Pop~III stars are seen to have flatter spectra with no prominent absorption lines, whereas lower-mass Pop~III stars show prominent H\,\textsc{i} Lyman series and He\,\textsc{ii} Balmer series absorption lines. The lightest stars $<10$\,M$_{\astrosun}$ are seen to produce few high energy Lyman-band photons.}
    \label{fig:lifetime_spectra}
\end{figure}

So far our discussions of Lyman-band emission focused mostly on the integrated Lyman-band emission but we also find interesting variations within the Lyman-band spectra. Fig.~\ref{fig:lifetime_spectra} shows $\varepsilon(\nu;M)$ against Pop~III stellar mass. The spectral shape of the Lyman-band emissivity is found to vary over the mass range. The highest mass stars ($M > 50$\,M$_{\astrosun}$) have fairly flat emission spectra, whereas the lower-mass stars ($ < 10$\,M$_{\astrosun}$) have very little emission at the highest frequencies. Higher energy Lyman-band photons play a key role in Ly\,$\alpha$ heating, as the injected photons from their cascades can cool the IGM \citep{Chuzhoy_2006, Reis_2021, Mittal_2021} reducing the overall heating effect. Hence, this suggests that the Ly\,$\alpha$ heating may be more efficient for lower-mass Pop~III stars due to their low emission of such higher energy photons, this effect is automatically taken into account in our 21-cm signal simulations.

In the spectra shown in Fig.~\ref{fig:lifetime_spectra}, a series of horizontal lines of low emission are apparent. These are absorption lines from the outer layers of the stellar atmospheres. Two separate sets of emission lines are visible, the H\,\textsc{i} Lyman lines and the He\,\textsc{ii} Balmer lines. The He\,\textsc{ii} Balmer lines are only seen between stars of $5$\,M$_{\astrosun}$ to $100$\,M$_{\astrosun}$ and are typically fainter than the H\,\textsc{i} Lyman lines. The absence of the He\,\textsc{ii} lines in the spectra of the lowest-mass stars is likely due to the low temperatures of these stars meaning they contain negligible He\,\textsc{ii} in their outer atmospheres. For the H\,\textsc{i} Lyman lines on the other hand the absorption troughs are most prominent in the lowest-mass stars being very broad and deep, with the lines getting fainter in more massive stars as the rising stellar temperatures ionize atomic hydrogen and so reduces the prevalence of H\,\textsc{i} in the outer atmospheres of the stars. 

These variations between the lifetime emission spectra of various masses of Pop~III stars are what leads to the IMF Lyman-band photon emissivities per baryon having different spectral shapes, as was previously seen in Fig.~\ref{fig:epsilons}. The Salpeter IMF has prominent Lyman absorption lines and weak high-frequency emission due to the majority of its emission coming from low-mass stars, whereas the other example IMFs have relatively flat spectra with weak absorption lines as we see is typical of high-mass Pop~III stars.

\section{Additional Instantaneous versus Non-Instantaneous Emission Comparisons}\label{app:IE_NIE}

\begin{figure*}
    \centering
    \includegraphics{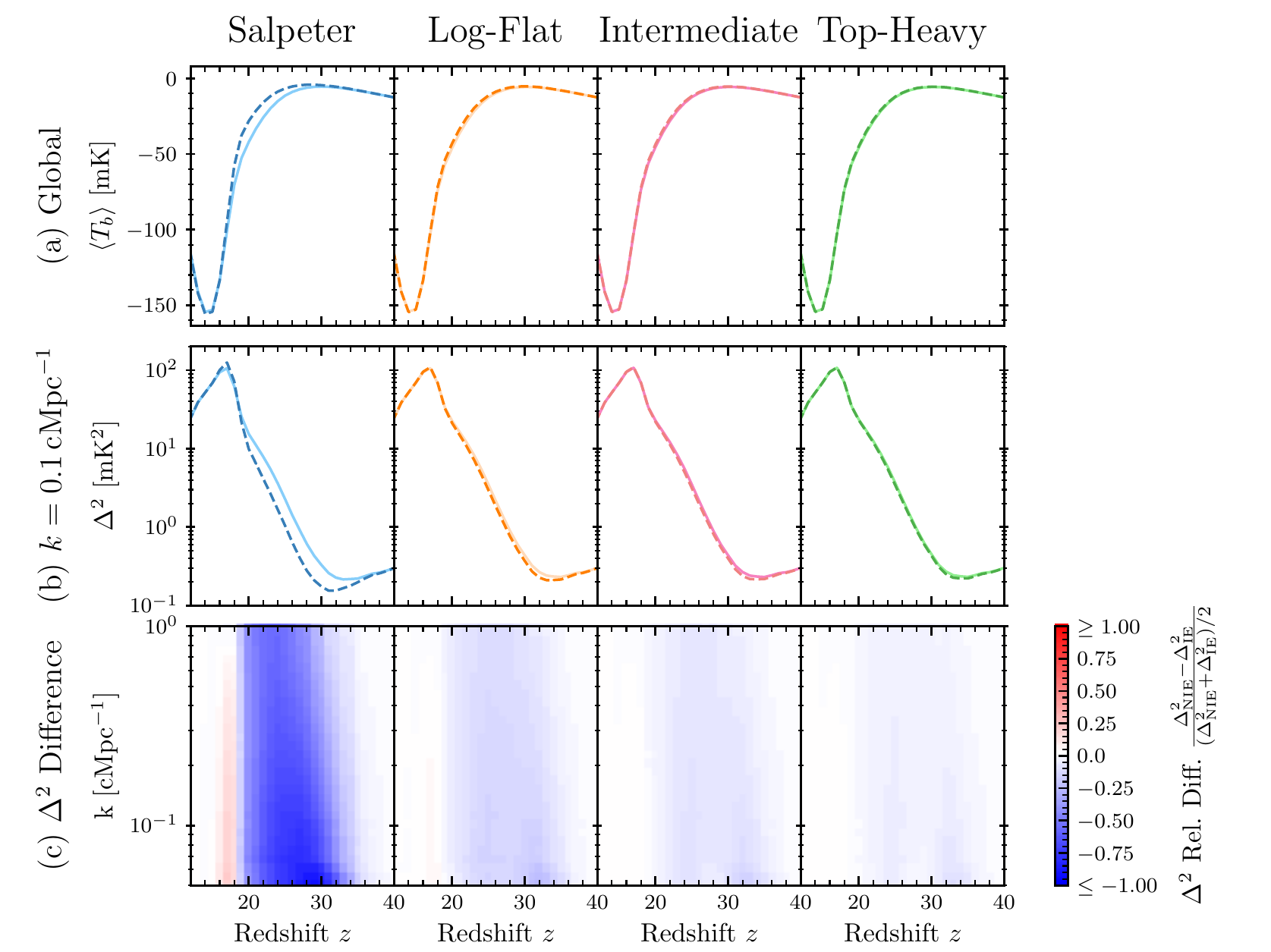}
    \caption{Comparison of the global 21-cm signal (top, a), power spectrum (middle, b), and the relative difference in the power spectra  (bottom panels, c) predicted for each of our example IMFs under the instantaneous (lighter solid line) and non-instantaneous stellar emission (darker dashed line) models. From left to right we show the results for the  Salpeter, Log-Flat, Intermediate and Top-Heavy IMFs. The largest difference between the 21-cm signals computed with and without the instantaneous emission approximation are seen for the Salpeter IMF, with the approximation having a much smaller impact on the 21-cm signals of the other IMFs. For all IMFs the relative differences in the power spectrum (bottom) are most prominent at cosmic dawn around redshift $32$ reaching 119 per cent at $z = 31$, 41 per cent at z = $33$, 31 per cent at z = $33$, and 22 per cent at z = $33$, for the Salpeter, Log-Flat, Intermediate, and Top-Heavy IMFs respectively. These modest differences suggest that simulating non-instantaneous stellar emission is necessary for accurate cosmic dawn 21-cm power spectrum predictions.}
    \label{fig:IE_NIE_comp_full}
\end{figure*}

Fig.~\ref{fig:assumptions} depicted the 21-cm signal computed for the Salpeter IMF with and without the assumption of instantaneous stellar emission. Here we provide an extended version of that figure, Fig.~\ref{fig:IE_NIE_comp_full}, which includes a similar comparison for the Log-Flat, Intermediate, and Top-Heavy IMFs alongside the previously depicted results for the Salpeter IMF. 

\bsp	
\label{lastpage}
\end{document}